\documentclass[11pt]{article}

\usepackage{amsmath}
\usepackage{amssymb}
\usepackage{amsthm}
\usepackage{amsfonts}
\usepackage{amstext}
\usepackage{amsopn}
\usepackage{amsxtra}
\usepackage[latin1]{inputenc}
\usepackage{mathrsfs}
\usepackage{dsfont}
\usepackage{graphicx}
\usepackage{xr}
\newcommand{\refI}[1]{{\rm I.}\ref{#1}}
\newcommand\R{{\ensuremath {\mathbb R} }}
\newcommand\C{{\ensuremath {\mathbb C} }}
\newcommand\N{{\ensuremath {\mathbb N} }}
\newcommand\Z{{\ensuremath {\mathbb Z} }}

\newcommand\1{{\ensuremath {\mathds 1} }}

\renewcommand\phi{\varphi}

\newcommand{\gH}{\mathfrak{H}}
\newcommand{\gS}{\mathfrak{S}}

\newcommand{\cP}{\mathcal{P}}
\newcommand{\cN}{\mathcal{N}}
\newcommand{\cD}{\mathcal{D}}
\newcommand{\cX}{\mathcal{X}}
\newcommand{\cB}{\mathcal{B}}
\newcommand{\cL}{\mathcal{L}}

\newcommand{\cF}{\mathcal F}
\newcommand{\cJ}{\mathcal J}
\newcommand{\cH}{\mathcal H}
\newcommand{\cI}{\mathcal I}
\newcommand{\cM}{\mathcal M}

\newcommand{\cR}{\mathcal R}
\newcommand{\tr}{{\rm tr}\,}
\newcommand{\cE}{\mathcal{E}}
\newcommand{\cC}{\mathcal{C}}

\newcommand{\cK}{\mathcal{K}}

\newcommand{\oE}{\overline{E}}
\newcommand{\uE}{\underline{E}}
\newcommand{\oF}{\overline{F}}
\newcommand{\uF}{\underline{F}}
\newcommand\ii{{\ensuremath {\infty}}}
\newcommand\pscal[1]{{\ensuremath{\left\langle #1 \right\rangle}}}
\newcommand{\norm}[1]{ \left| \! \left| #1 \right| \! \right| }

\newtheorem{thm}{Theorem}
\newtheorem{lemma}{Lemma}

\newtheorem{prop}[lemma]{Proposition}
\newtheorem{definition}{Definition}
\newtheorem{remark}{Remark}

\date{December 18, 2008}
\begin{document}

\title{The Thermodynamic Limit of Quantum Coulomb Systems\\ Part II. Applications}
\author{C. HAINZL, M. LEWIN and J. P. SOLOVEJ} 

\begin{center}
 \bf \Large The Thermodynamic Limit of Quantum\\ Coulomb Systems

\medskip

Part II. Applications
\end{center}

\medskip

\begin{center}
 \large Christian HAINZL$^a$, Mathieu LEWIN$^b$ \& Jan Philip SOLOVEJ$^c$
\end{center}

\medskip

\begin{center}
\small

 $^a$Department of Mathematics, UAB, Birmingham, AL 35294-1170, USA. 

\texttt{hainzl@math.uab.edu}

\medskip

$^b$CNRS \& Department of Mathematics UMR8088, University of Cergy-Pontoise, 2 avenue Adolphe Chauvin, 95302 Cergy-Pontoise Cedex, FRANCE. 

\texttt{Mathieu.Lewin@math.cnrs.fr}

\medskip

$^c$University of Copenhagen, Department of Mathematics, Universitetsparken 5, 2100 Copenhagen, DENMARK. 

\texttt{solovej@math.ku.dk}
\end{center}

\medskip

\begin{center}
 \it December 18, 2008
\end{center}

\medskip

\begin{abstract}
In a previous paper \cite{1}, we have developed a general theory of thermodynamic limits. We apply it here to three different Coulomb quantum systems, for which we prove the convergence of the free energy per unit volume.

The first system is the crystal for which the nuclei are classical particles arranged periodically in space and only the electrons are quantum particles. We recover and generalize a previous result of Fefferman. In the second example, both the nuclei and the electrons are quantum particles, submitted to a periodic magnetic field. We thereby extend a seminal result of Lieb and Lebowitz. Finally, in our last example we take again classical nuclei but optimize their position. To our knowledge such a system was never treated before.

The verification of the assumptions introduced in \cite{1} uses several tools which have been introduced before in the study of large quantum systems. In particular, an electrostatic inequality of Graf and Schenker is one main ingredient of our new approach.
\end{abstract}

\bigskip

\tableofcontents

\section*{Introduction}\addcontentsline{toc}{section}{Introduction}
In a previous paper\footnote{Equations or results with reference $n$ in the first paper \cite{1} will be denoted as ${\rm I}.n$.} \cite{1}, we have developed a general theory of thermodynamic limits. We have considered an abstract functional $E:\Omega\mapsto E(\Omega)\in\R$ defined on all open bounded subsets of $\R^3$ and given some general conditions allowing us to prove that 
\begin{equation}
 E(\Omega_n)\sim_{n\to\ii}\bar e|\Omega_n|
\label{thermo_limit}
\end{equation}
as $n\to\ii$ for all `regular' sequences $\{\Omega_n\}_n$ such that $|\Omega_n|\to\ii$.
In the present work we apply this general theory to three different Coulomb quantum systems. In all cases, we prove a behaviour similar to \eqref{thermo_limit} for both the grand canonical ground state energy and the free energy at temperature $\beta^{-1}$.

At first we consider the crystal, which we treat in details. We arrange classical nuclei periodically in space and only consider the electrons as quantum particles. The system is very rigid as the nuclei are fixed on the periodic lattice and cannot move. 

Property \eqref{thermo_limit} was proved for the crystal by Fefferman in \cite{F}. Our result is more general concerning the assumptions we put on the sequence $\{\Omega_n\}_n$. It is interesting to note that we are also able to prove the existence of the limit when the periodic lattice is perturbed locally (by adding defects or moving some nuclei) or in the Hartree-Fock approximation.

The second system that we treat is the case where nuclei are also considered as quantum particles. This model was considered by Lieb and Lebowitz in \cite{LL} who where the first to prove the existence of the thermodynamic limit for a Coulomb quantum system. Thanks to our new method, we are able to generalize their result by adding a constant (or even periodic) magnetic field. The system is then no more invariant under rotations, which was a crucial property used in the proof of \cite{LL}.

Eventually, we consider a third system where we impose again that the nuclei are classical particles but we do not fix their positions in space and rather optimize them. The existence of the thermodynamic limit for this model seems to be completely new.

\medskip

In the general framework developed in \cite{1}, we have proposed several general assumptions on the energy $E$ in order to obtain a behavior like \eqref{thermo_limit}. They were denoted by \textbf{(A1)}--\textbf{(A6)}, see the details in \cite{1}. In the three examples treated in the present paper, the difficulty in verifying these properties can vary significantly.

Assumption \textbf{(A2)} is the stability of matter
\begin{description}
\item[(A2)] $\forall\Omega,\quad E(\Omega)\geq -\kappa|\Omega|$
\end{description}
which as we have already explained in \cite{1} has been one of the main subject of investigation in the last decades, see, e.g., the reviews  \cite{Lieb1,Lieb2,l-heisen,Loss,Solovej_rev}. In this paper we shall give a detailed proof of stability for our three models although some parts were already known before. 

Another important property is a sort of continuity property
\begin{description}
\item[(A4)] $\forall \Omega'\subset\Omega,\quad E(\Omega)\leq E(\Omega')+\kappa|\Omega\setminus\Omega'|+\text{error}$
\end{description}
which essentially says that a small decrease of $\Omega$ will not decrease too much the energy. A similar property was used and proved in the crystal case by Fefferman, see \cite[Lemma 2]{F}. Verifying \textbf{(A4)} for our two last models is obvious as the energy is decreasing with respect to $\Omega$ (we simply construct a trial state for $\Omega$ by taking the ground state in $\Omega'$ and the vacuum state in $\Omega\setminus\Omega'$). But in the crystal case the verification of \textbf{(A4)} is much more involved. This difficulty will actually be the source of some regularity assumptions on our sequence $\{\Omega_n\}$ (like the so-called \emph{cone property}) which we need for the crystal and not for our two other models. It is not surprising that we need more assumptions on domains for the system which is the most `rigid'.

A crucial property which was considered in \cite{1} is
\begin{description}
\item[(A5)] $\displaystyle E(\Omega)\geq \frac{1}{|\triangle|}\int_{SO(3)}\,dR\int_{\R^3}\,du\, E\big(\Omega\cap (R\triangle+u)\big)-\text{error}$
\end{description}
which compares the energy of $\Omega$ with the energy of the reference set $\triangle$, averaged over rotations and translations of $\triangle$ inside $\Omega$. An inequality of this form was first remarked and used by Conlon, Lieb and Yau \cite{CLY1,CLY2}, for systems interacting with the Yukawa potential and $\triangle$ being a cube. For Coulomb interactions, it was proved by Graf and Schenker \cite{GS,G} in which case $\triangle$ is chosen to be a tetrahedron. This inequality of Graf and Schenker is recalled below in Theorem \ref{Prop_Graf_Schenker} as it is the main tool of our new approach. 

Actually, to get the existence of the thermodynamic limit for any `regular' sequence $\{\Omega_n\}$, we used in \cite{1} a more precise assumption \textbf{(A6)} which essentially says that the interaction potential is two-body, as this is the case in our three examples. In practice the main ideas of the proof of \textbf{(A5)} and \textbf{(A6)} were essentially already contained in \cite{GS} and it is not much more difficult to prove \textbf{(A6)} than \textbf{(A5)} for our examples.

However, our assumption \textbf{(A6)} in \cite{1} contains the strong subadditivity of the entropy
\begin{description}
\item[(A6.6)] $S(\Omega_1\cup\Omega_2\cup\Omega_3)+S(\Omega_2)\leq S(\Omega_1\cup\Omega_2)+S(\Omega_2\cup\Omega_3)$
\end{description}
for all disjoint subsets $\Omega_1$, $\Omega_2$ and $\Omega_3$ and this was not considered in \cite{GS}.
Conjectured by Lanford and Robinson \cite{LR} the strong subadditivity (SSA) of the entropy in the quantum mechanical case was proved by Lieb and Ruskai in \cite{LR1,LR2}. The fact that SSA is very important in the thermodynamic limit was first remarked by Robinson and Ruelle in \cite{RR}, see also Wehrl \cite{Wehrl}. In all these references, SSA is usually stated using the formalism of partial traces of density matrices. But this is not fully appropriate for the examples we want to treat. For this reason, we have written a detailed appendix where we recall how to localize in Fock space and prove the SSA of the entropy within this formalism. We use this theory in our three examples.

\medskip

The paper is organized as follows. 
In the first section, we recall two important inequalities for classical Coulomb systems. The first was proved by Lieb and Yau \cite{LY}, generalizing previous results of Baxter \cite{Baxter} and Onsager \cite{Onsager}. It essentially allows us to bound from below the full $N$-body Coulomb potential by a simple one-body term where each electron only sees its closest nucleus. We shall use this inequality to prove stability of matter. Notice the precise estimate of \cite{LY} which contains some residual from the interaction between the nuclei is used in our second and third examples. The second inequality which is recalled in the first section is the one proved by Graf and Schenker \cite{GS} which plays an important role as explained before.
In the second part of the first section, we state a result concerning the stability of matter for electrons. We also recall the Lieb-Thirring inequality \cite{LT} which is a useful tool when studying fermions. Finally, we state another inequality which we will use for bosons.

In the second section, we define our three examples properly and state our main results. We also give a rather detailed sketch of the proof of the existence of the thermodynamic limit in the crystal case. We provide this as an illustration of the techniques used to verify the assumptions of \cite{1}.

For the sake of clarity, we have gathered many of the technical proofs in Section 3. Indeed for the second and third examples, we usually only give the details which are different from the crystal case.

Lastly, as mentioned above, the last part is an appendix devoted to the presentation of localization in Fock space and the strong subadditivity of the entropy in a rather general setting.

Let us mention that the results of this paper and of \cite{1} have been summarized in \cite{proc}.

\bigskip

\noindent\textbf{Acknowledgment.} We would like to thank Robert Seiringer for useful discussions and advice. M.L. acknowledges support from the ANR project ``ACCQUAREL'' of the French ministry of research.

\section{Preliminaries}
This section is dedicated to the presentation of preliminary results which we shall need later.
We let $G=\R^3\rtimes SO(3)$ the group of translations and rotations which acts on $\R^3$, and let $d\lambda$ denote its Haar measure.

\subsection{Two inequalities for classical Coulomb systems}
We start this section by presenting two inequalities which will be useful throughout the paper. 

\subsubsection{The Lieb-Yau inequality}
The first is due to Lieb and Yau \cite{LY} and it gives a lower bound on the full many-body Coulomb interaction of classical particles in terms of the nearest nucleus of each electron. It generalizes a previous result of Baxter \cite{Baxter}. We shall use it to prove stability of matter, i.e. \textbf{(A2)} as was defined in \cite{1}.
\begin{thm}[Lieb-Yau Inequality]\label{Thm_Lieb_Yau} Given $z>0$ and $x_1,...,x_N\in\R^3$, $R_1,...,R_K\in\R^3$, we have
\begin{multline}
 \frac12\sum_{1\leq i\neq j\leq N}\frac{1}{|x_i-x_j|}-\sum_{i\geq1}\sum_{k=1}^K\frac{z}{|x_i-R_k|}+\frac12\sum_{1\leq k\neq k'\leq K}\frac{z^2}{|R_k-R_{k'}|}\\
\geq -\sum_{i=1}^N\frac{z+\sqrt{2z}+1/2}{\delta_R(x_i)}+\frac{z^2}4\sum_{k=1}^K\frac{1}{\delta_R(R_k)}
\label{eq:Lieb-Yau}
\end{multline}
where $\delta_R(x)=\inf\left(|x-R|,\ R\in\{R_k\}_{k=1}^K\setminus\{x\}\right)$.
\end{thm}

See \cite[Thm. 6]{LY} for an even more precise statement. Baxter \cite{Baxter} proved \eqref{eq:Lieb-Yau} without the last nuclear repulsion term in \eqref{eq:Lieb-Yau} and with $z+\sqrt{2z}+1/2$ replaced by $1+2z$.

\subsubsection{The Graf-Schenker inequality}
An interesting electrostatic inequality taking the form of \textbf{(A5)} was proved by Graf and Schenker \cite{GS,G}. Inspired by previous works by Conlon, Lieb and Yau \cite{CLY1,CLY2}, the Graf-Schenker inequality will be the basis of the proof of our assumptions \textbf{(A5)} and \textbf{(A6)} in the different applications. In his proof \cite{F} of the existence of the thermodynamic limit for the crystal, Fefferman also proved a lower bound on the free energy, but he used balls of different size to cover a fixed domain, with an average over translations only, and made a crucial use of the kinetic energy to control error terms. See, e.g. \cite{Hugues} for another description of Fefferman's method.

Let $\triangle$ be an open simplex (or tetrahedron), i.e. a set of the form
$$\triangle=\{x\in\R^3\ |\ a_i\cdot x< c_i\}$$
for some $(a_1,...,a_4)\in(\R^3)^4$ and $(c_1,...,c_4)\in\R^4$ such that
$$\sum_{i=1}^4a_i=0,\qquad |\det(a_i,a_j,a_k)|=1$$
for all $1\leq i< j<k\leq 4$.
The following result is contained in \cite{GS,G}:
\begin{thm}[Graf-Schenker Inequality]\label{Prop_Graf_Schenker} Let $\triangle$ be as above. There exists a constant $C$ (depending on $\triangle$) such that for any $N\in\N$, $z_1,...,z_N\in\R$, $x_1,...,x_N\in\R^3$ and any $\ell\geq0$,
\begin{multline}
\sum_{1\leq i<j\leq N} \frac{z_iz_j}{|x_i-x_j|}\geq \frac{1}{|\ell\triangle|}\int_{G} d\lambda(g)\sum_{1\leq i<j\leq N}\frac{z_iz_j\1_{g\ell\triangle}(x_i)\1_{g\ell\triangle}(x_j)}{|x_i-x_j|}\\
-\frac{C}{\ell}\sum_{i=1}^Nz_i^2.
\label{eq:Graf-Schenker}
\end{multline}
\end{thm}

Theorem \ref{Prop_Graf_Schenker} contains the main idea of the proof of our assumptions \textbf{(A5)} and \textbf{(A6)} for Coulomb systems. Like in \cite{GS}, it will however need to be generalized in order to localize the kinetic energy also, see Lemma \ref{GS_2} below.

\begin{remark}\rm 
For the Graf-Schenker inequality \eqref{eq:Graf-Schenker}, it is essential that the potential is $1/|x|$ (see the proof in \cite{GS}). On the other hand for the Lieb-Yau inequality \eqref{eq:Lieb-Yau}, it is essential that  the potential is the fundamental solution of the Laplacian. Hence only three-dimensional Coulomb systems will be considered in the present paper.
\end{remark}

\subsection{The stability of matter}

In this section, we state the stability of matter for electronic systems. 

\subsubsection*{Notation}
Let us first fix some notation.
If $h$ is an operator on a Hilbert space $\gH$ we will use the
notation $\sum_j h_j$ (it is often denoted
$d\Gamma(h)$) for its second quantization to an operator on
the fermionic Fock space $\cF(\gH):=\bigoplus_{N=0}^\ii\bigwedge_1^N\gH$. More precisely,
$$
\sum_j h_j=
\bigoplus_{N=0}^\infty\sum_{j=1}^N \underbrace{1\otimes\cdots\otimes1}_{\hbox{$j-1$ terms}}\otimes\,
h \otimes\underbrace{1\otimes\cdots\otimes1}_{\hbox{$N-j$ terms}}.
$$
Likewise if $W$ is an operator on $\gH\wedge\gH$, we write
$\frac{1}{2}\sum_{i\ne j}W_{ij}$ for its second quantization in $\cF(\gH)$.

A system of pointwise (classical) nuclei in $\Omega\subset\R^3$ is a set 
$$\cK=\{(R_k,z_k)\ |\ k=1,\ldots K\}\subseteq\Omega\times[0,\ii)$$
where $R_k$ and $z_k$ are respectively the positions and charges of the nuclei.
For any such $\cK\subseteq\Omega\times[0,\ii)$, we define the Coulomb potential acting on the electronic Fock space as
\begin{equation}
V_\cK:=
\frac12\sum_{1\leq i\neq j}\frac{1}{|x_i-x_j|}-\sum_{i\geq1}\sum_{k=1}^K\frac{z_k}{|x_i-R_k|}+\frac12\sum_{k\neq k'=1}^K\frac{z_kz_{k'}}{|R_k-R_{k'}|}
\end{equation}
and $V_\cK=+\ii$ if $R_k=R_{k'}$ for some $k\neq k'$ with $z_kz_{k'}\neq0$. Then, we choose some magnetic potential $A\in L^2_{\rm loc}(\R^3)$ such that ${\rm div}A=0$ in the distributional sense. We introduce the kinetic operator
$$T(A):=(-i\nabla +A(x))^2$$
with Dirichlet boundary conditions on $\Omega$. We shall use the shorthand notation $T:=T(0)$. In the whole paper, we work within a system of units in which the Planck constant $\hbar$ and the charge of the electron are set to one, whereas the mass of electrons is set to $1/2$.
We may define the associated Coulomb Hamiltonian acting on $\cF(L^2(\Omega))$ as
\begin{equation}
H_{\Omega,\cK}:=\sum_{j} T(A)_j+ V_{\cK},
\label{form_Coulomb_Hamiltonian_general}
\end{equation}
Since $H_{\Omega,\cK}$ commutes with the number operator $\cN$, we can also write
$$H_{\Omega,\cK}=\bigoplus_{N=0}^\ii H^N_{\Omega,\cK},$$
where each $H^N_{\Omega,\cK}$ acts on the $N$-body space $\bigwedge_1^NL^2(\Omega)$. We do not consider the spin variable for simplicity.
Unless specified, we use the notation `$\tr$' to denote the trace in the Fock space $\cF(L^2(\Omega))$.

\begin{remark}\rm
The Hamiltonian $H_{\Omega,\cK}$ may be defined as the Friedrichs extension associated with the quadratic form with form domain $\cF(H^1_0(\Omega))\cap D(\cN)$. The quadratic form is bounded below (by Stability of Matter, a lower bound is given by $-\kappa|\Omega|$ as we will see below).
\end{remark}

\begin{remark}\rm
We will always confine all the quantum particles in the domain $\Omega$. For the ground state energy another possibility would be to only consider nuclei which are in $\Omega$ and let the electrons live in the whole space, similarly to what was done in \cite{CLL1}.  However at positive temperature, confining the quantum particles is mandatory to have a bounded-below free energy. For this reason, we will always restrict ourselves to the case of confined quantum particles.
\end{remark}

\subsubsection{Stability of matter for electrons}
We now state and prove the stability of matter for a quantum system confined to a domain $\Omega$. The nuclei will be described differently in each application and therefore we limit ourselves to classical pointwise nuclei in this section. Stability of matter was proved first by Dyson and Lenard in \cite{DL1,DL2}. We give a proof based on the simpler proof by Lieb and Thirring in \cite{LT}, and the Lieb-Yau inequality \eqref{eq:Lieb-Yau}.

In the whole paper we will denote by 
\begin{multline}
 \cD(\Omega):=\Bigg\{\Gamma\in\cB(\cF(L^2(\Omega)))\ |\ \Gamma=\Gamma^*,\ \Gamma\geq0,\ \tr(\Gamma)=1,\\ \tr(\cN\Gamma)<\ii,\ \tr\left(\Gamma\sum_iT_i\right)<\ii\Bigg\}
\label{def:density_matrices}
\end{multline}
the convex set of all density matrices acting on $\cF(L^2(\Omega))$ and having a finite kinetic energy and a finite average particle number. We will always consider normal states $\omega$, i.e. states arising from a density matrix $\Gamma\in\cD(\Omega)$, $\omega(A)=\tr(A\Gamma)$.
Let us recall that every state $\Gamma\in\cD(\Omega)$ has a corresponding density $\rho_\Gamma$ which is defined as the unique function in $L^1(\Omega)$ such that 
$$\forall V\in L^\ii(\Omega),\qquad\tr\left\{\Gamma\left(\sum_iV_i\right)\right\}=\int_{\Omega}\rho_\Gamma(x)V(x)\,dx.$$
 If $\Gamma$ is a pure state, i.e. $\Gamma=|\Psi\rangle\langle\Psi|$ for some $\Psi\in\cF(L^2(\Omega))$, we use the notation $\rho_\Psi:=\rho_{|\Psi\rangle\langle\Psi|}$.
As in \cite{1}, we denote by $\cM$ is the set of all the open and bounded subsets of $\R^3$. The following result gives the stability of matter for a system composed of quantum electrons and an arbitrary number of classical nuclei with charge $\leq z$. We use the notation $x_+=\max(x,0)$.

\begin{thm}\label{stability_of_matter} Let us fix some $z>0$ and some $A\in L^2_{\rm loc}(\R^3)$ with ${\rm div}A=0$ in the distributional sense.

\smallskip

\noindent$\bullet$ For any $\Omega\in\cM$, we have the formula
\begin{equation}
\inf_{\substack{\cK\subseteq\Omega\times[0,z],\\ \#\cK<\ii}}\inf\sigma_{\cF(L^2(\Omega))}(H_{\Omega,\cK})=\inf_{\substack{\cK\subseteq\Omega\times\{z\},\\ \#\cK<\ii}}\inf\sigma_{\cF(L^2(\Omega))}(H_{\Omega,\cK}).
\label{formula_Daubechies_Lieb}
\end{equation}

\smallskip

\noindent$\bullet$ \emph{(Stability of Matter with Nuclear Repulsion).} There exists a constant $C>0$ (independent of $A$) such that the following holds for all $\Omega\in\cM$ and all $R=\{R_k\}_{k=1}^K\subset\Omega$
\begin{equation}
\inf\sigma_{\cF(L^2(\Omega))}(H_{\Omega,R\times\{z\}})\geq -C |\Omega|+\frac{z^2}{8}\sum_{k=1}^K\frac{1}{\delta_R(R_k)}.
\label{stability_of_matter_equation}
\end{equation}
A similar inequality holds true if $T(A)$ is replaced by $\lambda T(A)$ for some $\lambda>0$, with a constant $C$ depending on $\lambda$.

\smallskip

\noindent$\bullet$ \emph{(Stability of Matter for Electronic Free Energy).} There exists a constant $C>0$ (independent of $A$) such that the following holds for all $\Omega\in\cM$, $\beta>0$ and $\mu\in\R$
\begin{multline}
\inf_{\substack{\cK\subseteq\Omega\times[0,z],\\ \#\cK<\ii}}\left(-\frac1\beta\log\tr_{\cF(L^2(\Omega))}\left(e^{-\beta(H_{\Omega,\cK}-\mu\cN)}\right)\right)\\
\geq -C \left(1+\beta^{-5/2}+\mu_+^{5/2}\right)|\Omega|.
\label{stability_of_matter_FE_equation}
\end{multline}
\end{thm}

\begin{remark}\rm 
Formula \eqref{formula_Daubechies_Lieb} is due to Daubechies and Lieb \cite{DL}. It means that nuclei always prefer to have the highest possible charge.
\end{remark}

\begin{remark}\rm 
Estimate \eqref{stability_of_matter_equation} was first proved by Dyson and Lenard in \cite{DL1,DL2} when $A\equiv0$ (see Theorem 5 in \cite{DL1}), without the last nuclear repulsion term. 
The proof given by Lieb and Thirring in \cite{LT} does not seem to give a lower bound independent of the number and location of the classical nuclei. To prove \eqref{stability_of_matter_equation}, we use instead the Lieb-Yau estimate \eqref{eq:Lieb-Yau}.
\end{remark}

\begin{remark}\rm 
Of course \eqref{formula_Daubechies_Lieb} and \eqref{stability_of_matter_equation} imply that
\begin{equation}
 \inf_{\substack{\cK\subseteq\Omega\times[0,z],\\ \#\cK<\ii}}\inf\sigma_{\cF(L^2(\Omega))}(H_{\Omega,\cK})\geq -C|\Omega|.
\label{stability_of_matter_equation_bis}
\end{equation}
which is the original result of Dyson and Lenard (when $A=0$).

As explained in \cite{Loss}, a simple proof of \eqref{stability_of_matter_equation_bis} is to use directly stability of relativistic matter as proved by Lieb and Yau in \cite{LY} (it is itself based on \eqref{eq:Lieb-Yau}). We give a detailed proof of Theorem \ref{stability_of_matter} in Section \ref{sec:stability_of_matter} for the reader's convenience and because we shall need the more precise estimate \eqref{stability_of_matter_equation}.
\end{remark}

\subsubsection{The Lieb-Thirring inequality}
One very important tool which we shall use throughout the paper for Fermions (and in particular in the proof of Theorem \ref{stability_of_matter} given in Section \ref{sec:stability_of_matter}) is the Lieb-Thirring inequality \cite{LT}.
The usual formulation of this inequality is
\begin{equation}
\tr_{L^2(\Omega)}\left(-\Delta+V\right)_-\leq C'_{\rm LT}\int_{\Omega}V_-^{5/2}
\label{Lieb-Thirring_first_formulation}
\end{equation}
where we have used the notation $x_-=-\min(x,0)$. The constant $C'_{\rm LT}$ does not depend on the domain $\Omega\subseteq\R^3$.

We shall often make use of another version of the same inequality \cite{LT}. Let $\Psi$ be a (fermionic) common eigenvector of $\sum_j T_j$ and $\cN$, $\cN\Psi=N\Psi$. The inequality reads
\begin{equation}
\pscal{\left(\sum_{j=1}^NT_j\right)\Psi,\Psi}\geq C_{\rm LT}\int_{\Omega}\rho_\Psi(x)^{5/3}dx\geq C_{\rm LT} \pscal{\cN^{5/3}\Psi,\Psi}|\Omega|^{-2/3}.
\label{eq_Lieb_Thirring_rho}
\end{equation}
This can also be written
\begin{equation}
\sum_jT_j\geq C_{\rm LT} \cN^{5/3}|\Omega|^{-2/3}
\label{estim_Lieb_Thirring_kinetic}
\end{equation}
as operators on the Fock space $\cF(\Omega)$, since $\sum_jT_j$ and $\cN$ commute. The same formulas hold if $A\neq0$, by the diamagnetic inequality \cite{Lieb-Loss}.

The above inequality \eqref{estim_Lieb_Thirring_kinetic} will always be used to control terms of the form $-C\cN$, as one has for any $\mu\geq0$
$$C_{\rm LT} \cN^{5/3}|\Omega|^{-2/3}-\mu\cN\geq-\left(\frac{3}{5C_{\rm LT}}\right)^{3/2}\mu^{5/2}|\Omega|.$$

\subsubsection{An inequality for repelling particles}
Next we provide an inequality which will replace the Lieb-Thirring inequality for bosons, assuming that there is an additional repulsive term (which we will usually get from the last term of the Lieb-Yau inequality \eqref{eq:Lieb-Yau}).
\begin{prop}[An inequality for repelling particles]
\label{lemma:estim_nuclei}
Let $\epsilon>0$. There exists a constant $c>0$ (depending on $\epsilon$) such that the following inequality holds for any $N\geq0$
\begin{equation}
\sum_{i=1}^N\left(T(A)_{x_i}+\epsilon\max_{k\neq  i}|x_i-x_k|^{-1}\right)\geq cN\min\left\{\frac{N}{|\Omega|},\frac{N^{1/3}}{|\Omega|^{1/3}}\right\}
\label{estim_nuclei}
\end{equation}
as an operator acting on $L^2(\Omega^N,\R)$. 
\end{prop}

The proof of Proposition \ref{lemma:estim_nuclei} is provided in Section \ref{proof_estim_nuclei}. The estimate \eqref{estim_nuclei} may serve like \eqref{estim_Lieb_Thirring_kinetic} to control a term of the form $-CN$, as one has
$$cN\min\left\{\frac{N}{|\Omega|},\frac{N^{1/3}}{|\Omega|^{1/3}}\right\}-\mu N\geq-\frac{\mu^2}c\max\left(1,\frac{\mu^2}{c^2}\right)|\Omega|.$$

\section{Thermodynamic limit of three quantum systems}

In this section, we apply the general formalism of \cite{1} to different quantum systems. Details for the proofs are given in the next section.

\subsection{The perturbed crystal case}\label{crystal}
In this section, we do not consider any magnetic field, $A\equiv0$.
Let $\Lambda$ be a fixed discrete subgroup of $\R^3$ with bounded and open fundamental domain $W$. Let $\{(R_k,z_k)\}_{k=1}^K\subseteq W\times [0,\ii)$ be a finite set of nuclei in $W$ such that $k\neq k'\Rightarrow R_k\neq R_{k'}$. We then let 
$$\cL:=\{(\lambda +R_k,z_k),\ k=1,...,K,\ \lambda\in\Lambda\}\subset\R^3\times[0,\bar z]$$
where $\bar z=\max z_k$. The set $\cL$ describes our infinite periodic lattice of classical nuclei. 

\begin{figure}[h]
\centering
\includegraphics[width=6cm]{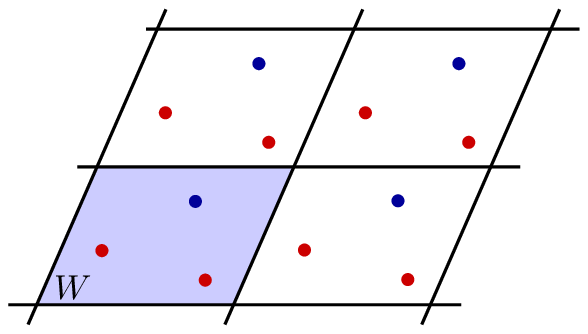}
\label{fig:crystal}
\caption{The crystal $\cL$ without defect.}
\end{figure}

In the following, we shall allow deformations\footnote{Our result was already announced without deformations in \cite{proc}.} of $\cL$ and we consider a set of nuclei of the form
\begin{equation}
\cL':= (Id+D)\cL\cup \cL_{\rm d}
\label{perturbed_lattice}
\end{equation}
where $D:\cL\to\R^3\times\R$ is the deformation and $\cL_{\rm d}=\{(R_i,z_i)\}_{i\in I}\subseteq \R^3\times[0,\ii)$ are some defects ($I$ is any countable set). Essentially our results will be valid when $D$ and the charges of $\cL_{\rm d}$ decay sufficiently fast at infinity (for instance when they have a bounded support). We will write $D(R,z)=(D_{\rm dis}(R),D_{\rm ch}(z))$ where $D_{\rm dis}$ is the displacement of the nuclei in $\cL$ and $D_{\rm ch}$ is the associated change of charge. The precise assumptions which we shall need are the following:
\begin{equation}
\sup_{x\in\R^3}\left\{\sum_{(R,z)\in\cL,\ R\in B(x,L)}|D_{\rm dis}(R)|\right\} =o\left(L^{6/5}\right),
\label{hyp_D1}
\end{equation}
\begin{equation}
\sup_{x\in\R^3}\left\{\sum_{i\in I,\ R_i\in B(x,L)}|z_i|+ \sum_{(R,z)\in\cL,\ R\in B(x,L)}|D_{\rm ch}(z)|\right\}=o\left(L\right),
\label{hyp_D2}
\end{equation}
\begin{equation}
\sup_{(R,z)\in\cL}|D(R,z)|<\ii\quad \text{and}\quad \inf_{\substack{(R,z)\in\cL',\\ (R',z')\in\cL',\ R'\neq R}}|R-R'|=\delta>0.
\label{hyp_D3}
\end{equation}
The function $D$ and the defect set $\cL_{\rm d}$ will be fixed in the following but the thermodynamic limit will be proved to be independent of them. Conditions \eqref{hyp_D1} and \eqref{hyp_D2} are probably not optimal. We have not tried to optimize them.

We define the Schrödinger ground state energy in the domain $\Omega\in\cM$ with Dirichlet boundary conditions:
\begin{equation}
\boxed{E^{\rm Sch}_{\cL'}(\Omega)=\inf_{\Gamma\in\cD(\Omega)}\tr_{\cF(L^2(\Omega))}(H_{\Omega,\cL'}\Gamma).}
\label{Sch} 
\end{equation}
We recall that $H_{\Omega,\cL'}$ is the grand-canonical Schrödinger Coulomb operator with Dirichlet boundary condition on $\Omega$ and nuclei in $\cL'\cap(\Omega\times[0,\ii))$, see \eqref{form_Coulomb_Hamiltonian_general}. We do not consider any magnetic field, $A\equiv0$.
We define the Schrödinger free energy at temperature $\beta^{-1}>0$ and chemical potential $\mu$ in the domain $\Omega$ as
\begin{equation}
\boxed{F^{\rm Sch}_{\cL'}(\Omega,\beta,\mu)= -\frac{1}{\beta}\log\tr_{\cF(L^2(\Omega))}\left(e^{-\beta (H_{\Omega,\cL'}-\mu\cN)}\right).}
\label{Sch_FE}
\end{equation}
Notice the formula can be written as a minimization problem
\begin{equation}
F^{\rm Sch}_{\cL'}(\Omega,\beta,\mu) =\inf_{\Gamma\in\cD(\Omega)}\tr_{\cF(L^2(\Omega))}\left[(H_{\Omega,\cL'} -\mu\cN)\Gamma+\frac{1}{\beta}\Gamma\log\Gamma\right].\label{Sch_FE_bis}
\end{equation}
We shall also be able to consider the Hartree-Fock approximation for the crystal. The Hartree-Fock ground state energy is defined as (see \cite{BLS})
\begin{equation}
\boxed{E^{\rm HF}_{\cL'}(\Omega)=\inf_{\substack{\Gamma\in\cD(\Omega)\\ \text{quasi-free}}}\tr_{\cF(L^2(\Omega))}(H_{\Omega,\cL'}\Gamma).}
\label{HF}
\end{equation}
We recall \cite{BLS} that a quasi-free state $\Gamma$ is a state in $\cD(\Omega)$ satisfying Wick's Theorem, i.e. for any $N\geq1$
$$\tr(\Gamma e_1\cdots e_{2N-1})=0,$$
$$\tr(\Gamma e_1\cdots e_{2N})={\sum_{\pi}}'\varepsilon(\pi)\tr(\Gamma e_{\pi(1)}e_{\pi(2)})\cdots\tr(\Gamma e_{\pi(2N-1)}e_{\pi(2N)}).$$
Here the sum is taken over all permutations $\pi\in\gS(2N)$ such that $\pi(1)<\pi(3)<\cdots\pi(2N-1)$ and $\pi(2j-1)<\pi(2j)$ for all $1\leq j\leq N$. The $e_i$'s are either an annihilation operator $c_i$ or a creation operator $c_i^\dagger$ in a chosen basis of $L^2(\Omega)$. In particular, the following equality is assumed to hold
$$\tr(\Gamma e_1e_2e_3e_4)=\tr(\Gamma e_{1}e_{2})\tr(\Gamma e_{3}e_{4})-\tr(\Gamma e_{1}e_{3})\tr(\Gamma e_{2}e_{4})+\tr(\Gamma e_{1}e_{4})\tr(\Gamma e_{2}e_{3}).$$
When applied to compute the Hartree-Fock energy $\tr(H_{\Omega,\cL'}\Gamma)$, the three terms of the right hand side respectively yield to the direct, the exchange and the pairing energy.
Similarly to \eqref{Sch_FE}, the Hartree-Fock free energy at temperature $\beta^{-1}>0$ and chemical potential $\mu$ in the domain $\Omega$ is defined as \cite{BLS}
\begin{equation}
\boxed{F^{\rm HF}_{\cL'}(\Omega,\beta,\mu)  =  \inf_{\substack{\Gamma\in\cD(\Omega)\\ \text{quasi-free}}}\tr_{\cF(L^2(\Omega))}\left[(H_{\Omega,\cL'} -\mu\cN)\Gamma+\frac{1}{\beta}\Gamma\log\Gamma\right].}
\label{HF_FE}
\end{equation}

\medskip

The following result is easily proved:
\begin{thm}[Existence of Ground States] Assume that $\cL'$ is defined as above and let $\Omega\in\cM$ be a bounded open set in $\R^3$. Then the above minimization problems \eqref{Sch}, \eqref{Sch_FE_bis}, \eqref{HF} and \eqref{HF_FE} all possess a minimizer.

For the Schrödinger ground state minimizing \eqref{Sch}, one can take a pure state $\Gamma=|\Psi\rangle\langle\Psi|$ where $\Psi$ is an eigenvector of the number operator: 
$$H_{\Omega,\cL'}\Psi=E_{\cL'}^{\rm Sch}(\Omega)\Psi\quad \text{and}\quad \cN\Psi=N\Psi,$$
i.e. $\Psi\in\bigwedge_1^NH^2(\Omega)\subseteq\cF(L^2(\Omega))$.

The Schrödinger ground state minimizing \eqref{Sch_FE_bis} is unique. It is the Gibbs state 
$$\Gamma=Z^{-1}{\rm exp}\left(-\beta(H_{\Omega,\cL'}-\mu\cN)\right)$$
where $Z=\tr_{\cF(L^2(\Omega))}{\rm exp}\left(-\beta(H_{\Omega,\cL'}-\mu\cN)\right)$. 
\end{thm}

In the study of the thermodynamic limit, the first step is to prove stability of matter:
\begin{thm}[Stability in the crystal case]\label{stability_cristal} Assume that $\cL'$ is defined as above. There exists a real constant $\kappa$ (independent of $\cL'$, $\beta>0$ and $\mu\in\R$) such that the following holds for all $\Omega\in\cM$
$$E^{\rm Sch}_{\cL'}(\Omega)\geq -\kappa |\Omega|,\qquad F^{\rm Sch}_{\cL'}(\Omega,\beta,\mu)\geq -\kappa\left(1+\beta^{-5/2}+\mu_+^{5/2}\right)|\Omega|.$$
\end{thm}
\begin{proof}
 This is an obvious consequence of Theorem \ref{stability_of_matter}.
\end{proof}

\begin{figure}[h]
\centering
\includegraphics[width=6cm]{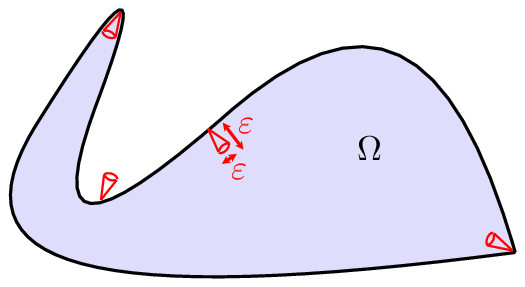}
\label{image_cone_pty}
\caption{Cone property.}
\end{figure}

To measure the regularity of the sequence of domains $\{\Omega_n\}$ in the thermodynamic limit, we shall need the
\begin{definition}[Cone property]\label{def_cone_pty} An open bounded set $\Omega$ is said to have the \emph{$\varepsilon$-cone property} if for any $x\in\Omega$ there is a unit vector $a_x\in\R^3$ such that 
$$
\{y\in\R^3\ |\  (x-y)\cdot a_x> (1-\varepsilon^2) |x-y|,\
|x-y|<\varepsilon\}
\subseteq\Omega.
$$
We denote by $\cC_\varepsilon$ the set of all $\Omega\in\cM$ such that both $\Omega$ and $\R^3\setminus\Omega$ have the $\varepsilon$-cone property. Note that $\Omega\in\cC_\varepsilon \Longrightarrow \forall \lambda\geq1,\ \lambda\Omega\in\cC_\varepsilon$.
\end{definition}

Clearly, any open convex set belongs to $\cC_\varepsilon$ for some small enough $\varepsilon>0$. We recall that $\cR_\eta\subset\cM$ contains all the subsets having the Fisher $\eta$-regularity property (Definition \refI{def_regular}).

Our main result is the following
\begin{thm}[Thermodynamic limit for the perturbed crystal]\label{limit_crystal} There exist two constants $\bar e^{\rm Sch}\leq \bar e^{\rm HF}$ and two functions $\bar f^{\rm Sch}\leq\bar f^{\rm HF}:(0,\ii)\times\R\mapsto\R$, independent of $D$ and $\cL_{\rm d}$ appearing in \eqref{perturbed_lattice}, such that the following holds: for any sequence $\{\Omega_n\}_{n\geq1}\subseteq \cR_{\eta}\cap\cC_{\varepsilon}$ with $|\Omega_n|\to\ii$ and ${\rm diam}(\Omega_n)|\Omega_n|^{-1/3}\leq C$, with $\eta(t)=a|t|$ for some $a>0$ and $\varepsilon>0$,
\begin{equation}
\lim_{n\to\ii}\frac{E^{\rm Sch}_{\cL'}(\Omega_n)}{|\Omega_n|}=\bar e^{\rm Sch},\qquad \lim_{n\to\ii}\frac{E^{\rm HF}_{\cL'}(\Omega_n)}{|\Omega_n|}=\bar e^{\rm HF},
\label{limit_energy}
\end{equation}
\begin{equation}
\lim_{n\to\ii}\frac{F^{\rm Sch}_{\cL'}(\Omega_n,\beta,\mu)}{|\Omega_n|}=\bar f^{\rm Sch}(\beta,\mu),\quad \lim_{n\to\ii}\frac{F^{\rm HF}_{\cL'}(\Omega_n,\beta,\mu)}{|\Omega_n|}=\bar f^{\rm HF}(\beta,\mu)
\label{limit_FE}
\end{equation}
for all $\beta>0,\ \mu\in\R$.
\end{thm}

\begin{remark}\rm
 The existence of the thermodynamic limit in the Schrödinger case and for the periodic case (perfect crystal) was already proved by C.~Fefferman in \cite{F}. Our result is more general: we allow any sequence $\Omega_n$ tending to infinity and which is regular in the sense that $\{\Omega_n\}_{n\geq1}\subseteq \cR_\eta\cap\cC_\varepsilon$. In \cite{F}, $\Omega_n=\ell_n(\Omega+x_n)$ where $\ell_n\to\ii$, $\Omega$ is a fixed convex set with a non-empty interior and $x_n$ is any sequence in $\R^3$. These sets are always in $\cR_\eta\cap\cC_\varepsilon$ for some $\eta$ and $\varepsilon$ by Lemma \refI{regularity_convex}. Moreover, we provide a precise statement for the case where the nuclei are arranged on a perturbation $\cL'$ of a periodic lattice $\cL$.
\end{remark}

\begin{remark}\rm
The thermodynamic limit in the Hartree-Fock case for the crystal was investigated by I. Catto, C. Le Bris and P.-L. Lions in \cite{CLL1}. There, a periodic Hartree-Fock energy $\cE^{\rm HF}_{\rm per}$ was proposed to describe the infinite periodic crystal, but only some hints were given in favour of the equality $\bar e^{\rm HF}=\inf\{\cE^{\rm HF}_{\rm per}\}$. The latter equality was however proven in \cite{CLL1,CDL} in the case where the exchange term is neglected, yielding the so-called \cite{Solovej2} reduced Hartree-Fock model.
\end{remark}

\begin{remark}\rm
 As recalled in \cite{1}, we know from \cite[Appendix A p. 385]{LL} and \cite[Lemma 1]{Fisher} that if each set $\Omega_n$ of the considered sequence is connected, then automatically ${\rm diam}(\Omega_n)|\Omega_n|^{-1/3}\leq C$ for all $n$.
\end{remark}

\begin{remark}\rm
It can be shown that in the thermodynamic limit the system is essentially neutral, i.e. one has $N=Z+o(N)=Z+o(|\Omega|)$ where $N$ is the number of electrons and $Z$ is the total charge of the nuclei in $\Omega$. This implies that $\bar f^{\rm Sch}$ takes the form $$\bar f^{\rm Sch}(\beta,\mu)=\phi(\beta)-\rho_{\rm nuc}\mu$$
where $\phi$ is concave and $\rho_{\rm nuc}$ is the density of charge of the nuclei, i.e. 
$$\rho_{\rm nuc}:=|W|^{-1}\sum_{(R,z)\in W\times\R}z.$$
\end{remark}

\begin{remark}\rm
In Theorem \ref{limit_crystal}, the negatively charged particles are electrons, i.e. fermions. The same theorem (with slight modifications to the assumptions \eqref{hyp_D1}, \eqref{hyp_D2} and \eqref{hyp_D3} on the perturbation of the crystal) holds in the Schrödinger case when the negatively charged particles are assumed to be bosons. We shall not give a detailed proof of this assertion but we will mention the necessary adaptations throughout the proof.
\end{remark}
\medskip

\subsubsection*{Proof of Theorem \ref{limit_crystal}}
For the convenience of the reader, we give here the main steps of the proof of Theorem \ref{limit_crystal}. Details can be found in Section \ref{proof_limit_crystal}.
We only write the proof for the Schrödinger case, the arguments being exactly the same for the Hartree-Fock case.

\subsubsection*{Step 1. A priori upper bounds.}
Notice Theorem \ref{stability_cristal} implies that the energy and the free energy of the deformed crystal satisfy the stability of matter \textbf{(A2)}  introduced in \cite{1}. 

\begin{remark}\rm \textbf{(Bosonic case)}. For bosons, it is necessary to adapt the proof of Theorem \ref{stability_cristal} using Proposition \ref{lemma:estim_nuclei}. 
\end{remark}

The following Proposition tells us that the energy and free energy per unit volume are bounded from above, for regular domains.

\begin{prop}[Upper bound for energy and density]\label{upper_bd_cristal}
Assume that $a,\varepsilon>0$, $\eta(t)=a|t|$ and that $\cL'$ is chosen as before. Then there exists a constant $\kappa'$ such that for all $\Omega\in \cR_\eta\cap\cC_\varepsilon$
\begin{equation}
E^{\rm Sch}_{\cL'}(\Omega)\leq\kappa'|\Omega|,\qquad F^{\rm Sch}_{\cL'}(\Omega,\beta,\mu)\leq \kappa'|\Omega|.
\label{stability_of_matter_equation2b}
\end{equation}
Furthermore, consider a set $\Omega\in \cR_\eta\cap\cC_\varepsilon$ and let $\Gamma\in\cB(\cF(H^1_0(\Omega)))$ be a ground state for $F^{\rm Sch}_{\cL'}(\Omega,\beta,\mu)$ or for $E^{\rm Sch}_{\cL'}(\Omega)$. Then 
\begin{equation}
\int_\Omega\rho_\Gamma(x)^{5/3}dx\leq \kappa'|\Omega|\quad \text{and}\quad \int_{\Omega}\rho_\Gamma(x)\,dx=\tr(\cN\Gamma)\leq \kappa'|\Omega|.
\label{estim_GS_LTb}
\end{equation}
\end{prop}

Proposition \ref{upper_bd_cristal} is proved in Section \ref{proof_lemma_stability_cristal}. This is done by a convenient choice of a trial state.

\begin{remark}\rm
Using \eqref{stability_of_matter_equation}, it can be easily seen that $\Omega\in\cM\mapsto E^{\rm Sch}_{\cL'}(\Omega)|\Omega|^{-1}$ is \emph{not bounded from above} on the whole class of open bounded sets $\cM$. It suffices to take a very elongated domain containing lots of nuclei.
\end{remark}
\begin{remark} \rm\textbf{(Bosonic case).} The bound \eqref{stability_of_matter_equation2b} and the second bound in \eqref{estim_GS_LTb} hold also for bosons.
 
\end{remark}

\subsubsection*{Step 2. Reduction to the periodic case.}
The second step consists in proving that the thermodynamic limit of the perturbed crystal $\cL'$ is the same as the one for the perfect crystal $\cL$ (i.e. with $\cL_{\rm d}=\emptyset$ and $D\equiv0$).
\begin{prop}[Reduction to the periodic case]\label{reduc_periodic} Assume that $a,\varepsilon>0$, $\eta(t)=a|t|$  and that $\cL'$ is chosen as before. Assume that  $\{\Omega_n\}_{n\geq1}$ is a sequence of $\cR_\eta\cap\cC_\varepsilon$ such that $|\Omega_n|\to\ii$. Then
$$E^{\rm Sch}_{\cL'}(\Omega_n)=E^{\rm Sch}_{\cL}(\Omega_n)+o(|\Omega_n|),$$
$$F^{\rm Sch}_{\cL'}(\Omega_n,\beta,\mu)=F^{\rm Sch}_{\cL}(\Omega_n,\beta,\mu)+o(|\Omega_n|)$$
for any fixed $\beta>0$ and $\mu\in\R$.
\end{prop}

The proof of Proposition \ref{reduc_periodic}, given in Section \ref{proof_reduc_periodic}. It is mainly based on the Lieb-Thirring estimates on the density of the electrons in Proposition \ref{upper_bd_cristal}.

\subsubsection*{{Step 3. Introduction of an auxiliary problem and proof of \textbf{(A4)}}.} From now on, we take $D=0$ and $\cL_{\rm d}=\emptyset$, i.e. we prove the existence of the thermodynamic limit for the periodic crystal $\cL$. However, because of some localization issues, the proof that $E^{\rm Sch}_\cL$ and $F^{\rm Sch}_\cL$ satisfy assumptions \textbf{(A5)} and \textbf{(A6)} is a bit tedious. Hence, for the sake of simplicity we shall introduce an auxiliary problem for which this task will be easier.

Consider a set $\Omega\in\cM$ and let $\cK=\{(R_k,z_k)\}_{k=1}^K=\cL\cap(\Omega\times\R)$ be the collection of all the nuclei of the periodic lattice $\cL$ which are inside $\Omega$. Fix a small enough positive real number $\delta'\leq \min\{|R-R'|,\ (R,z),(R',z')\in\cL,\ R\neq R'\}/10$. Reordering the indices of the nuclei if necessary, we can assume that ${\rm d}(R_k,\partial\Omega)>\delta'$ for all $k=1...K'\leq K$ and ${\rm d}(R_k,\partial\Omega)\leq\delta'$ for all $k=K'+1...K$. We now define auxiliary problems consisting in optimizing the charges of the nuclei which are at a distance less than $\delta'$ of the boundary of $\Omega$:
$$\uE(\Omega):=\inf_{\substack{z'_k\in[0,z_k],\\ k=K'+1...K}}\ \inf_{\Gamma\in\cD(\Omega)}\tr(H_{\Omega,\cK_{z'}}\Gamma),$$
$$\oE(\Omega):=\sup_{\substack{z'_k\in[0,z_k],\\ k=K'+1...K}}\ \inf_{\Gamma\in\cD(\Omega)}\tr(H_{\Omega,\cK_{z'}}\Gamma),$$
$$\cK_{z'}=\{(R_k,z_k),\ k=1...K'\}\cup\{(R_k,z'_k),\ k=K'+1...K\}.$$
We introduce similar definitions for the non-zero temperature case, $\uF(\Omega,\beta,\mu)$ and $\oF(\Omega,\beta,\mu)$.
It is clear that 
$$\uE(\Omega)\leq E^{\rm Sch}_\cL(\Omega)\leq\oE(\Omega)\quad \text{and}\quad \uF(\Omega,\beta,\mu)\leq F^{\rm Sch}_\cL(\Omega,\beta,\mu)\leq\oF(\Omega,\beta,\mu).$$

The idea of the proof is now to show that $\uE$ and $\uF(\cdot,\beta,\mu)$ satisfy \textbf{(A1)}--\textbf{(A6)} introduced in \cite{1}. 

Let us split the unit cube $C=[-1/2,1/2]^3$ of $\R^3$ in 24 open tetrahedrons by considering all the planes passing through the center of $C$ and an edge or a diagonal of a face. We denote by $\triangle'$ one of the so-obtained simplices. The other 23 simplices can be obtained as $\omega\triangle'$ where $\omega\in O$, the group of order 24 consisting of the pure rotations of the octahedral (or hexahedral) group, i.e. the symmetry group of the cube. We then fix some vector $v$ such that $0\in\triangle:=\triangle'+v$ and let $\Gamma:=g_v^{-1}(\Z^3\rtimes O) g_v$ with $g_v=(v,Id)\in G$. Clearly $\Gamma$ is a discrete subgroup of $G$ with compact quotient $G/\Gamma$ and $\triangle$ defines a $\Gamma$-tiling of $\R^3$, as defined in \cite[Section \ref{section_general_domains}]{1}. 

We have defined our reference set $\triangle$.
Next we fix $\eta(t)=a|t|$, $\varepsilon>0$ and choose $\cR:=(\cR_\eta\cap\cC_\varepsilon)$. We will assume that $a$ is large enough and that $\varepsilon$ is small enough such that $\triangle\in\cR$. 
Hence \textbf{(P1)} and \textbf{(P3)} are fulfilled.
We define $\cR'$ as $\cR':=\cR_{\eta'}\cap\cC_{\varepsilon'}$, where $\eta'=m\tilde\eta$, $m\geq1$ being the constant given by Proposition \refI{reg_inner_approx}. Since any union of simplices of the tiling obviously satisfies the $\varepsilon'$-cone property when $\varepsilon'$ is chosen small enough, it is then clear that \textbf{(P5)} is satisfied.

Also it is very easy to generalize Theorem \ref{stability_cristal} and Proposition \ref{upper_bd_cristal} to $\uE$ and $\uF(\cdot,\beta,\mu)$, giving that \textbf{(A2)} and \eqref{stability_of_matter_equation2b} are satisfied. It remains to prove that \textbf{(A3)}--\textbf{(A6)} are true. Here we start by stating the following proposition, which in particular implies that $\uE$ and $\uF(\cdot,\beta,\mu)$ satisfy the continuity property \textbf{(A4)}:
\begin{prop}[Controlling charge variations at the boundary]\label{prop_dipoles} Fix some $\eta\in\cE$ with $\eta(t)=a|t|$ and some $\varepsilon>0$. Let $\Omega,\Omega'\in\cR_\eta\cap\cC_\varepsilon$ such that $\Omega'\subseteq\Omega$ and ${\rm d}(\partial\Omega,\partial\Omega')>10\delta'$. Then there exists a constant $C$ (independent of $\Omega'$ and $\Omega$ but depending on $a$, $\varepsilon$, $\beta>0$ and $\mu\in\R$) such that
\begin{equation}
\oE(\Omega)\leq \uE(\Omega')+C|\Omega\setminus\Omega'|+o(|\Omega|),
\label{estim_dipoles_general1} 
\end{equation}
\begin{equation}
\oF(\Omega,\beta,\mu)\leq \uF(\Omega',\beta,\mu)+C|\Omega\setminus\Omega'|+o(|\Omega|).
\label{estim_dipoles_general2} 
\end{equation}
\end{prop}

\begin{figure}[h]
\centering
\includegraphics[width=9cm]{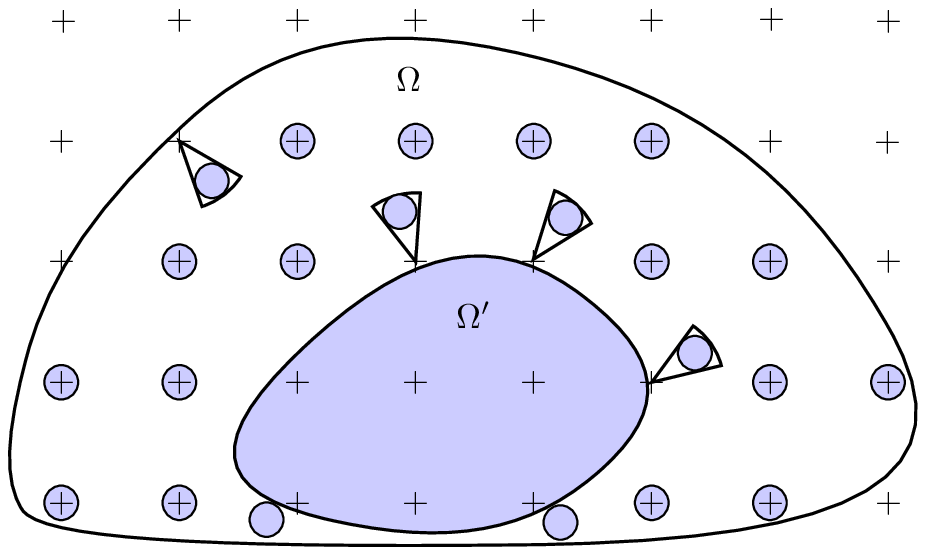}
\caption{Proof of the Continuity Property (A4).}
\label{image_preuve2}
\end{figure}

Fefferman already proved in \cite[Lemma 2]{F} a similar result. A different proof based on stability of matter is provided in Section \ref{proof_prop_dipoles}.
For the energy, the main idea of this proof (see Figure \ref{image_preuve2}) is to construct a trial state in $\Omega$ by taking the ground state of $\uE(\Omega')$ inside $\Omega'$ and by placing radial electrons around each nucleus in $\Omega\setminus\Omega'$ such that their mutual interaction cancels, yielding only an energy proportional to $|\Omega\setminus\Omega'|$. This will not be possible when a nucleus is too close to the boundary of $\Omega\setminus\Omega'$. In this case, thanks to the $\varepsilon$-cone property, we can place a ball close but not on top of the nucleus, creating a dipole. Notice the cone property is used both inside the boundary of $\Omega$ and outside the one of $\Omega'$. Also for the nuclei close to the boundary of $\Omega'$ there is a difference of charge between the problems $\uE(\Omega')$ and $\oE(\Omega)$ which we shall need to compensate by adding electrons outside $\Omega'$, creating also some dipoles. The difficult task compared to the proof of Theorem \ref{stability_of_matter} is then the estimate of the interaction between the system in $\Omega'$ and all the so-defined dipoles. This is done by means of a specific version of stability of matter, see Lemma \ref{stability_dipoles}.

\begin{remark}\rm \textbf{(Bosonic case)} Proposition \ref{prop_dipoles} is also true for bosons. 
\end{remark}

The generality of Proposition \ref{prop_dipoles} also allows us to prove the existence of the thermodynamic limit for the original functions $E^{\rm Sch}_\cL$ and $F^{\rm Sch}_\cL(\cdot,\beta,\mu)$, \emph{assuming we have proved that $\uE$ and $\uF$ satisfy} \textbf{(A3)}--\textbf{(A6)}. We explain that now. Consider a sequence $\{\Omega_n\}_{n\geq1} \subseteq\cR_{\eta}\cap\cC_{\varepsilon}$ with $|\Omega_n|\to\ii$ and choose some sequence $\ell_n\to\ii$ such that $\ell_n|\Omega_n|^{-1/3}\to0$. By \textbf{(P5)} (or Proposition \refI{reg_inner_approx}), we know that the inner approximation $A_n:=A_{0,\ell_n,Id}(\Omega_n)$ of $\Omega_n$ at a distance $\delta=10\delta'$ is a sequence of domains which satisfies $\{A_n\}\subseteq\cR'=\cR_{\eta'}\cap\cC_{\varepsilon'}$ and $|A_n|=|\Omega_n|+o(|\Omega_n|)$.
Equation \eqref{estim_dipoles_general1} gives us that
$$\uE(\Omega_n)\leq E^{\rm Sch}_\cL(\Omega_n)\leq \oE(\Omega_n)\leq \uE(A_n)+o(|\Omega_n|).$$
When $\uE$ satisfies \textbf{(A3)}--\textbf{(A6)} for any fixed $\eta$ and $\varepsilon$, we have by Theorem \refI{limit_general} $\lim_{n\to\ii}\uE(\Omega_n)|\Omega_n|^{-1}=\lim_{n\to\ii}\uE(A_n)|A_n|^{-1}=\bar e$ for some constant $\bar e$.
Hence $\lim_{n\to\ii}E^{\rm Sch}_\cL(\Omega_n)|\Omega_n|^{-1}=\bar e$. The argument is the same for $F(\cdot,\beta,\mu)$.

In conclusion, Theorem \ref{limit_crystal} will be proved provided we can show that $\uE$ and $\uF$ satisfy \textbf{(A3)}--\textbf{(A6)}. Notice Proposition \ref{prop_dipoles} already tells us in particular that $\uE$ and $\uF$ satisfy \textbf{(A4)}.

\subsubsection*{Step 4. Proof of the Translation Invariance in Average \textbf{(A3)}.} We prove that $\uE$ satisfies \textbf{(A3)}. Notice that the energy is $\Lambda$-periodic: $\forall \lambda\in\Lambda,\ \uE(\Omega+\lambda)=\uE(\Omega)$.
Let us introduce
$$\hat{e}(\Omega)=\frac{1}{|W|}\int_W\frac{\uE(\Omega+u)}{|\Omega|}du$$
where $W$ is the fundamental domain of $\Lambda$. Using that $\uE(\Omega)|\Omega|^{-1}$ is bounded uniformly for all $\Omega\in\cR_\eta\cap\cC_\varepsilon$, it is then easy to prove that 
$$\left||B(0,L)|^{-1}\int_{B(0,L)}\frac{\uE(\Omega+u)}{|\Omega|}du-\hat{e}(\Omega)\right|\leq \frac{C}{L}$$
where $C$ depends on $\eta$ and $\varepsilon$.
The same holds for the free energy $\uF(\cdot,\beta,\mu)$ with the definition
$$\hat{f}(\Omega,\beta,\mu)=\frac{1}{|W|}\int_W\frac{\uF(\Omega+u,\beta,\mu)}{|\Omega|}du.$$

\bigskip

It now remains to prove that $\uE$ and $\uF(\cdot,\beta,\mu)$ satisfy \textbf{(A5)} and \textbf{(A6)}. To this end, we shall use a localization method in Fock spaces which is explained in the Appendix. 

\subsubsection*{Step 5. Proof of \textbf{(A5)}.}
We start with the proof of \textbf{(A5)} for the energy $\uE$ and the free energy $\uF(\cdot,\beta,\mu)$.
Let $\theta^\ell=(\1_{\ell \triangle}*j)^{1/2}$, where $\triangle$ is the simplex introduced above which defines a
$\Gamma$-tiling and $j$ is a smooth non-negative radial function of compact support in $B(0,\delta'/10)$ with $\int j=1$. 
It was proved in \cite[p. 225]{GS} that $\theta^\ell$ is a $C^1$ function. Note that $\theta^\ell$ localizes near the simplex $\ell \triangle$ (at a distance independent of $\ell$) and that $\int(\theta^\ell)^2=|\ell \triangle|$.

We have 
$$\forall x\in\R^3,\qquad \frac1{|\ell\triangle|}\int d\lambda(g) \theta^\ell(g^{-1}x)^2=1$$
and $\theta^\ell(g^{-1}x)=(\1_{g\ell \triangle}*j)^{1/2}$ since $j$ is radial. Thus we obtain by the IMS localization formula for the kinetic energy on the one-body space
$$T=\frac1{|\ell\triangle|}\int d\lambda(g) \bigg((g\theta^\ell) T(g\theta^\ell)-|\nabla (g\theta^\ell)|^2\bigg)$$
where we have used the notation $gf(x)=f(g^{-1}x)$.
Now
$$\frac1{|\ell\triangle|}\int d\lambda(g) |\nabla (g\theta^\ell)|^2=\frac1{|\ell\triangle|}\int d\lambda(g) |g\nabla \theta^\ell|^2=\frac{\int|\nabla\theta^\ell|^2}{|\ell\triangle|}=O(\ell^{-1}).$$
Hence
\begin{equation}
\sum_i T_i\geq \frac1{|\ell\triangle|}\int d\lambda(g) \left(\sum_i (g\theta^\ell)_iT_i(g\theta^\ell)_i\right)-C\frac{\cN}{\ell}.
\label{estim_below_kinetic_IMS}
\end{equation}
For the Coulomb potential, we use the following adaptation of Proposition \ref{Prop_Graf_Schenker}, proved in \cite[Lemma 6]{GS}:
\begin{lemma}[Graf-Schenker Inequality with Smooth Localization]\label{GS_2} Assume that $\theta^\ell$ is defined as above. Then there exists a constant $C$ and a $\ell_0>0$ such that, for any $N\in\N$, $z_1,...,z_N\in\R$, $x_i\in\R^3$ and any $\ell\geq \ell_0$,
\begin{multline}
\sum_{1\leq i<j\leq N} \frac{z_iz_j}{|x_i-x_j|}\geq \frac{1-C/\ell}{|\ell\triangle|}\int_{G} d\lambda(g)\sum_{1\leq i<j\leq N}\frac{z_iz_j(g\theta^\ell)(x_i)^2(g\theta^\ell)(x_j)^2}{|x_i-x_j|}\\
+\frac{C}{\ell}\sum_{1\leq i<j\leq N}z_iz_jW(x_i-x_j)-\frac{C}{\ell}\sum_{i=1}^Nz_i^2
\end{multline}
where 
$W(x)=\frac{1}{|x|(1+|x|)}.$
\end{lemma}
Now, let us consider an $\Omega\in\cM$ and choose the charges close to the boundary of $\Omega$ to realize the infimum of $\uE(\Omega)$. Denoting $H_\Omega$ the so-obtained Hamiltonian acting on $\cF(L^2(\Omega))$, we obtain
\begin{equation}
H_\Omega\geq \frac{1-C/\ell}{|\ell\triangle|}\int_{G} d\lambda(g)H_{g\theta^{\ell}}\\
+\frac{C}{\ell}\left(\sum_iT_i+\mathbb{W}-\cN\right)-\frac{CK}{\ell} 
\label{estim_below_A5}
\end{equation}
where $K$ is the number of nuclei in $\Omega$, which (by the geometry of the crystal) is bounded above by a constant times $|\Omega|_{\rm r}$, the lowest volume of regular sets containing $\Omega$:
$$|\Omega|_{\rm r}=\inf\{|\tilde\Omega|,\ \tilde\Omega\supset\Omega,\ \tilde\Omega\in\cR_{\eta}\cap\cC_{\varepsilon}\}.$$
The operator $\mathbb{W}$ is the second quantization of the operator $W$ appearing in Lemma \ref{GS_2}, with the appropriate charges of the nuclei. 
The Hamiltonian $H_{g\theta^{\ell}}$ is defined as
\begin{multline}
H_{g\theta^{\ell}}=\sum_i(g\theta^\ell Tg\theta^\ell)_i-\sum_i\sum_{(R,z)\in\cL\cap(\Omega\times\R)}\frac{z(g\theta^\ell)(R)^2(g\theta^\ell)(x_i)^2}{|x_i-R|}\\
+ \frac12\sum_{i\neq j}\frac{(g\theta^\ell)(x_i)^2(g\theta^\ell)(x_j)^2}{|x_i-x_j|}+{\rm C}_{g\theta^{\ell}}, \label{localized_hamil_A5}
\end{multline}
with ${\rm C}_{g\theta^{\ell}}$ denoting the Coulomb interaction of the nuclei in ${g\theta^{\ell}}$
$${\rm C}_{g\theta^{\ell}}=\frac12\sum_{\substack{(R,z),(R',z')\in\cL\cap(\Omega\times\R),\\ R\neq R'}}\frac{(g\theta^\ell)(R)^2(g\theta^\ell)(R')^2}{|R-R'|}.$$
Notice the charges have been changed close to the boundary of $\Omega\cap (g\ell\triangle)$ due to the multiplication by the smooth cut-off function. The operator $\mathbb{W}$ satisfies a (weak) version of stability of matter. This is because we know from \cite[Thm A.1]{CLY1} that
\begin{multline}
 \frac12\sum_{i=1}^N(-\Delta)_{x_i}-\sum_{i=1}^N\sum_{k=1}^Kz_kY_\nu(x_i-R_k)+\frac12\sum_{i\neq j}Y_\nu(x_i-x_j)\\+\frac12\sum_{k\neq l}z_kz_lY_\nu(R_k-R_l)\geq -C(N+K)
\label{stability_Yukawa}
\end{multline}
where $Y_\nu(x)=e^{-\nu|x|}|x|^{-1}$ is the Yukawa potential and $C$ does not depend on $\nu$. 
Next, it suffices to remark like in \cite[p. 222]{GS} that
\begin{equation}
W(x)=\int_0^\ii e^{-\nu}Y_\nu(x).
\label{def_potential_Graf-Schenker} 
\end{equation}
Next one can estimate the number of nuclei inside $\Omega$ by $C|\Omega|_{\rm r}$ and obtain
\begin{equation}
\frac12\sum_iT_i+\mathbb{W}-\cN\geq-C|\Omega|_{\rm r}
\label{stability_W}
\end{equation}
where we have used one more time \eqref{estim_Lieb_Thirring_kinetic} to control $-\cN$. We obtain
\begin{equation}
H_\Omega\geq \frac{1-C/\ell}{|\ell\triangle|}\int_{G} d\lambda(g)H_{g\theta^{\ell}}\\
+\frac{C}{2\ell}\left(\sum_iT_i\right)-\frac{C|\Omega|_{\rm r}}{\ell}.
\label{estim_below_A5bis}
\end{equation}

Taking the expectation of this Hamiltonian on a ground state and using the theory of localization in Fock space as developed in the Appendix, one can prove the following lemma, whose proof is explained in Section \ref{proof_satisf_A5}.

\begin{lemma}\label{satisf_A5}
The energy $\Omega\mapsto\uE(\Omega)$ and the free energy $\Omega\mapsto\uF(\Omega,\beta,\mu)$ both satisfy \textnormal{\textbf{(A5)}}.
\end{lemma}

\begin{remark}\rm \textbf{(Bosonic case)} 
Equation \eqref{stability_Yukawa} is also true for bosons, provided the nuclei have a smallest distance from each other as this is the case for the crystal. One may also retain some part of the interaction between the negatively charged particles by the Lieb-Yau inequality. Next the Lieb-Thirring inequality used in \eqref{stability_W} has to be replaced by Proposition \ref{lemma:estim_nuclei}.
\end{remark}

\subsubsection*{Step 6. Proof of \textbf{(A6)}.}
We end the proof of Theorem \ref{limit_crystal} by proving \textbf{(A6)} for the free energy $\uF(\cdot,\beta,\mu)$. We do not write the proof for the energy $\uE(\Omega)$ for which the argument is even simpler. Consider a regular set $\Omega\in \cR=(\cR_\eta\cap\cC_\varepsilon)$ and fix some $g\in G$. We introduce as before the partition of unity $(\Theta^\ell_{g\mu})_{\mu\in\Gamma}$ where this time
\begin{equation}
 \Theta^\ell_g:=\left(\1_{\ell g\triangle}\ast j\right)^{1/2}.
\label{def_localization_A6}
\end{equation}
Note that $\sum_{\mu\in\Gamma}\big(\Theta_{g\mu}^\ell\big)^2=1$.
\begin{remark}\label{rmk:theta}\rm 
Notice $\ell$ and $g$ have been exchanged compared to the previous section to fit to the formalism of \textbf{(A6)}. More precisely
$$\left(g\theta^\ell(x)\right)^2=\int_{\R^3}\1_{g\ell\triangle}(y)j(x-y)\,dy,\ \ \left(\Theta_g^\ell(x)\right)^2=\int_{\R^3}\1_{\ell g\triangle}(y)j(x-y)\,dy.$$ 
\end{remark}

Let us choose the charges of the nuclei to minimize $\uF(\Omega,\beta,\mu)$. 
By the IMS localization formula, we have for the global Hamiltonian on $\Omega$
\begin{equation}
H_\Omega=\sum_{\mu\in\Gamma}H(g\mu) + \frac{1}{2}\sum_{\substack{\mu,\nu\in\Gamma\\ \mu\neq\nu}} I(g\mu,g\nu)
- \sum_i\sum_{\mu\in\Gamma}|\nabla \Theta^\ell_{g\mu}|^2_i
\label{decomp_H_Omega}
\end{equation}
where we have introduced the operators acting on the Fock space $\cF(L^2(\Omega))$
\begin{multline}
H(g\mu)= \sum_i(\Theta^\ell_{g\mu} T \Theta^\ell_{g\mu})_i-\sum_i\sum_{(R,z)\in\cK}\frac{z\Theta^\ell_{g\mu}(R)^2\Theta^\ell_{g\mu}(x_i)^2}{|x_i-R|}\\
+ \frac12\sum_{i\neq j}\frac{\Theta^\ell_{g\mu}(x_i)^2\Theta^\ell_{g\mu}(x_j)^2}{|x_i-x_j|}+\frac12\sum_{\substack{(R,z),(R',z')\in\cK,\\ R\neq R'}}\frac{\Theta^\ell_{g\mu}(R)^2\Theta^\ell_{g\mu}(R')^2}{|R-R'|}
\label{def_Hgmu}
\end{multline}
(compare this formula with \eqref{localized_hamil_A5}) and
\begin{multline}
I(g\mu,g\nu)=-\sum_i\sum_{(R,z)\in\cK}\frac{z\Theta^\ell_{g\mu}(R)^2\Theta^\ell_{g\nu}(x_i)^2+z\Theta^\ell_{g\nu}(R)^2\Theta^\ell_{g\mu}(x_i)^2}{|x_i-R|}\\
+\sum_{i\neq j}\frac{\Theta^\ell_{g\mu}(x_i)^2\Theta^\ell_{g\nu}(x_j)^2}{|x_i-x_j|}+\sum_{\substack{(R,z),(R',z')\in\cK,\\ R\neq R'}}\frac{\Theta^\ell_{g\mu}(R)^2\Theta^\ell_{g\nu}(R')^2}{|R-R'|}
\end{multline}
with $\cK\subseteq \Omega\times\R$ denoting the positions and (optimized) charges of the nuclei of the lattice $\cL$ inside $\Omega$. Notice in the above formulas, $\mu$ sums over a tiling whereas $g$ allows to translate and rotate this tiling.

Let us now consider a ground state $\omega$ (with density matrix $\Gamma$ and density of charge  $\rho_\omega$) for $\uF(\Omega,\beta,\mu)$. 
For any $\cP\subseteq\Gamma$, we introduce like in the Appendix, Equation \eqref{def_q_P},
\begin{equation}
 q_\cP:=\left(\sum_{\mu\in\cP}(\Theta^\ell_{g\mu})^2\right)^{1/2}
\label{def_q_cP}
\end{equation}
and denote by $\omega_\cP^g:=\omega_{q_\cP}$ the $q_\cP$-localized state associated with $\omega$. The theory of localization in Fock space is developed in Appendix, Section \ref{sec:localization}, in which the reader will find a precise definition of $\omega_{q_\cP}$. Let us now define the functions appearing in \textbf{(A6)}:
$$E^\Omega_\ell(g\mu):=\omega\left( H(g\mu)\right)-\beta^{-1}S(\omega^g_{\{\mu\}}),$$
$$I^\Omega_\ell(g\mu,g\nu):=\omega\left(I(g\mu,g\nu)\right),$$
$$s^\Omega_\ell(g,\cP):=\beta^{-1}\left(S(\omega_\cP^g)-\sum_{\mu\in\cP}S(\omega^g_{\{\mu\}})\right).$$

It is then possible to prove the following lemma. Details are given in Section \ref{proof_satisf_A6}.
\begin{lemma}\label{satisf_A6}
With the above definitions, Assumption \textnormal{\textbf{(A6)}} is satisfied.
\end{lemma}

\subsection{Quantum nuclei and electrons in a periodic magnetic field}
In this section we present our second example in which we come back to the simpler model of quantum nuclei. For simplicity we assume that there is only one kind of nuclei which are all bosons of charge $z$ with no spin. We could consider a finite number of different species of particles, each having its statistics, spin and charge.\footnote{The fermionic case is simpler than the bosonic case thanks to the Lieb-Thirring inequality. It can be treated using essentially the same method as before. We consider here bosons to give an illustration of the use of Proposition \ref{lemma:estim_nuclei}.} Also we could consider the (generalized) Hartree-Fock model but we do not write an explicit result.

Let $\Omega\in\cM$. We define the $K$-body bosonic space of nuclei and the $N$-body fermionic space of the electrons as
$$\cF^{\rm nuc}_K=\bigotimes_1^K\!{}_sL^2(\Omega),\qquad\cF^{\rm el}_N=\bigwedge_1^NL^2(\Omega),\qquad \cF^{\rm nuc}_0=\cF^{\rm el}_0=\C.$$
The subscript $s$ means that we take the symmetric tensor product. The total Fock space is 
$$\cF:=\bigoplus_{N,K\geq0}\cF^{\rm nuc}_K\otimes\cF^{\rm el}_N.$$
The total, electronic and nucleic number operators read respectively 
$$\cN:=\bigoplus_{N,K\geq0}(K,N),\qquad \cN_e:=\bigoplus_{N,K\geq0}N,\qquad \cN_p:=\bigoplus_{N,K\geq0}K.$$
The Coulomb many-body quantum Hamiltonian is defined on $\cF$ as
\begin{multline}
H_\Omega=\sum_jT(A)_{x_j}+\frac1M\sum_kT(A)_{R_k}-\sum_i\sum_k\frac{z}{|x_i-R_k|}\\
+\sum_{i<j}\frac{1}{|x_i-x_j|}+\sum_{k<\ell}\frac{z^2}{|R_k-R_\ell|}, 
\end{multline}
where $x_i$ and $R_k$ denote respectively the positions of the electrons and the nuclei. The number $M$ is (twice) the mass of the nuclei (the mass of the electrons is normalized to $1/2$). The operator $T(A)=(-i\nabla+A(x))^2$ is the Dirichlet magnetic Laplacian on $L^2(\Omega)$.
As before, we assume that $A\in L^2_{\rm loc}(\R^3)$ and that ${\rm div}(A)=0$ in the distributional sense. We also assume that there exists some discrete subgroup $\Gamma$ of $\R^3$ with compact fundamental domain $W=\R^3/\Gamma$, such that $B=\nabla\times A$ is a $\Gamma$-periodic distribution. The most simple example is of course the case of a constant magnetic field $B(x)=B_0$ which is $\Gamma$-periodic for instance for $\Gamma=\Z^3$.

The grand-canonical energy is defined as
$$\boxed{E(\Omega)=\inf\sigma_{\cF}(H_\Omega)}$$
and the free energy at temperature $\beta^{-1}>0$ and chemical potential $\mu=(\mu_1,\mu_2)\in\R^2$ reads
$$\boxed{F(\Omega,\beta,\mu)=-\frac1\beta \log\left(\tr_{\cF}\left(e^{-\beta(H_\Omega-\mu\cdot\cN)} \right)\right).}$$

An important result is the stability of matter:
\begin{thm}[Stability of matter for quantum nuclei and electrons in a periodic magnetic field]\label{thm_stability_LL}  Let $z>0$ and $A\in L^2_{\rm loc}(\R^3,\R^3)$ be such that ${\rm div}(A)=0$ in the distributional sense and $B=\nabla\times A$ is $\Gamma$-periodic.  

Assume that $\beta>0$ and $\mu\in\R^2$ Then there exists a constant $\kappa$ (depending on $\beta$, $\mu$ and $z$ but independent of $A$) such that for all $\Omega\in\cM$,
$$E(\Omega)\geq-\kappa|\Omega|,\qquad F(\Omega,\beta,\mu)\geq -\kappa|\Omega|.$$
\end{thm}
Whereas the stability of matter for the energy is an obvious consequence of Theorem \ref{stability_of_matter}, the proof for the free energy relies on Proposition \ref{lemma:estim_nuclei}. It is explained in Section \ref{proof_thm_stability_LL}.
Our main result is the
\begin{thm}[Thermodynamic limit for quantum nuclei and electrons in a periodic magnetic field]\label{thm_LL} Let $z>0$ and $A\in L^2_{\rm loc}(\R^3,\R^3)$ be such that ${\rm div}(A)=0$ in the distributional sense and $B=\nabla\times A$ is $\Gamma$-periodic.  

There exist a constant $\bar e$ and a function $\bar f:(0,\ii)\times\R^2\to \R$ such that the following holds: for any sequence $\{\Omega_n\}_{n\geq1}\subseteq \cR_\eta$ with $|\Omega_n|\to\ii$, ${\rm diam}(\Omega_n)|\Omega_n|^{-1/3}\leq C$,  and where $\eta(t)=a|t|^b\in\cE$ for some $a>0$ and $b\in(0,1]$,
$$\lim_{n\to\ii}\frac{E(\Omega_n)}{|\Omega_n|}=\bar e\qquad \text{and}\qquad \lim_{n\to\ii}\frac{F(\Omega_n,\beta,\mu)}{|\Omega_n|}=\bar f(\beta,\mu).$$ 
\end{thm}

\smallskip

We outline a proof based on our new general method in Section \ref{sec_proof_LL}.

\begin{remark}\rm
 Theorem \ref{thm_LL} was proved by Lieb and Lebowitz in \cite{LL} in the case $A\equiv0$. Their proof which is based on the rotation invariance cannot be adapted when $A\neq0$.
\end{remark}

\begin{remark}\rm
 Notice we can prove the existence of the thermodynamic limit for a larger class of sequences $\{\Omega_n\}_{n\geq1}$ than for the crystal case: we do not need the cone property and we do not need that $b=1$. In \cite{LL}, an even weaker condition is assumed to hold for $\{\Omega_n\}_{n\geq1}$. 
\end{remark}

\subsection{Movable classical nuclei}\label{sec:classical_nuclei}
In this section, we state the existence of the thermodynamic limit for our third and last example, a system composed of quantum electrons and classical nuclei, whose position is optimized. Surprisingly, this does not seem to have already been done in the literature. We do not consider any magnetic field in this section, $A\equiv0$.

We recall that for any set of nuclei $\cK=\{(R_k,z_k)\}\subset\Omega\times[0,\ii)$ in a domain $\Omega\in\cM$, the associated grand-canonical Coulomb Hamiltonian acting on $\cF(L^2(\Omega))$ is denoted by $H_{\Omega,\cK}$, see \eqref{form_Coulomb_Hamiltonian_general}.
We consider a system where we put arbitrarily many nuclei of charge $z$ in $\Omega$ and optimize both their number and positions. The associated ground state energy is
\begin{equation}
 \boxed{E(\Omega) =  \inf_{\substack{\cK\subseteq\Omega\times\{z\},\\ \#\cK<\ii}}\inf\sigma_{\cF(L^2(\Omega))}(H_{\Omega,\cK}).}
\label{nucl_E}
\end{equation}
Another problem consists in optimizing also the charges of the nuclei inside $\Omega$, limiting their higher value to $z$. This leads to the following definition
\begin{equation}
\boxed{\uE(\Omega) =  \inf_{\substack{\cK\subseteq\Omega\times[0,z],\\ \#\cK<\ii}}\inf\sigma_{\cF(L^2(\Omega))}(H_{\Omega,\cK}).}
\label{nucl_E2}
\end{equation}
By \eqref{formula_Daubechies_Lieb} in Theorem \ref{stability_of_matter}, we know that
\begin{equation}
 E(\Omega)=\uE(\Omega).
\label{Lieb_Daubechies_mov}
\end{equation}

We consider a chemical potential $\mu=(\mu_1,\mu_2)\in\R^2$. The free energy corresponding to \eqref{nucl_E} is
\begin{equation}
\fbox{$\begin{array}{l}
\displaystyle F(\Omega,\beta,\mu)  =  -\frac1{\beta}\log\Biggl(\sum_{K\geq0}\frac{1}{K!}\int_{\Omega^K}dR_1\cdots dR_K\times\\
\displaystyle\qquad\qquad\qquad\qquad\qquad \times\tr_{\cF(L^2(\Omega))}\left(e^{-\beta\left(H_{\Omega,\{(R_i,z)\}}-\mu_1\cN-\mu_2K\right)}\right)\Biggr).       
      \end{array}$}
\label{nucl_FE}
\end{equation}
Similarly, the free energy corresponding to \eqref{nucl_E2} is
\begin{equation}
\fbox{$\begin{array}{l}
\displaystyle \uF(\Omega,\beta,\mu)  =  -\frac1{\beta}\log\Biggl(\sum_{K\geq0}\frac{1}{K!}\int_{(\Omega\times[0,z])^K}dR_1\cdots dR_K\, dz_1\cdots dz_K\times\\
\displaystyle \qquad\qquad\qquad\qquad\qquad\times\tr_{\cF(L^2(\Omega))}\left(e^{-\beta\left(H_{\Omega,\{(R_i,z_i)\}}-\mu_1\cN-\mu_2K\right)}\right)\Biggr),
      \end{array}$}
\label{nucl_FE2}
\end{equation}
but we only have $\uF(\Omega,\beta,\mu)\leq F(\Omega,\beta,\mu)$.
As usual, the first step is to prove stability of matter:
\begin{thm}[Stability of Matter for Optimized Classical Nuclei]\label{stability_class_nuclei}
Let $z>0$, $\beta>0$ and $\mu\in\R^2$. There exists a constant $\kappa$ such that
\begin{equation}
E(\Omega)=\uE(\Omega)\geq-\kappa|\Omega|,\qquad  F(\Omega,\beta,\mu)\geq\uF(\Omega,\beta,\mu)\geq-\kappa|\Omega|
\label{stability_class_nuclei_eq}
\end{equation}
for all $\Omega\in\cM$.
\end{thm}
\begin{proof}
For the energy $\uE(\Omega)$, this is contained in Theorem \ref{stability_of_matter}. For the free energy, we use \eqref{stability_of_matter_FE_equation} for any fixed configuration $(R_k,z_k)_{k=1}^K$ of the nuclei:
$$\tr_{\cF(L^2(\Omega))}\left(e^{-\beta\left(H_{\Omega,\{(R_k,z_k)\}}-\mu_1\cN\right)}\right)\leq e^{C|\Omega|}$$
for some constant $C$ independent of $(R_k,z_k)_{k=1}^K$. Hence we obtain
\begin{equation*}
 \uF(\Omega,\beta,\mu)  \geq  -\frac1{\beta}\log\biggl(\sum_{K\geq0}\frac{|\Omega|^Kz^K}{K!}e^{C|\Omega|+\beta\mu_2K}\biggr)=-\frac{\left(C+ze^{+\beta\mu_2}\right)|\Omega|}{\beta}.
\end{equation*}
\end{proof}

For the thermodynamic limit we give in Section \ref{sec_proof_mov} the proof of the following result.
\begin{thm}[Thermodynamic limit for movable nuclei]\label{thm_mov}
Let be $z>0$  and $\mu\in\R^2$.
There exists a constant $\bar e$ and two functions $\bar{f},\bar{g}:(0,\ii)\times\R^2\to \R$ such that the following holds: for any sequence $\{\Omega_n\}_{n\geq1}\subseteq \cR_\eta$ with $|\Omega_n|\to\ii$, ${\rm diam}(\Omega_n)|\Omega_n|^{-1/3}\leq C$,  and where $\eta(t)=a|t|^b\in\cE$ for some $a>0$ and $b\in(0,1]$,
$$\lim_{n\to\ii}\!\frac{E(\Omega_n)}{|\Omega_n|}=\bar e,\   \lim_{n\to\ii}\!\frac{F(\Omega_n,\beta,\mu)}{|\Omega_n|}=\bar f(\beta,\mu)\  \text{and}\  \lim_{n\to\ii}\!\frac{\uF(\Omega_n,\beta,\mu)}{|\Omega_n|}=\bar{g}(\beta,\mu).$$ 
\end{thm}

\section{Proofs}
\subsection{The stability of matter: proof of Theorem \ref{stability_of_matter}}\label{sec:stability_of_matter}
\paragraph*{Step 1. Proof of \eqref{formula_Daubechies_Lieb}.}
We start by proving \eqref{formula_Daubechies_Lieb}, following ideas from Daubechies-Lieb \cite{DL}. For simplicity, we use the notation of Section \ref{sec:classical_nuclei} and introduce:
\begin{equation}
 E(\Omega) =  \inf_{\substack{\cK\subseteq\Omega\times\{z\},\\ \#\cK<\ii}}\inf\sigma_{\cF(L^2(\Omega))}(H_{\Omega,\cK}),
\label{nucl_Ebis}
\end{equation}
\begin{equation}
 \uE(\Omega) =  \inf_{\substack{\cK\subseteq\Omega\times[0,z],\\ \#\cK<\ii}}\inf\sigma_{\cF(L^2(\Omega))}(H_{\Omega,\cK}).
\label{nucl_E2bis}
\end{equation}
It is clear that $\uE(\Omega)\leq E(\Omega)$ for all $\Omega\in\cM$.
Fix now a positive number $K$ and some positions $R_1,...,R_K\in\Omega$ with $R_k\neq R_{k'}$ when $k\neq k'$. We introduce the ground state energy when the nuclei have charges $(z_1,...,z_K)\in[0,z]^K$:
$$f(z_1,...,z_K)=\inf\sigma_{\cF(L^2(\Omega))}\left(H_{\Omega,\{(R_k,z_k)\}}\right).$$
Notice $H_{\Omega,\{(R_i,z_i)\}}$ depends linearly on each $z_k$ separately, hence $f$ is a concave function in each variable separately. By induction, one proves that for any $z_1,...,z_K\in[0,z]^K$, there exists $z'_1,...,z'_K\in\{0,z\}^K$ such that
$$f(z_1,...,z_K)\geq f(z'_1,...,z'_K).$$
Of course taking some charges equal to zero is the same as removing the associated nuclei,
$H_{\Omega,\{(R_k,z'_k)\}}=H_{\Omega,\{(R_k,z)\ |\ z'_k\neq0\}},$
therefore $f(z'_1,...,z'_K)\geq E(\Omega)$ which eventually shows that $\uE(\Omega)\geq E(\Omega)$.

\paragraph*{Step 2. The stability of matter.} We now prove \eqref{stability_of_matter_equation}. 
By the Lieb-Yau inequality \eqref{eq:Lieb-Yau}, we have
\begin{equation*}
 H_{R\times\{z\}}\geq \sum_jT(A)_j-\sum_i\frac{z+\sqrt{2z}+1/2}{\delta_R(x_i)}+\frac{z^2}{4}\sum_k\frac1{\delta_R(R_k)}.
\end{equation*}

\begin{lemma}\label{lemma:stability_LL}
There exists a constant $C>0$ such that for any $R\subset\R^3$, 
 \begin{equation}
H^R_\Omega:=\frac12\sum_jT(A)_j-\sum_i\frac{z+\sqrt{2z}+1/2}{\delta_R(x_i)}+\frac{z^2}{8}\sum_k\frac1{\delta_R(R_k)}\geq-C|\Omega|.
\label{eq:stability_LL}
\end{equation}
\end{lemma}

\smallskip

Using \eqref{eq:stability_LL}, we obtain
\begin{equation}
 H_{R\times\{z\}}\geq -C|\Omega|+\frac12 \sum_jT(A)_j+\frac{z^2}{8}\sum_k\frac1{\delta_R(R_k)}.
\label{estim_Hamil_Lieb_Yau}
\end{equation}
This clearly yields \eqref{stability_of_matter_equation}, as $\sum_jT(A)_j\geq0$.

\begin{proof}[Proof of Lemma \ref{lemma:stability_LL}]
Let $R=\{R_1,...,R_K\}$ be some positions of the nuclei in $\R^3$. We use the notation $W(x):=(z+\sqrt{2z}+1/2)\delta_R(x)^{-1}$ and fix some constant $\delta>0$. 
For any $k$, we introduce $D_k:=\delta_R(R_k)$, the distance to the closest nucleus.
Next we define the following three sets
$$X_\delta=\bigcup_{k\ |\ D_k\leq 4\delta}B(R_k,\delta),\quad X'_\delta=\bigcup_{k\ |\ D_k> 4\delta}B(R_k,\delta),\quad Y_\delta=\R^3\setminus(X_\delta\cup X'_\delta).$$
Then by the definition of $W$
\begin{multline}
H^R_\Omega\geq \sum_j\left(\frac{T(A)}{4} -W(\1_{X_\delta}+\1_{X_\delta'})\right)_j+\frac{z^2}{8}\sum_{k=1}^K\frac{1}{D_k}\\
+\sum_j\frac{T(A)_j}{4}-\frac{(z+ \sqrt{2z}+1/2)}\delta\cN.
\end{multline}
By the Lieb-Thirring inequality \eqref{Lieb-Thirring_first_formulation} and assuming $\delta\leq1$, we have
\begin{multline*}
\sum_j\left(\frac{T(A)}{8} -W\1_{X_\delta}\right)_j \geq -C\int_{X_\delta}|W(x)|^{5/2}dx\\
 \geq  -C\sum_{k\ |\ D_k\leq 4\delta}\int_{B(R_k,1)}\frac{(z+ \sqrt{2z} +1/2)^{5/2}}{|x-R_k|^{5/2}}dx=-C'n_\delta
\end{multline*}
where $n_\delta:=\#\{k\ |\ D_k\leq4\delta\}$. Then, by the definition of $X_\delta$,
\begin{equation}
\sum_j\left(\frac{T(A)}{8} -W\1_{X_\delta}\right)_j+\frac{z^2}{8}\sum_{k=1}^K\frac{1}{D_k}\geq \left(-C' + \frac{z^2}{32\delta}\right)n_\delta\geq 0
\end{equation}
if we choose $\delta\leq z^{2}/(32C')$.

We now estimate $\sum_j\left(T(A)/8 -W\1_{X'_\delta}\right)_j$. Let $0\leq\theta\leq1$ be a smooth radial function which vanishes outside the ball $B(0,2)$ and is equal to 1 in $B(0,1)$. Let us define $\theta_k(x)=\theta((x-R_k)/\delta)$, whose support is contained in $B(R_k,2\delta)$. By definition of $X'_\delta$, all the $B(R_k,2\delta)$ are disjoint when $k$ varies in
$\cI_\delta:=\{k\ |\ D_k >4\delta\}$.
Then, let $0\leq \Theta\leq 1$ be such that $\sum_{k\in\cI_\delta}\theta_k^2+\Theta^2=1$. We notice that 
$$\theta_k\frac{1}{|x-R_k|}\theta_k\leq C\left(\int_{B(0,2\delta)}\frac{dx}{|x|^{3/2}}\right)^{2/3}\theta_k(-\Delta)\theta_k$$
by the critical Sobolev injection of $H^1(\R^3)$ in $L^6(\R^3)$. Hence by the diamagnetic inequality \cite{Lieb-Loss},
$$\theta_k\frac{1}{|x-R_k|}\theta_k\leq C\left(\int_{B(0,2\delta)}\frac{dx}{|x|^{3/2}}\right)^{2/3}\theta_kT(A)\theta_k.$$
Thus, we may find a fixed small enough $\delta$ such that
$$\theta_k\left(\frac{T(A)}{8}-W\1_{X'_\delta}\right)\theta_k\geq0$$
for any $k\in\cI_\delta$. Using that
$$T(A)\geq \sum_{k\in\cI_\delta}\theta_kT(A)\theta_k+\Theta T(A)\Theta -\frac{C}{\delta^2}$$
for some constant $C$ depending only on $\theta$, we infer
\begin{equation}
\sum_j\left(\frac{T(A)_j}{8} -W\1_{X'_\delta}\right)_j \geq  -\frac{C}{\delta^2}\cN
\label{estim_kinetic_local_Sobolev}
\end{equation}
where we recall that $\delta$ is now a fixed constant. Hence we have proved that
\begin{equation*}
H^R_\Omega\geq \sum_j\frac{T(A)_j}{4}-C\cN
\label{derniere_estim_Hamiltonien}
\end{equation*}
for some other constant $C$ depending on $z$. Using the Lieb-Thirring inequality \eqref{estim_Lieb_Thirring_kinetic} we obtain
\begin{equation*}
H^R_\Omega\geq C_{\rm LT}\cN^{5/3}|\Omega|^{-2/3}-C\cN
\label{toute_derniere_estim_Hamiltonien}
\end{equation*}
which gives \eqref{eq:stability_LL} when optimized over $N$.
\end{proof}

\paragraph*{Step 3. The stability of matter for the free energy.} We end this section by proving \eqref{stability_of_matter_FE_equation}.
By \eqref{stability_of_matter_equation_bis} (with $T(A)$ replaced by $T(A)/2$), we have for any $\cK\subset\Omega\times[0,z]$
\begin{equation}
H_{\Omega,\cK}\geq \frac12 \sum_jT(A)_j-C|\Omega|.
\label{estim_Hamil_stability_with_kinetic_any_charge}
\end{equation}
Using again the Lieb-Thirring inequality \eqref{estim_Lieb_Thirring_kinetic}, we obtain for any $\mu\in\R$
\begin{equation}
H_{\Omega,\cK}-\mu\cN\geq \frac14 \sum_jT(A)_j-C(1+\mu_+^{5/2})|\Omega|
\label{estim_Hamil_stability_with_kinetic_any_charge2}
\end{equation}
where the constant $C$ is independent of $\mu$. It is then a consequence of Peierls' inequality \cite[Prop. 2.5.5]{Ruelle} that
\begin{multline}
-\frac1{\beta}\log\tr_{\cF(L^2(\Omega))}\left(e^{-\beta(H_{\Omega,\cK}-\mu\cN)} \right)\geq -\frac1\beta\log\tr_{\cF(L^2(\Omega))}\left(e^{-\beta(\sum_jT(A)_j)/4}\right)\\-C(1+\mu_+^{5/2})|\Omega|. 
\label{estimate_free_energy_TA2}
\end{multline}
Notice the following equality
\begin{equation}
 \tr_{\cF(L^2(\Omega))}\left(e^{-\beta(\sum_jT(A)_j)/4}\right)=\prod_{k\geq1}\left(1+e^{-\beta(\lambda_k(A))/4}\right)
\label{formula_trace_product}
\end{equation}
where $(\lambda_k(A))$ are the eigenvalues of $T(A)$ on $\Omega$ with Dirichlet boundary conditions and from which we infer
\begin{equation}
 \log\tr_{\cF(L^2(\Omega))}\left(e^{-\beta(\sum_jT(A)_j)/4}\right)=\tr_{L^2(\Omega)}f(T(A))
\label{formula_free_energy_TA}
\end{equation}
with $f(t)=\log(1+e^{-\beta t/4})$. By the diamagnetic inequality \cite{Lieb-Loss,Simon} we have
\begin{equation}
\tr_{L^2(\Omega)}f(T(A))\leq \tr_{L^2(\Omega)}\left(e^{-\beta T(A)/4}\right)\leq \tr_{L^2(\Omega)}\left(e^{-\beta T(0)/4}\right).
\label{estimate_free_energy_TA}
\end{equation}

We then prove the following lemma whose proof follows that of an estimate of Li and Yau \cite{Li-Yau} (see also \cite[Thm 12.3]{Lieb-Loss}):
\begin{lemma}\label{Li-Yau} Let $\Omega$ be a bounded open set in $\R^n$, and $f:[0,\ii)\to\R$ a convex function such that $t\mapsto t^{(n-2)/2}f(t)$ is in $L^1([0,\ii),\R)$. We use the notation $-\Delta$ for the Dirichlet Laplacian on $\Omega$ and $-\Delta_f$ for the free Laplacian acting on the whole space $\R^3$. Then $f(-\Delta)$ is a trace-class operator on $L^2(\Omega)$ and
\begin{equation}
\tr_{L^2(\Omega)}\left(f(-\Delta)\right)\leq |\Omega|(2\pi)^{-n}\int_{\R^n}f(|p|^2)dp=\tr_{L^2(\R^3)}(\1_\Omega(x)f(-\Delta_f)).
\label{inequality_Li_Yau}
\end{equation}
\end{lemma}
\begin{proof}
Let $(\phi_i)$ be an orthonormal system of eigenfunctions of $-\Delta$. Then, using Jensen's inequality,
\begin{eqnarray*}
\tr_{L^2(\Omega)}\left(f(-\Delta)\right) & = & \sum_if(\pscal{(-\Delta)\phi_i,\phi_i}) = \sum_if\left(\int_{\R^3}|p|^2|\widehat{\phi_i}(p)|^2dp\right)\\
 & \leq & \int_{\R^3}f\left(|p|^2\right)\sum_i|\widehat{\phi_i}(p)|^2dp.
\end{eqnarray*}
Now, introducing $e_p(x)=(2\pi)^{-n/2}e^{ip\cdot x}\1_\Omega(x)$, we have
$$\sum_i|\widehat{\phi_i}(p)|^2=\sum_i|\pscal{\phi_i,e_p}|^2\leq \norm{e_p}_{L^2(\Omega)}^2=(2\pi)^{-n}|\Omega|$$
which ends the proof of Lemma \ref{Li-Yau}.
\end{proof}

\medskip

We now use this to finish the proof of \eqref{stability_of_matter_FE_equation}. We have
$$\log\tr_{\cF(L^2(\Omega))}\left(e^{-\beta\sum_jT(0)_j/4}\right)\leq \beta^{-3/2}(2\pi)^{-3}|\Omega|\int_{\R^3}e^{-|p|^2/4}dp$$
and therefore, by \eqref{estimate_free_energy_TA2}, \eqref{formula_free_energy_TA} and \eqref{estimate_free_energy_TA}
$$-\frac1{\beta}\log\tr_{\cF(L^2(\Omega))}\left(e^{-\beta(H_{\Omega,\cK}-\mu\cN)} \right)\geq -C\left(1+\beta^{-5/2}+\mu_+^{5/2}\right)|\Omega|.$$
This ends the proof of Theorem \ref{stability_of_matter}.\qed

\subsection{Proof of Proposition \ref{lemma:estim_nuclei}: an estimate for repelling particles}\label{proof_estim_nuclei}
We cut the space into cubes $Q_j$, $j\in {\bf Z}^3$ of side length $a>0$ (to be chosen below), 
and get a lower bound on the energy by introducing Neumann boundary
conditions on the boundaries of the $Q_j$. 

Let us consider a specific
cube $Q$ and the intersection $Q\cap\Omega$. Assume that we have $N'$
particles in this cube. Consider the operator \eqref{estim_nuclei}
\begin{equation}
\sum_{i=1}^{N'}\left(T(A)_{x_i}+\epsilon\max_{k\neq  i}|x_i-x_k|^{-1}\right)
\label{estim_nuclei2}
\end{equation}
restricted to the cube $Q$ with Neumann
boundary conditions on $\partial Q$ and Dirichlet boundary conditions on $\partial \Omega$. The ground state energy of this so-defined operator is bounded below by 0 if $N'=0$ and if $N'>0$ by
$$
N'\lambda(A)+(N'-1)\epsilon a^{-1}
$$
where $\lambda(A)$ is the lowest eigenvalue of $T(A)$ on $Q\cap\Omega$. By the diamagnetic inequality, we have $\lambda(A)\geq\lambda(0)$.

We now estimate $\lambda(0)$. 
For all functions $u$ we have a
Sobolev inequality
$$
\left(\int_Q|u|^6\right)^{1/3}\leq C_1 \int_Q|\nabla u|^2+C_0|Q|^{-2/3}\int_Q|u|^2.
$$
Using Hölder's inequality, we get if $u$ has its support in $\Omega$ that 
$$
\int_Q|u|^2\leq\left(\int_Q|u|^6\right)^{1/3}|Q\cap\Omega|^{2/3}.
$$
Thus if $|Q\cap\Omega|\leq
(1/2C_0)^{3/2}|Q|$ then 
\begin{eqnarray*}
 (2C_1)^{-1}|Q\cap\Omega|^{-2/3}\int_Q|u|^2 & \leq & C_1^{-1}
|Q\cap\Omega|^{-2/3}\left(1-C_0\frac{|Q\cap\Omega|^{2/3}}{|Q|^{2/3}}\right)\int_Q|u|^2\\
 & \leq & \int_Q|\nabla u|^2.
\end{eqnarray*}
Hence we conclude that if $|Q\cap\Omega|\leq
(1/2C_0)^{3/2}|Q|$ then 
\begin{equation}
\lambda(A)\geq\lambda(0)\geq (2C_1)^{-1}|Q\cap\Omega|^{-2/3}\geq 
(2C_1)^{-1}|Q|^{-2/3}=(2C_1)^{-1}a^{-2}.
\label{estim_lambda_no1} 
\end{equation}

We therefore have two cases. If $|Q\cap\Omega|\geq
(1/2C_0)^{3/2}|Q|$ then we estimate $\lambda(A)\geq0$ and find that 
the energy in the cube is bounded below by
\begin{align*}
 \epsilon(N'  a^{-1} -  a^{-1})&\geq \epsilon\left(N' a^{-1} - C (|Q\cap
\Omega|/|Q|)a^{-1}\right)\\
&=\epsilon\left(N' a^{-1} - C |Q\cap
\Omega|a^{-4}\right).
\end{align*}
On the other hand if $|Q\cap\Omega|\leq
(1/2C_0)^{3/2}|Q|$ we use \eqref{estim_lambda_no1} and obtain 
that the energy in the cube is bounded below by
$$
(N'-1)_+\min\{\epsilon a^{-1},(2C_1)^{-1}a^{-2}\}\geq c'N'\min\{a^{-1},a^{-2}\}. 
$$
If we sum these estimates over all cubes we obtain the total lower bound
$$
c'N\min\{a^{-1},a^{-2}\}- C |\Omega|a^{-4}.
$$
Optimizing in $a$ gives the claimed result.
\qed

\subsection{The crystal case: details for the proof of Theorem \ref{limit_crystal}}\label{proof_limit_crystal}
\subsubsection{Proof of Proposition \ref{upper_bd_cristal}: upper bounds}\label{proof_lemma_stability_cristal}
We start by proving \eqref{stability_of_matter_equation2b} for the energy.
Let be $\Omega\in\cR_\eta\cap\cC_\varepsilon$ and consider $r=\min(\delta/10,\varepsilon/10)$, where we recall that $\delta$ was defined in \eqref{hyp_D3}. We use the notation $\cL'\cap\Omega\times\R:=\{(R_k,z_k)\}_{k=1}^K$ with $\text{d}(R_k,R_k')>\delta$ for all $k\neq k'$. Since $\Omega$ has the $\varepsilon$-cone property, we know that each point $R_k\in\Omega$, has a ball $B_k$ of radius $r$ located at a distance at most $\delta/3$ which is completely contained in $\Omega$. When the ball centered at $R_k$ is already contained in $\Omega$, $B(R_k,r)\subseteq\Omega$, we simply choose this ball:  $B_k=B(R_k,r)$. We denote by $K_1$ the set of all the $k$'s which satisfy this property and by $K_2=\{1,...,K\}\setminus K_1$ its complementary. By definition of $r$, none of the balls $B_k$ intersect. Since $\Omega$ has an $\eta$-regular boundary, it is clear that
$r^3\#K_2\leq C|\Omega|^{2/3}r$ and on the other hand $r^3\#K_1\leq C|\Omega|$.

Let $\bar\Psi$ be a normalized eigenvector of $\sum_jT_j$ in $\cF(H^1_0(B(0,r)))$ with total momentum zero and such that $\cN\bar\Psi=([\bar z]+1)\bar\Psi$. In each Fock space $\cF(L^2(B_k))$, we can then consider the following state of total charge $z_k$
$$\Gamma_k=\left(1-\frac{z_k}{[\bar z]+1}\right)|0_k\rangle\langle0_k| + \frac{z_k}{[\bar z]+1}|\bar\Psi(\cdot-R_k)\rangle\langle\bar\Psi(\cdot-R_k)|$$
where $|0_k\rangle$ is the vacuum of $\cF(L^2(B_k))$.

By the definition of the balls $B_k$'s, we have  $L^2(\cup_kB_k)=\bigoplus_{k}L^2(B_k)$ which translates on the Fock spaces as
$$\cF(L^2(\Omega))\supseteq\cF(L^2(\cup_kB_k))\simeq \bigotimes_k\cF(L^2(B_k)).$$
Thus we can formally take $\Gamma=\bigotimes_k\Gamma_k$ as a test state for $H_{\Omega,\cL'}$ in $\Omega$. By Newton's theorem, the energy is then the sum of the energies in each ball, plus the interaction energy between classical dipoles placed at each $R_k$ with $k\in K_2$, i.e.
$$E^{\rm Sch}_{\cL'}(\Omega)\leq \tr\left(H_{\Omega,\cL'}\Gamma\right)\leq C|\Omega|+ \sum_{k\neq k'\in K_2}\frac{C}{|R_k-R_{k'}|^3}$$
for some constant $C$. For any fixed $k$, the last term attains its minimum if all the other $R_{k'}$ are distributed regularly in a ball of radius proportional to $(\# K_2)^{1/3}$. Thus
\begin{eqnarray}
\sum_{k\neq k'\in K_2}\frac{1}{|R_k-R_{k'}|^3}  & \leq & C(\#K_2)\int_{B(\delta/3,(\# K_2)^{1/3})}\frac{dr}{|r|^3}\nonumber\\
 & \leq &  C'(\#K_2)\log(\#K_2) \leq C''|\Omega|^{2/3}\log(|\Omega|),\label{estim_interac_dipoles}
\end{eqnarray}
which proves \eqref{stability_of_matter_equation2b} for the energy.

We now prove \eqref{estim_GS_LTb}. By Theorem \ref{stability_of_matter}, we have
$$H_{\Omega,\cL'}\geq \frac12\sum_jT_j-C|\Omega|.$$
Hence any ground state $\Gamma$ of $H_{\Omega,\cL'}$ satisfies
\begin{equation}
\frac{1}{2} \tr\left[\left(\sum_jT_j\right)\Gamma\right]-C|\Omega|\leq \tr(H_{\Omega,\cL'}\Gamma)\leq C'|\Omega|.
\label{final_eq_stability}
\end{equation}
By the Lieb-Thirring inequality \eqref{estim_Lieb_Thirring_kinetic},
$$C_{\rm LT}\tr(\cN\Gamma)^{5/3}|\Omega|^{-2/3}\leq  \tr\left[\left(\sum_jT_j\right)\Gamma\right]\ \ \text{and}\ \  \int_\Omega\rho_\Gamma=\tr(\cN\Gamma)\leq C'|\Omega|$$
for some constant $C'$. This finishes the proof for the energy since by \eqref{final_eq_stability}
$$C_{\rm LT}\int\rho_\Gamma^{5/3}\leq \tr\left[\left(\sum_jT_j\right)\Gamma\right]\leq C''|\Omega|.$$

For the free energy, since $x\log x\leq 0$ for all $x\in[0,1]$, we can use a ground state $\Gamma$ for $E^{\rm Sch}_{\cL'}(\Omega)$ as trial state to obtain
$$F^{\rm Sch}_{\cL'}(\Omega,\beta,\mu)\leq E^{\rm Sch}_{\cL'}(\Omega)-\mu\tr(\cN\Gamma)\leq (C'-\min(\mu,0))|\Omega|$$
by Theorem \ref{stability_of_matter}.
The proof of \eqref{estim_GS_LTb} is the same as for the energy. \qed

\subsubsection{Proof of Proposition \ref{reduc_periodic}: reduction to periodic case}\label{proof_reduc_periodic}
For the energy, we only prove the inequality
\begin{equation}
E^{\rm Sch}_{\cL}(\Omega_n)\geq E^{\rm Sch}_{\cL'}(\Omega_n)+o(|\Omega_n|),
\label{reduc_equ_cristal}
\end{equation}
the other one being obtained by the same argument. Let $\Psi_n$ be a ground state of $H_{\cL\cap\Omega_n\times\R}$ in $\cF(H^1_0(\Omega_n))$. By \eqref{estim_GS_LTb}, we know that there exists a constant $C$ independent of $n$ such that
$$\int\rho_{\Psi_n}\leq C|\Omega_n|,\qquad \int\rho_{\Psi_n}^{5/3}\leq C|\Omega_n|.$$
Then, we write
\begin{equation}
H_{\Omega_n,\cL}=H_{\Omega_n,\cL'}+\sum_{i}(W_n(x_i)+W_n'(x_i))+c_n
\end{equation}
where
$$W_n(x)=\sum_{(R,z)\in\cL\cap\Omega_n\times\R}\frac{D_{\rm ch}(z)}{|x-R|}+\sum_{(R',z')\in \cL_{\rm d}}\frac{z'}{|x-R'|}$$
$$W_n'(x)=\sum_{(R,z)\in\cL\cap\Omega_n\times\R}(z+D_{\rm ch}(z))\left(\frac{1}{|x-R-D_{\rm dis}(R)|}-\frac{1}{|x-R|}\right)$$
and
\begin{align*}
c_n=&\sum_{\substack{(R,z)\in\cL,\ (R',z')\in\cL\\ R\neq R'}}(z+D_{\rm ch}(z))(z'+D_{\rm ch}(z'))\times\\
&\qquad\qquad\qquad\times\left(\frac1{|R-R'|}-\frac1{|R-R'+D_{\rm dis}(R)-D_{\rm dis}(R')|}\right)\\
&\quad- \sum_{\substack{(R,z)\in\cL,\ (R',z')\in\cL\\ R\neq R'}}\frac{zD_{\rm ch}(z')+(z'+D_{\rm ch}(z'))D_{\rm ch}(z)}{|R-R'|}\\
&\quad-\sum_{(R,z)\in\cL,\ (R',z')\in\cL_{\rm d}}\frac{zz'}{|R-R'|}-\sum_{(R,z)\in\cL_{\rm d},\ (R',z')\in\cL_{\rm d}}\frac{zz'}{|R-R'|}.
\end{align*}
Applied on $\Psi_n$, this gives
$$E^{\rm Sch}_{\cL}(\Omega_n)\geq E^{\rm Sch}_{\cL'}(\Omega_n)+\int(W_n+W'_n)\rho_{\Psi_n}+c_n.$$
Notice that for any $R\in\R^3$,
\begin{eqnarray*}
\int\frac{\rho_{\Psi_n}(x)}{|x-R|}dx & = & \int_{|x-R|\leq\lambda}\frac{\rho_{\Psi_n}(x)}{|x-R|}dx+\int_{|x-R|>\lambda}\frac{\rho_{\Psi_n}(x)}{|x-R|}dx\\
 & \leq & (2\pi)^{2/5}\lambda^{1/5}\left(\int\rho_{\Psi_n}^{5/3}\right)^{3/5}+\frac{\int\rho_{\Psi_n}}{\lambda}\\
 & \leq & C(\lambda^{1/5}|\Omega_n|^{3/5}+|\Omega_n|\lambda^{-1})\leq C|\Omega_n|^{2/3}
\end{eqnarray*}
by \eqref{estim_GS_LTb} and optimizing over $\lambda$. By assumption, we know that the smallest ball $B_n$ containing $\Omega_n$ satisfies $|\Omega_n|/|B_n|\geq \delta'$ for some $\delta'>0$ independent of $n$. Thus, by \eqref{hyp_D2},
\begin{eqnarray*}
\int W_n\rho_{\Psi_n} & \leq &  |\Omega_n|^{2/3}\left(\sum_{(R,z)\in\cL\cap\Omega_n\times\R}|D_{\rm ch}(z)|+\sum_{(R',z')\in \cL_{\rm d}}|z'|\right)\\
 & \leq & |\Omega_n|^{2/3}o(|B_n|^{1/3})\leq o(|\Omega_n|).
\end{eqnarray*}

We argue similarly for $W'_n$: let $\lambda$ be small enough such that all the balls of radius $\lambda$ centered at $R$ with $(R,z)\in\cL$ (resp. $R+D_{\rm dis}(R)$) never intersect (we shall actually take $\lambda\to0$ later on). We compute
\begin{align}
& \int_{B(R,\lambda)\cup B(R+D_{\rm dis}(R),\lambda)}\rho_{\Psi_n}(x)\left|\frac{1}{|x-R-D_{\rm dis}(R)|}-\frac{1}{|x-R|}\right|dx\nonumber\\
& \qquad\qquad\leq  4\left(\int_{B(R,\lambda)\cup B(R+D_{\rm dis}(R),\lambda)}\rho_{\Psi_n}^{5/3}\right)^{3/5}
\left(\int_{B(R,\lambda)}\frac{dx}{|x-R|^{5/2}}\right)^{2/5}\nonumber\\
&\qquad\qquad \leq \frac{12}5\lambda^{1/5}\left(\int_{B(R,\lambda)}\rho_{\Psi_n}^{5/3}+\int_{B(R+D_{\rm dis}(R),\lambda)}\rho_{\Psi_n}^{5/3}\right) 
+\frac85(2\pi)\lambda^{1/5}\label{eq_dipole_estim_1}
\end{align}
where we have used Young's inequality.
Then we notice that there exists a constant such that (see, e.g., the proof of \cite[Lemma 9]{ML})
\begin{align}
\left|\frac{1}{|x-R-D_{\rm dis}(R)|}-\frac{1}{|x-R|}\right| & \leq  \frac{C|D_{\rm dis}(R)|}{|x-R-D_{\rm dis}(R)|\times |x-R|}\label{estim_dipole_potential}\\
 & \leq  \frac{C|D_{\rm dis}(R)|}2 \left(\frac1{|x-R|^2}+\frac1{|x-R-D_{\rm dis}(R)|^2}\right)\nonumber
\end{align}
hence
\begin{multline}
 \int_{|x-R|>\lambda,\ |x-R-D_{\rm dis}(R)|>\lambda}\rho_{\Psi_n}(x)\left|\frac{1}{|x-R-D_{\rm dis}(R)|}-\frac{1}{|x-R|}\right|dx\\
\leq {C'|D_{\rm dis}(R)|}\left(\int_{\Omega_n}\rho_{\Psi_n}^{5/3}\right)^{3/5}\left(\int_\lambda^\ii\frac{dt}{t^3}\right)^{2/5}
 \leq C''|D_{\rm dis}(R)|\lambda^{-1/5}|\Omega_n|^{3/5}.\label{eq_dipole_estim_2}
\end{multline}
Summing \eqref{eq_dipole_estim_1} and \eqref{eq_dipole_estim_2} over $R$ and using both \eqref{estim_GS_LTb} and that the number of nuclei inside $\Omega_n$ is proportional to $|\Omega_n|$ (by the regularity properties of $\Omega_n$ and the fact that there is a smallest distance between the nuclei), we find
\begin{equation}
\int W'_n\rho_{\Psi_n}\leq C|\Omega_n|(\lambda^{1/5}+\alpha(|\Omega_n|)\lambda^{-1/5})
\label{eq_dipole_estim_3}
\end{equation}
where $\alpha$ is some function satisfying $\lim_{t\to\ii}\alpha(t)=0$. Choosing for instance $\lambda^{1/5}=\sqrt{\alpha(|\Omega_n|)}$ which converges to 0 as $n\to\ii$, we obtain that the right hand side of \eqref{eq_dipole_estim_3} is a $o(|\Omega_n|)$.

The proof that $c_n=o(|\Omega_n|)$ is much simpler thanks to the estimates
\begin{equation}
\sum_{\substack{(R',z')\in\cL\cap\Omega_n\times\R\\ R'\neq R}}\frac{1}{|R-R'|}\leq C|\Omega_n|^{2/3},\quad \sum_{\substack{(R',z')\in\cL\cap\Omega_n\times\R\\ R'\neq R}}\frac{1}{|R-R'|^2}\leq C|\Omega_n|^{1/3}
\label{estim_sums_nuclei}
\end{equation}
for any $R$ such that $(R,z)\in\cL$ for some $z$. This ends the proof of \eqref{reduc_equ_cristal}.

The proof of the same estimate for the free energy 
$$F^{\rm Sch}_{\cL}(\Omega_n,\beta,\mu)\geq F^{\rm Sch}_{\cL'}(\Omega_n,\beta,\mu)+o(|\Omega_n|)$$
is an easy adaptation of the above arguments, thanks to \eqref{estim_GS_LTb}.
\qed

\subsubsection{Proof of Proposition \ref{prop_dipoles}: dipole argument}\label{proof_prop_dipoles}
Let us consider two sets $\Omega'\subseteq\Omega$ in $\cR_\eta\cap\cC_\varepsilon$ with $\text{d}(\partial\Omega,\partial\Omega')>\delta:=10\delta'$. This clearly implies that $\Omega\setminus\Omega'$ satisfies the $\varepsilon$-cone property for $\varepsilon$ small enough. We assume that the nuclei in $\Omega'$ which are at a distance less than $\delta'$ from $\partial\Omega'$ all have a charge which realizes the minimum in the definition of $\uE(\Omega')$. Similarly we choose the charges of the nuclei in $\Omega$ at a distance less than $\delta'$ from $\partial\Omega$ to realize the supremum in $\oE(\Omega)$. We can do this because the boundaries of $\Omega'$ and $\Omega$ are far enough such that the two sets of nuclei whose charges are optimized are disjoint.

Because $\Omega\setminus\Omega'$ satisfies the $\varepsilon$-cone property, we know that for any nucleus $R$ in $\Omega\setminus\Omega'$, there exists a ball $B(R+a_R,r)$ with small radius $r\leq\delta'$, contained in $\Omega\setminus\Omega'$, with $|a_R|\leq \varepsilon$, such that none of these balls intersect, and such that $R$ is the closest nucleus from $R+a_R$. Whenever possible, we simply take the ball centered at the nucleus, $a_R=0$. This is not possible for the nuclei which are a distance less than $r$ to the boundary of $\Omega\setminus\Omega'$. Let us denote by $\cJ_1$ the set of all these nuclei.
Because $\Omega',\Omega\in\cR_{\eta}$, we know that $\#\cJ_1\leq C |\Omega|^{2/3}$. 

Now, denote by $\cJ_2$ the set of nuclei whose charge was optimized to realize the minimum of $\uE(\Omega')$, i.e. those which are inside $\Omega'$, at a distance at most $\delta'$ to the boundary of $\Omega'$. Their charge $z'$ differs from the charge $z\geq z'$ of the original nucleus in $\cL$. Any such nucleus is at a distance at most $\delta'$ to a point on the boundary of $\Omega'$, at which there is a cone of size $\epsilon$ pointing in $\Omega\setminus\Omega'$. Thus we can find a ball $B(R+a_R,r)$ with $|a_R|\leq 2\delta'$ and small radius $r\leq\delta'$ contained in $\Omega\setminus\Omega'$ and such that $R$ is the closest nucleus of the center of this ball (since $\delta'$ was chosen small enough). Also $\#\cJ_2\leq C |\Omega'|^{2/3}$.

Then, let $\Gamma':=|\Psi'\rangle\langle\Psi'|$ with $\Psi'\in\cF(H^1_0(\Omega'))$ be a pure ground state of $\uE(\Omega')$. For each nucleus $(R,z)$ in $\Omega\setminus\Omega'$, we can choose a state $\Gamma_{R}$ of the Fock space $\cF(H^1_0(B(R+a_R,r))$ with total momentum zero and charge $q_R:=-z$ like in the proof of Theorem \ref{stability_of_matter}. For each nucleus $(R,z')$ with $R\in\cJ_2$ and $(R,z)\in\cL$, we also choose a state $\Gamma_{R}$ of the Fock space $\cF(H^1_0(B(R+a_R,r))$ with total momentum zero and charge $-q_R=z-z'$.
Doing so, we create (see Figure \ref{image_preuve2})
\begin{itemize}
\item $O(|\Omega\setminus\Omega'|)$ neutral systems of size $r$;
\item $O(|\Omega|^{2/3})$ dipoles of size $|a_R|\leq \max\{\varepsilon,2\delta'\}$, all located at a distance at most $r$ to the boundary of $\Omega\setminus\Omega'$, of charges $q_R\leq\bar z=\max\{z,\ (R,z)\in\cL\}$.
\end{itemize}

Following the proof of Theorem \ref{stability_of_matter}, we can choose as a trial state the tensor product $\Gamma:=\Gamma'\bigotimes_{(R,z)}\Gamma_{(R,z)}$, the full Fock space being the tensor product of the local Fock spaces,
$$\cF\left(L^2\left(\Omega'\cup_{(R,z)}B(R+a_R,r)\right)\right)\simeq\cF\left(L^2(\Omega')\right)\bigotimes_{(R,z)}\cF\left(L^2(B(R+a_R,r))\right).$$
The energy of our trial state for $\oE(\Omega)$ (with the nuclear charges optimized at the boundary of $\Omega$) can then be estimated by
\begin{multline}
\uE(\Omega')+\kappa|\Omega\setminus\Omega'|+C|\Omega|^{2/3}+\pscal{\Psi',\left(\sum_iW_i\right)\Psi'}\\
-\sum_{(R,z)\in\Omega'\times\R\cap \cL}zW(R)+o(|\Omega|).
\label{estim_energy_dipoles_argument}
\end{multline}
The second term is an estimate of the energy of all the radial electrons inside $\Omega\setminus\Omega'$, whereas the third term is the kinetic energy of the electrons used to create the dipoles at the boundary of $\Omega\setminus\Omega'$. The fourth term is the interaction between the electrons of the ground state $\Psi'$ inside $\Omega'$ and the dipoles created at the boundary (by Newton's theorem, a classical dipole is seen by the particles in $\Omega'$):
$$W(x)=-\sum_{R\in\cJ=\cJ_1\cup\cJ_2}W_{R}(x),\quad W_{R}(x)=q_R\left(\frac{1}{|R-x|}-\frac{1}{|R+ a_R-x|}\right)$$
with the convention that $W_R(R)=-q_R|a_R|^{-1}$.
The fifth term of \eqref{estim_energy_dipoles_argument} is the interaction energy between the dipoles and the nuclei inside $\Omega'$. 
The last term of \eqref{estim_energy_dipoles_argument} contains the self-interaction of the dipoles which is a $o(|\Omega|)$ as already shown in the proof of Theorem \ref{stability_of_matter}, see \eqref{estim_interac_dipoles}. 

We want to prove that 
\begin{equation}
 \pscal{\Psi',\left(\sum_iW_i\right)\Psi'}-\sum_{(R,z)\in\Omega'\times\R\cap\cL}zW(R)\leq o(|\Omega|).
\label{but_dipole}
\end{equation}
To this end, we shall use the following  
\begin{lemma}[Stability of Matter with $K$ nuclei and $M$ dipoles]\label{stability_dipoles}
Assume that we have $K+M$ nuclei, located at $\{R_k\}_{k=1}^K$ and $\{y_m^+\}_{m=1}^M$ on a lattice, where each $R_k$ has a positive charge $0\leq z_k\leq z$ and each $y_m^+$ a charge $0\leq q_m^+\leq q$. Assume that close to each $y_m^+$ there is a negative point charge $0\geq q_m^-\geq-q$ located at $y_m^-$ with the property that
$y_m^\pm$ is closer to $y_m^\mp$ than to any other fixed $R_k$ or $y_\ell^\pm$.  Let $E(M,K,N)$ denote the ground state of $N$ electrons and the above classical charges.  Then
\begin{equation*}
 E(N,K,M)\geq -C(1+z^2)N-z^2K-C(1+z^2+q^2)M-Cq^{5/2}M.
\end{equation*}
where $C$ depends on the the smallest distance between the classical charges but not on $q$ and $z$.
\end{lemma}

Before proving the lemma, we show how to use it to derive \eqref{but_dipole}.
We have for all $0<\varepsilon<1$
\begin{multline*}
\varepsilon \uE(\Omega')-\left\langle\Psi,\left(\sum_iW_i\right)\Psi\right\rangle +\sum_{(R,z)\in\cL\cap\Omega'\times\R}zW(R)=-\varepsilon^{-1}U\\
+\varepsilon \left\langle\Psi,\left[H_{\Omega'}-\varepsilon^{-1}\left(\sum_iW_i\right)
+\varepsilon^{-1}\sum_{(R,z)\in\cL\cap\Omega'\times\R}zW(R)
+\varepsilon^{-2}U\right]\Psi\right\rangle
\end{multline*}
where $H_{\Omega'}$ is the Coulomb Hamiltonian in $\Omega'$ with charges realizing the infimum of $\uE(\Omega')$ and  $U$ is the total Coulomb interaction between the dipoles, which is a $o(|\Omega|)$ as already mentioned before.
Note that the operator 
$$H_{\Omega'}-\varepsilon^{-1}\left(\sum_iW_i\right)
+\varepsilon^{-1}\sum_{(R,z)\in\cL\cap(\Omega'\times\R)}zW(R)
+\varepsilon^{-2}U
$$
corresponds to the system in $\Omega'$ with the dipoles added with
$-\varepsilon^{-1}$ times their original charges.
Using the lemma and the fact that $M=O(|\Omega|^{2/3})$, $K=O(|\Omega|)$ by the regularity properties of $\Omega$ and $\Omega'$, we conclude that 
this operator is bounded below by
\begin{multline}
\inf_N \left\{CN^{5/3}|\Omega|^{-2/3}-CN-C|\Omega|-C(1+\varepsilon^{-5/2})M\right\}\\
\geq -C|\Omega|-C\varepsilon^{-5/2}|\Omega|^{2/3}.
\end{multline}
Returning to our estimate above and using $\uE(\Omega')\leq C|\Omega|$, we find that 
\begin{multline}
 \pscal{\Psi',\left(\sum_iW_i\right)\Psi'}-\sum_{(R,z)\in\Omega'\times\R}zW(R)\\
\leq C\varepsilon|\Omega|+C\varepsilon^{-3/2}|\Omega|^{2/3}
+\varepsilon^{-1}|\Omega|^{2/3}\log|\Omega|
\end{multline}
where we have used the estimate on the self-interaction between the dipoles, see \eqref{estim_interac_dipoles}.
It is clear that we can choose $\varepsilon$ such that this is
$o(|\Omega|)$.

\medskip

For the free energy, the proof of \textbf{(A4)} follows exactly the same lines: $\Gamma_{\Omega'}$ is chosen to be a ground state for $F^{\rm Sch}_\cL(\Omega',\beta,\mu)$ and the other $\Gamma_{R}$ are kept the same. The only additional ingredient which is needed is the additivity of the entropy for tensor products, which reads
\begin{eqnarray*}
\tr_{\cF(L^2(\Omega))} (\Gamma\log\Gamma) & = & \tr_{\cF(L^2(\Omega'))}(\Gamma_{\Omega'}\log\Gamma_{\Omega'})\\
 & & \qquad+\sum_{(R,z)}\tr_{\cF(L^2(B_{(R,z)}))}(\Gamma_{R}\log\Gamma_{R})\\
 & \leq & \tr_{\cF(L^2(\Omega'))}(\Gamma_{\Omega'}\log\Gamma_{\Omega'}).
\end{eqnarray*}

\begin{proof}[Proof of Lemma \ref{stability_dipoles}]
The importance in this version of stability of matter is that we keep
track of the dependence of the charges $q$ and $z$.  Unfortunately, as
far as we can see, the lemma does not follow from known versions of
stability of matter. We shall prove it using a comparison with the Yukawa potential.
More precisely, we have for all $\nu\geq0$
$$\frac{1}{|p|^2}-\frac1{|p|^2+\nu^2}\geq0\qquad\text{and}\qquad\lim_{x\to0}\left(\frac{1}{|x|}-\frac{e^{-\nu|x|}}{|x|}\right)=\nu,$$
hence for any $x_i\in\R^3$ and any $q_i\in\R$,
\begin{equation}
\sum_{i<j}\frac{q_iq_j}{|x_i-x_j|}\geq \sum_{i<j}q_iq_jY_\nu(x_i-x_j)-\frac\nu2\sum_{i}q_i^2.
\label{estim_Coulomb_Yukawa} 
\end{equation}
Let us use this inequality to estimate the the total Coulomb potential $V_C$ of our system of electrons, nuclei and dipoles. We denote by $x_i$ the positions of the electrons.
\begin{align*}
V_C\geq& \sum_{1\leq i<j\leq N}Y_\nu(x_i-x_j)-\frac\nu2(N+z^2K+q^2M)\\
&\quad-\sum_{i=1}^N\sum_{k=1}^KzY_\nu(x_i-R_k)-\sum_{i=1}^N\sum_{m=1}^MqY_\nu(x_i-y_m^+)\\
&\quad-\sum_{m=1}^M\sum_{k=1}^KqzY_\nu(y_m^--R_k)-\sum_{m=1}^M\sum_{\ell=1}^Mq^2Y_\nu(y_m^--y_\ell^+).
\end{align*}
On the right we have ignored several positive terms.
Next we note that since the nuclei are placed on a lattice $\cL$, we have for all $x$ and with a constant $C$ depending on the lattice and of $\nu$
\begin{equation}
 \sum_{k=1}^KY_\nu(x-R_k)\leq  \max_{R\in\cL}|x-R|^{-1}+\frac{C}{\nu^2}
\label{estim_sum_Yukawa}
\end{equation}
where $C$ is independent of $\nu$ but depends on the smallest distance $\delta'$ between the nuclei of the lattice $\cL$. Indeed let us define $\delta=\delta'/10$. We have assuming $y\in\cL$ and $|y-x|\geq\delta$
\begin{equation*}
\int_{B(y,\delta)}\frac{e^{-\nu(|y-z|-|y-x|)}|x-y|}{|x-z|}\,dz \geq \frac12\int_{B(y,\delta)}e^{-\nu(|y-z|-|y-x|)}\,dz.
\end{equation*}
Next it is clear that $|y-z|\leq|y-x|$ for $z$ in a subset of $B(y,\delta)$ of non zero measure. Hence we obtain the estimate
$$Y_\nu(x-y)\leq C\int_{B(y,\delta)}Y_\nu(x-z)\,dz$$
whenever $|y-x|\geq\delta$. Summing over $\cL$ we infer
\begin{equation}
\sum_{(R,z)\in\cL}Y_\nu(x-R)\leq \max_{R\in\cL}|x-R|^{-1}+C\int_{\R^3}Y_\nu(r)\,dr
\label{estim_sum_Yukawa2}
\end{equation}
which yields \eqref{estim_sum_Yukawa}.

We may argue like in the proof of Theorem \ref{stability_of_matter}, Eq. \eqref{estim_kinetic_local_Sobolev}, using the Sobolev inequality in small balls around each nucleus to get
$$\sum_{i=1}^N\frac{(-\Delta)_{x_i}}{4}-\sum_{i=1}^Nz\max_k|x_i-R_k|^{-1}\geq -Cz^2N.$$
Using $Y_\nu(x)\geq \frac{1}{|x|}-\nu$ by the convexity of $r\mapsto e^{-\nu r}$ and choosing for instance $\nu=1$ we arrive at
\begin{align}
\sum_{i=1}^N(-\Delta)_{x_i}+V_C\geq& \sum_{i=1}^N\frac{3(-\Delta)_{x_i}}{4}+\sum_{i=1}^N\max_{x_j\neq x_i}|x_i-x_j|^{-1}\nonumber\\
&\quad-C(1+z^2)N-z^2K-C(1+z^2+q^2)M\nonumber\\
&\quad-\sum_{i=1}^N\sum_{m=1}^MqY_\nu(x_i-y_m^+).\label{estim_with_repulsion}
\end{align}
\begin{remark}\rm
We shall not need the repulsive term $\sum_{i=1}^N\max_{x_j\neq x_i}|x_i-x_j|^{-1}$ in this proof but we highlight it as it is useful for bosons. 
\end{remark}

Let us introduce the following potential
$$W(x)=\sum_{m=1}^MY_\nu(x_i-y_m^+)$$
Denoting by $\delta_{\rm dip}(x):=\min|x-y_m^+|$ the closest distance of $x$ to the dipoles, we have similarly to \eqref{estim_sum_Yukawa2}
\begin{align*}
\sum_{m=1}^MY_\nu(x-y_m^+)&\leq C\delta_{\rm dip}(x)^{-1}\1_{\delta_{\rm dip}(x)\leq1}+C\int_{|r|\geq \delta_{\rm dip}(x)}Y_\nu(r)\,dr\1_{\delta_{\rm dip}(x)\geq1}\\
&\leq  C\delta_{\rm dip}(x)^{-1}\1_{\delta_{\rm dip}(x)\leq1}+Ce^{-\delta_{\rm dip}(x)}\1_{\delta_{\rm dip}(x)\geq1}.
\end{align*}
Hence it is clear that $W\in L^{5/2}(\R^3)$ and that
$$\int_{\R^3}W^{5/2}\leq CM$$
for a constant $C$ independent of $M$. By the Lieb-Thirring inequality we infer
\begin{equation}
 E(N,K,M)\geq -C(1+z^2)N-z^2K-C(1+z^2+q^2)M-Cq^{5/2}M.
\label{estim_dipole_final}
\end{equation}
\end{proof}

\begin{remark}\rm
Let us denote by $E_\nu(N,K)$ the ground state energy of $N$ particles (fermions or bosons) interacting via the Yukawa potential $Y_\nu$ with $K$ nuclei located at a distance at least $\delta'$ with each other. We have proved the estimate
$$E_\nu(N,K)\geq -\left(\nu+\frac{C}{\nu^2}z+Cz^2\right)N-\nu z^2 K$$
where $C$ depends only on $\delta'$. This $\nu$-dependent estimate is sufficient to get the stability of matter in the crystal case for the potential $W$ defined in \eqref{def_potential_Graf-Schenker} and which is used in the Graf-Schenker estimate with smooth localization.
\end{remark}

\begin{remark}\rm \textbf{(Bosonic case)}. For bosons, an estimate similar to \eqref{estim_dipole_final} is true. One needs to use Proposition \ref{lemma:estim_nuclei} and the repulsion term in \eqref{estim_with_repulsion}, instead of the Lieb-Thirring inequality.
\end{remark}

\subsubsection{Proof of \textbf{(A5)} and Lemma \ref{satisf_A5}}\label{proof_satisf_A5}
Let $\omega$ be a ground state for $\uE(\Omega)$, with one-body (resp.~two-body) density matrix $\gamma_1$ (resp. $\gamma_2$). We have
by \eqref{estim_below_A5bis}
\begin{equation}
\uE(\Omega)=\omega(H_\Omega)\geq \frac{1-C/\ell}{|\ell\triangle|}\int_{G} d\lambda(g)\omega(H_{g\theta^{\ell}})-\frac{C|\Omega|_{\rm r}}{\ell}.
\label{estim_below_energy_simplices} 
\end{equation}
Remark
\begin{eqnarray*}
 \omega(H_{g\theta^{\ell}}) & = & \tr\left(\left(T-\sum_{(R,z)\in\cL\cap(\Omega\times\R)}\frac{z(g\theta^\ell)(R)^2}{|x-R|}\right)(g\theta^\ell)\gamma_1(g\theta^\ell)\right)\\
 & &\qquad  +\frac12\tr\left((g\theta^\ell)\otimes(g\theta^\ell)\gamma_2(g\theta^\ell)\otimes(g\theta^\ell)\frac{1}{|x-y|}\right)+{\rm C}_{g\theta^\ell}\\
 & = & \omega_{g\theta^\ell}\left(\tilde H_{g\theta^\ell}\right),
\end{eqnarray*}
where $\omega_{g\theta^\ell}$ is the localized state as defined in the Appendix and
\begin{equation}
\tilde H_{g\theta^{\ell}}=\sum_iT_i-\sum_i\sum_{(R,z)\in\cL\cap(\Omega\times\R)}\frac{z(g\theta^\ell)(R)^2}{|x_i-R|}
+ \frac12\sum_{i\neq j}\frac{1}{|x_i-x_j|}+{\rm C}_{g\theta^{\ell}}.
\label{def_tilde_H_gmu} 
\end{equation}
Notice $\omega_{g\theta^\ell}$ lives over $\Omega\cap g\ell'\triangle$ where $\ell'=\ell+\delta''$ where $\delta''$ is chosen such that ${\rm Supp}(\theta^\ell)\subseteq \ell'\triangle$. The support of $j$ was chosen small enough to ensure that $\nabla\theta^\ell$ has its support at a distance at most $\delta'$ of the boundary of $\ell'\triangle$. Hence only the charges of the nuclei close to the boundary are changed and we have
$$\omega_{g\theta^\ell}\left(\tilde H_{g\theta^\ell}\right)\geq \uE(\Omega\cap g\ell'\triangle).$$
Now \eqref{estim_below_energy_simplices} gives
$$\uE(\Omega)\geq \frac{1-\alpha(\ell')}{|\ell'\triangle|}\int_{G} d\lambda(g)\uE(\Omega\cap g\ell'\triangle)-\frac{C|\Omega|_{\rm r}}{\ell'-\delta}$$
with
$$\alpha(\ell')=1-\left(1+\frac{\delta''}{\ell'-\delta''}\right)^3\left(1-\frac{C}{\ell'-\delta''}\right)$$
which proves \textbf{(A5)} for the energy.

For the free energy, choose $\omega$ a ground state for $\uF(\Omega,\beta,\mu)$ with density matrix $\Gamma$ and entropy $S(\omega)=-\tr_{\cF(L^2(\Omega))}(\Gamma\log\Gamma)$. Estimate \eqref{estim_below_A5bis} and the above computation show that
\begin{eqnarray*}
\uF(\Omega,\beta,\mu) & = & \omega(H_\Omega-\mu)-\beta^{-1}S(\omega)\\
 &\geq & (1-C/\ell)\left(\frac{1}{|\ell\triangle|}\int_{G} d\lambda(g)\omega_{g\theta^{\ell}}(\tilde H_{g\theta^{\ell}})-\beta^{-1}S(\omega)\right)\\
 &  & \qquad +\frac{C}{\ell}\left(\frac12\omega\left(\sum_iT_i-\mu\cN\right)-\beta^{-1}S(\omega)\right)
-\frac{C|\Omega|_{\rm r}}{\ell}.
\end{eqnarray*}
We have seen in the proof of Theorem \ref{stability_cristal} that 
$$\frac12\omega\left(\sum_iT_i-\mu\cN\right)-\beta^{-1}S(\omega)\geq -C|\Omega|$$
hence
\begin{equation}
\uF(\Omega,\beta,\mu)\geq  (1-C/\ell)\left(\frac{1}{|\ell\triangle|}\int_{G} d\lambda(g)\omega_{g\theta^{\ell}}(\tilde H_{g\theta^{\ell}})-\beta^{-1}S(\omega)\right) -\frac{C|\Omega|_{\rm r}}{\ell}. 
\label{estim_FE}
\end{equation}

\begin{lemma} \label{lemme_subadd}
One has the subadditivity property of the entropy:
\begin{equation}
 S(\omega) \leq \frac{1}{|\ell\triangle|}\int_{G} d\lambda(g) S(\omega_{g\theta^\ell}).
\label{estim_entropy_subadd}
\end{equation}
\end{lemma}

Using Lemma \ref{lemme_subadd}, we deduce that
\begin{eqnarray*}
\uF(\Omega,\beta,\mu) & \geq & \frac{1-\alpha(\ell')}{|\ell'\triangle|}\int_{G} d\lambda(g)\left(\omega_{g\theta^{\ell}}(\tilde H_{g\theta^{\ell}})-\beta^{-1}S(\omega_{g\theta^{\ell}})\right) -\frac{C|\Omega|_{\rm r}}{\ell'-\delta''}\\
& \geq & \frac{1-\alpha(\ell')}{|\ell'\triangle|}\int_{G} d\lambda(g)\uF(\Omega\cap g\ell'\triangle,\beta,\mu) -\frac{C|\Omega|_{\rm r}}{\ell'-\delta''},
\end{eqnarray*}
hence that $\uF(\cdot,\beta,\mu)$ satisfies condition \textbf{(A5)} which was defined in \cite{1}.

\begin{proof}[Proof of Lemma \ref{lemme_subadd}] Recall that there is a subgroup $\Gamma$ of $G$ such that $\ell\triangle$ defines a $\Gamma$-tiling of $\R^3$ and $G/\Gamma$ is compact. Then
$$\frac{1}{|\ell\triangle|}\int_{G} d\lambda(g) S(\omega_{g\theta^\ell})=\frac{1}{|\ell\triangle|}\int_{G/\Gamma} d\hat{\lambda}([g]) \left(\sum_{\mu\in\Gamma}S(\omega_{g\mu\theta^\ell})\right)$$
and
\begin{equation}
 \frac{1}{|\ell\triangle|}\int_{G/\Gamma} d\hat{\lambda}(g)=1.
\label{normalization_G_Gamma}
\end{equation}
Fix now some $[g]\in G/ \Gamma$ and consider the family $(\mu\theta^\ell_g)_{\mu\in\Gamma}$ with $\theta^\ell_g(x)=\theta^\ell(g^{-1}x)$. By construction $\sum_{\mu\in\Gamma} (\mu\theta^\ell_g)^2=1$.
Hence we are within the formalism of the Appendix, Section \ref{sec:SSA_quantum}, with $q_\mu:=\mu\theta^\ell_g$. The subadditivity of the entropy \eqref{subadd} implies
$$\forall [g]\in G/\Gamma,\qquad S(\omega) \leq \sum_{\mu\in\Gamma}S(\omega_{g\mu\theta^\ell}).$$
Thus \eqref{estim_entropy_subadd} is obtained integrating over $G/\Gamma$ and using \eqref{normalization_G_Gamma}.
\end{proof}

\subsubsection{Proof of \textbf{(A6)} and Lemma \ref{satisf_A6}}\label{proof_satisf_A6}
Recall that by Proposition \ref{upper_bd_cristal} there exists a constant such that
\begin{equation}
 \int\rho_\omega^{5/3}\leq C|\Omega|,\qquad \omega\left(\sum_i T_i\right)\leq C|\Omega|.
\label{LT_estim_A6}
\end{equation}

It is clear that \textbf{(A6.4)} and \textbf{(A6.5)} hold true. We know from Proposition \ref{prop_strong_subadd} in the Appendix that $s^\Omega_\ell(\cdot)$ satisfies the strong subadditivity property \textbf{(A6.6)}. We have by \eqref{decomp_H_Omega}
\begin{multline*}
\uF(\Omega,\beta,\mu) = \omega(H_\Omega)
 = \sum_{\mu\in\Gamma}E^\Omega_\ell(g\mu)+\frac12\sum_{\substack{\mu,\nu\in\Gamma\\, \mu\neq\nu}}I^\Omega_\ell(g\mu,g\nu)+s_\ell^\Omega(\Gamma)\\
 -\omega\left(\sum_i\sum_{\mu\in\Gamma}|\nabla \Theta^\ell_{g\mu}|^2_i\right).
\end{multline*}
Recall $\theta^\ell\neq\Theta^\ell_{g\mu}$, see Remark \ref{rmk:theta} above. Notice
\begin{multline}
\omega\left(\sum_i\sum_{\mu\in\Gamma}|\nabla \Theta^\ell_{g\mu}|^2_i\right) = \int_\Omega\rho_\omega\sum_{\mu\in\Gamma}|\nabla\Theta^\ell_{g\mu}|^2\\
 \leq  C|\Omega|^{3/5}\left|{\rm Supp}\left(\sum_{\mu\in\Gamma}|\nabla\Theta^\ell_{g\mu}|^2\right)\cap\Omega\right|^{2/5}
 \leq  C'\frac{|\Omega|}{\ell^{2/5}}
\label{estim_IMS_rest}
\end{multline}
where we have used the Lieb-Thirring estimate \eqref{LT_estim_A6} and the fact that since $\Omega$ is regular, the number of simplices intersecting $\Omega$ can be estimated by a constant times $|\Omega|/|\ell\triangle|$. This shows that $\uF(\cdot,\beta,\mu)$ satisfies \textbf{(A6.1)}.

Fix now some $\cP\subset\Gamma$ and recall that 
$$q_\cP=\left(\sum_{\mu\in\cP}\big(\Theta^\ell_{g\mu}\big)^2\right)^{1/2}$$
was defined in \eqref{def_q_cP}. The IMS formula reads
$$\sum_{\mu\in\cP}\Theta^\ell_{g\mu}T\Theta^\ell_{g\mu}=q_\cP Tq_\cP+\sum_{\mu\in\cP}|\nabla \Theta^\ell_{g\mu}|^2-|\nabla q_\cP|^2.$$
Then, notice that
$$|\nabla q_\cP|^2=\frac{\left|\sum_{\mu\in\cP}\Theta^\ell_{g\mu}\nabla\Theta^\ell_{g\mu} \right|^2}{\sum_{\mu\in\cP}(\Theta^\ell_{g\mu})^2}\leq \sum_{\mu\in\cP}|\nabla \Theta^\ell_{g\mu}|^2$$
Hence we have
$$\sum_{\mu\in\cP}\Theta^\ell_{g\mu}T\Theta^\ell_{g\mu}\geq q_\cP Tq_\cP.$$
We obtain
\begin{equation}
 \sum_{\mu\in\cP}E^\Omega_\ell(g\mu)+\frac12\sum_{\substack{\mu,\nu\in\cP\\, \mu\neq\nu}}I^\Omega_\ell(g\mu,g\nu)+s_\ell^\Omega(\cP)
=\omega(H(g,\cP))-\beta^{-1}S(\omega_\cP^g),
\end{equation}
where we have defined 
\begin{multline}
H(g,\cP) := \sum_i(q_\cP T q_\cP)_i-\sum_i\sum_{(R,z)\in\cK}\frac{zq_\cP(R)^2q_\cP(x_i)^2}{|x_i-R|}\\
+ \frac12\left(\sum_{i\neq j}\frac{q_\cP(x_i)^2q_\cP(x_j)^2}{|x_i-x_j|}+\sum_{\substack{(R,z),(R',z')\in\cK,\\ R\neq R'}}\frac{q_\cP(R)^2q_\cP(R')^2}{|R-R'|}\right).
\label{def_Hg_cP}
\end{multline}
Compare Formula \eqref{def_Hg_cP} with \eqref{def_Hgmu}: we have $H(g,\{\mu\})=H(g\mu)$.
Notice as before
$$\omega(H(g,\cP))=\omega_\cP^g(\tilde H(g,\cP))$$
where $\tilde H(g,\cP)$ is the Coulomb Hamiltonian defined on ${\rm Supp}(q_\cP)$ with the charges close to its boundary changed due to the multiplication by $q_\cP(R)^2$, similarly to \eqref{def_tilde_H_gmu}.
Hence
$$\omega(H(g,\cP))-\beta^{-1}S(\omega_\cP^g)\geq \uF\left(\Omega\cap\bigcup_{\mu\in\Gamma}\ell g\mu(1+\delta''/\ell)\triangle\; ,\;\beta\; ,\;\mu\right).$$
Therefore we obtain \textbf{(A6.2)} without any error:
\begin{multline}
 \sum_{\mu\in\cP}E^\Omega_\ell(g\mu)+\frac12\sum_{\substack{\mu,\nu\in\cP\\, \mu\neq\nu}}I^\Omega_\ell(g\mu,g\nu)+s_\ell^\Omega(\cP)\\
\geq \uF\left(\Omega\cap\bigcup_{\mu\in\Gamma}\ell g\mu(1+\delta''/\ell)\triangle\; ,\; \beta\; ,\;\mu\right).
\end{multline}

Let us now prove \textbf{(A6.3)}. Lemma \ref{GS_2} gives
\begin{equation}
\frac{1}{|\triangle|}\int_Gd\lambda(g)\sum_{\substack{\mu\in\Gamma\\ \mu\neq0}}I^\Omega_\ell(g\mu,g)\geq -\frac{C}{\ell|\triangle|}\int_Gd\lambda(g)I^\Omega_\ell(g,g)+\frac{C}{\ell}\omega\left(\mathbb{W}\right)\\
-\frac{C|\Omega|}{\ell}
\label{estim_below_proof_A6}
\end{equation}
where we have used that $\Omega$ is regular to estimate the number of nuclei inside $\Omega$ by a constant times $|\Omega|$.
Then we have by \eqref{LT_estim_A6} and the fact that $\mathbb{W}$ satisfies a version of stability of matter, see \eqref{stability_W},
$$\omega\left(\mathbb{W}\right)\geq -C|\Omega|-\omega\left(\sum_i T_i\right)\geq -C'|\Omega|.$$
Also by \eqref{estim_FE} in the proof of \textbf{(A5)}
$$\frac{1}{|\triangle|}\int_Gd\lambda(g)I^\Omega_\ell(g,g)\leq \frac{1}{|\triangle|}\int_Gd\lambda(g)\omega(H(g0))\leq \uF(\Omega,\beta,\mu)+\alpha(\ell)|\Omega|\leq C|\Omega|.$$
Hence we obtain \textbf{(A6.3)} from \eqref{estim_below_proof_A6}:
\begin{equation}
\frac{1}{|\triangle|}\int_Gd\lambda(g)\sum_{\substack{\mu\in\Gamma\\ \mu\neq0}}I^\Omega_\ell(g\mu,g)\geq -\frac{C}{\ell}|\Omega|.
\label{estim_below_proof_A62}
\end{equation}
This finishes the proof that $\uF(\cdot,\beta,\mu)$ satisfies \textbf{(A6)}.\qed

\subsection{Quantum nuclei: proof of Theorems \ref{thm_stability_LL} and \ref{thm_LL}}
\subsubsection{Proof of Theorem \ref{thm_stability_LL}: stability of matter}\label{proof_thm_stability_LL}
For the energy, this is an obvious consequence of Theorem \ref{stability_of_matter}. We only write the proof for the free energy. By Theorem \ref{stability_of_matter}, Equation \eqref{stability_of_matter_equation} with $T(A)$ replaced by $T(A)/4$, we have the following inequality
\begin{equation}
 H_\Omega\geq \frac1M\sum_kT(A)_{R_k}+\frac34\sum_jT(A)_{x_j}-C|\Omega|+\frac{z^2}{8}\frac1{\delta_R(R_k)}.
\label{estim_below_Hamil_LL}
\end{equation}
Now we use Proposition \ref{lemma:estim_nuclei}. Using \eqref{estim_below_Hamil_LL} and \eqref{estim_nuclei}, we obtain that
$$H_\Omega-\mu\cdot\cN\geq \frac12\sum_iT(A)_{x_i}+\frac1{2M}\sum_kT(A)_{R_k}-C|\Omega|+cK\min\left\{\frac{K}{|\Omega|},\frac{K^{1/3}}{|\Omega|^{1/3}}\right\}-\mu_1 K$$
where we have used the Lieb-Thirring estimate \eqref{estim_Lieb_Thirring_kinetic} to infer
$$\frac14\sum_iT(A)_{x_i}-\mu_2\cN\geq -C|\Omega|.$$
Clearly
$$K\left(c\min\left\{\frac{K}{|\Omega|},\frac{K^{1/3}}{|\Omega|^{1/3}}\right\}-\mu_1\right)\geq -C|\Omega|$$
hence
\begin{equation}
H_\Omega-\mu\cdot\cN\geq \frac12\sum_iT(A)_{x_i}+\frac1{2M}\sum_kT(A)_{R_k}-C|\Omega|. 
\label{bound_below_LL}
\end{equation}
By Peierls' inequality \cite[Prop. 2.5.5]{Ruelle}
\begin{eqnarray*}
F(\Omega,\beta,\mu) & \geq & -\frac1\beta \log\left(\tr_{\cF}\left(e^{-\frac{\beta}2\left(\sum_iT(A)_{x_i}+\sum_kT(A)_{R_k}/M\right)} \right)\right)-C'|\Omega|\\
 & = & -\frac1\beta \log\left(\tr_{\cF^{\rm el}}\left(e^{-\frac{\beta}2\sum_iT(A)_{x_i}} \right)\right)\\
 & & \qquad -\frac1\beta \log\left(\tr_{\cF^{\rm nuc}}\left(e^{-\frac{\beta}{2M}\sum_kT(A)_{R_k}} \right)\right)-C'|\Omega|
\end{eqnarray*}
We have for fermions and bosons respectively
$$\tr_{\cF^{\rm el}}\left(e^{-\frac\beta2\sum_iT(A)_{x_i}}\right)=\prod_{j\geq 1}\left(1+e^{-\beta\lambda_j(A)/2}\right),$$
$$\tr_{\cF^{\rm nuc}}\left(e^{-\frac{\beta}{2M}\sum_kT(A)_{R_k}}\right)=\prod_{j\geq 1}\left(1-e^{-\beta\lambda_j(A)/(2M)}\right)^{-1}$$
where $(\lambda_j(A))$ are the eigenvalues of $T(A)$ with Dirichlet boundary condition on $\Omega$. 
Hence
$$F(\Omega,\beta,\mu) \geq -\frac1\beta \tr_{L^2(\Omega)} f(T(A))-C'|\Omega|$$
where 
$$f(t)=\log(1+e^{-\beta t/2})-\log(1-e^{-\beta t/(2M)}).$$
By the diamagnetic inequality \cite{Lieb-Loss,Simon}, we have
$$\tr_{L^2(\Omega)}\log(1+e^{-\beta T(A)/2})\leq \tr_{L^2(\Omega)}e^{-\beta T(A)/2}\leq\tr_{L^2(\Omega)}e^{-\beta T(0)/2}$$
and
\begin{multline*}
-\tr_{L^2(\Omega)}\log(1-e^{-\beta T(A)/(2M)})  =  \sum_{n\geq0}\frac{\tr_{L^2(\Omega)}e^{-\beta(n+1) T(A)/(2M)}}{n+1}\\
\leq  \sum_{n\geq0}\frac{\tr_{L^2(\Omega)}e^{-\beta(n+1) T(0)/(2M)}}{n+1}=  -\tr_{L^2(\Omega)}\log(1-e^{-\beta T(0)/(2M)}).
\end{multline*}
Hence
$$\tr_{L^2(\Omega)} f(T(A))\leq \tr_{L^2(\Omega)} g(T(0))$$
with $g(t)=e^{-\beta t/2}-\log\left(1-e^{-\beta t/(2M)}\right)$. 
As $g$ is convex, we can apply Lemma \ref{Li-Yau} to obtain
$$\tr_{L^2(\Omega)}g(T(0)) \leq |\Omega|(2\pi)^{-3}\int_{\R^3}g(|p|^2)dp,$$
and finally get the desired bound
$$F(\Omega,\beta,\mu) \geq -\kappa|\Omega|.$$
\qed

\subsubsection{Proof of Theorem \ref{thm_LL}: thermodynamic limit}\label{sec_proof_LL}
As announced in the introduction, we only outline briefly the proof, which is much easier than for the crystal. 

Consider some function $\eta(t)=a|t|^b$ with $a>0$ and $b\in(0,1]$. Increasing $a$ if necessary, we may assume that the simplex defined before  $\triangle\in\cR_\eta$ and we simply define $\cR=\cR_\eta$. Take also $\cR'=\cR$. Clearly \textbf{(P1)}--\textbf{(P5)} are satisfied. Also \textbf{(A1)} holds true by convention. 

As the magnetic field $B=\nabla\times A$ is periodic with respect to some group $\Gamma$, we have that $E$ and $F$ are also $\Gamma$-periodic. Hence \textbf{(A3)} is obviously satisfied, see Step 4 in the proof for the crystal case.
The stability property \textbf{(A2)} was already proved in Theorem \ref{thm_stability_LL}.

The continuity property \textbf{(A4)} is easily verified as one has for any $\Omega',\Omega\in\cM$,
$$\Omega'\subseteq\Omega\Longrightarrow E(\Omega)\leq E(\Omega').$$

Hence only \textbf{(A5)} and \textbf{(A6)} are not obvious. They are proved by following closely the localization method of the proof for the crystal in the previous section. There is only one difference: for the crystal we were using a localization at the boundary of the simplices $g\ell\triangle$, at a finite distance. Here we use a localization at a distance $\sqrt{\ell}$. The reason is that for the bosons we cannot use the Lieb-Thirring inequality to get a bound on the error coming from the localization in the IMS formula as we did in \eqref{estim_IMS_rest}. A localization at a larger scale allows to control this error easily\footnote{Notice this choice of a localization at a distance $\sqrt{\ell}$ cannot be used for the crystal because we would change the charges of two many classical nuclei close to the boundary.}.
First we need the
\begin{lemma}[Upper bound on the number of particles]\label{bound_nb_part_quantum} There exists a constant $C'$ such that for any $\Omega\in\cM$ and any ground state $\Gamma$ for $E(\Omega)$ or for $F(\Omega,\beta,\mu)$,
\begin{equation}
\tr(\cN_e^{5/3}\Gamma)+\tr(\cN_p\Gamma)\leq C'|\Omega|,
\label{estim_nb_particles} 
\end{equation}
where we recall that $\cN_e$ and $\cN_p$ give respectively the number of electrons and nuclei.
\end{lemma}
\begin{proof}
The estimate on the number of electrons is obtained as usual by means of the Lieb-Thirring inequality. For the number of nuclei in the case of the free energy, we just use Theorem \ref{thm_stability_LL}
\begin{equation}
0\geq F(\Omega,\beta,\mu)\geq F(\Omega,\beta,\mu+(1,0))+\tr(\cN_p\Gamma)\geq -\kappa|\Omega|+\tr(\cN_p\Gamma).
\label{estim_free_energy_chemical_potential}
\end{equation}
For $E(\Omega)$, we use similarly \eqref{bound_below_LL} and obtain $0\geq E(\Omega)\geq \tr(\cN_p\Gamma)-C|\Omega|.$
\end{proof}
\begin{remark}\rm
Notice contrarily to the crystal case, we have a bound on the particle number for any domain $\Omega$, not only regular domains.
\end{remark}

We now prove \textbf{(A5)} for the energy $E(\Omega)$. The proof is the same for the free energy.
Let $\theta^\ell=(\1_{\ell \triangle}*j_\ell)^{1/2}$, where $\triangle$ is the simplex introduced above which defines a
$\Gamma$-tiling and $j_\ell(x)=\ell^{-3/4}j(x/\sqrt{\ell})$, with as for the crystal case $j$ being a smooth non-negative radial function of compact support in $B(0,1)$ with $\int j=1$. 
By the IMS localization formula, estimate \eqref{estim_below_kinetic_IMS} becomes 
\begin{multline}
\sum_iT(A)_{x_i}+\frac1M\sum_kT(A)_{R_k}
\geq \frac1{|\ell\triangle|}\int  \left(\sum_i (g\theta^\ell T(A)g\theta^\ell)_{x_i}\right)d\lambda(g)\\
+\frac1{|\ell\triangle|}\int \left(\sum_k\left(g\theta^\ell \frac{T(A)}Mg\theta^\ell\right)_{R_k}\right)d\lambda(g)-C\frac{\cN_e+\cN_p}{\ell^{3/2}}.
\label{estim_below_kinetic_IMS2}
\end{multline}

Following \cite{GS}, Lemma \ref{GS_2} becomes
\begin{multline}
\sum_{1\leq i<j\leq N} \frac{z_iz_j}{|x_i-x_j|}\geq \frac{1-C/\sqrt{\ell}}{|\ell\triangle|}\int_{G} d\lambda(g)\sum_{1\leq i<j\leq N}\frac{z_iz_j(g\theta^\ell)(x_i)^2(g\theta^\ell)(x_j)^2}{|x_i-x_j|}\\
+\frac{C}{\sqrt{\ell}}\sum_{1\leq i<j\leq N}z_iz_jW_{\ell^{-1/2}}(x_i-x_j)-\frac{C}{\ell}\sum_{i=1}^Nz_i^2
\end{multline}
where this time
$W_\mu(x)=\frac{1}{|x|(1+\mu|x|)}.$
Similarly to \eqref{estim_below_A5}, we obtain
\begin{multline}
H_\Omega\geq \frac{1-C/\sqrt{\ell}}{|\ell\triangle|}\int_{G} d\lambda(g)H_{g\theta^{\ell}}
+\frac{C}{\sqrt{\ell}}\Bigg(\sum_iT(A)_{x_i}+\frac1M\sum_kT(A)_{R_k}\\+\mathbb{W}_{\ell^{-1/2}}-C\frac{\cN_e+\cN_p}{\ell}\Bigg).
\label{estim_below_A52}
\end{multline}
The operator $\mathbb{W}_\mu$ is the second quantization of the operator $W_\mu$.
We remark as in \cite[p. 222]{GS} that $W_\mu(x)=\int_0^\ii e^{-\nu}Y_{\mu\nu}(x)$. Hence using again the stability of matter \eqref{stability_Yukawa} for the Yukawa potential (uniformly in the coefficient of the exponential), we get like in \eqref{stability_W}
\begin{equation}
\sum_iT(A)_{x_i}+\mathbb{W}_{\ell^{-1/2}}\geq -C(\cN_e+\cN_p).
\label{stab_Yukawa} 
\end{equation}
Taking the expectation value of \eqref{estim_below_A52} on a ground state and using both \eqref{stab_Yukawa} and \eqref{estim_nb_particles}, we easily obtain \textbf{(A5)}\footnote{Notice we even have an error depending on $|\Omega|$ and not on $|\Omega|_{\rm r}$.}.

The proof of \textbf{(A6)} is similar to the crystal case. Of course \eqref{def_localization_A6} is replaced by
\begin{equation}
 \Theta^\ell_g:=\left(\1_{\ell g\triangle}\ast j_\ell\right)^{1/2}
\label{def_localization_A62}
\end{equation}
and \eqref{estim_IMS_rest} is replaced by
\begin{equation}
\omega\left(\sum_i\sum_{\mu\in\Gamma}|\nabla \Theta^\ell_{g\mu}|^2_i\right) \leq \frac{C}{\ell}\omega(\cN_e+\cN_p)\leq\frac{C|\Omega|}{\ell},
\label{estim_IMS_rest2}
\end{equation}
where we have used that the localization is done on the scale $\sqrt{\ell}$. Notice also for the strong subadditivity of the entropy, we need to use a generalization to the case of several kinds of particles of different symmetries, as explained in Appendix, Section \ref{sec:SSA_quantum}.

\subsection{Movable classical nuclei: proof of Theorem \ref{thm_mov}}\label{sec_proof_mov}
The proof of Theorem \ref{thm_mov} is similar to that of Theorem \ref{thm_LL} and we shall not give all the details. In particular, for the energy thanks to equality \eqref{Lieb_Daubechies_mov} it suffices to prove the existence of the limit for $\uE(\cdot)$. This avoids some localization problems which were encountered in the crystal case.

Consider some function $\eta(t)=a|t|^b$ with $a>0$ and $b\in(0,1]$. Define like in the previous section the set of regular domains as $\cR=\cR_\eta\cup\{g\ell\triangle,\ g\in G,\ \ell\geq1\}$, where $\triangle$ is the simplex as defined before. Take also $\cR'=\cR$. Clearly \textbf{(P1)}--\textbf{(P5)} are satisfied. Also \textbf{(A1)} holds true by convention. 

It is clear that $E(=\uE)$, $F$ and $\uF$ are all translation-invariant. Hence \textbf{(A3)} is obviously satisfied.
The stability property \textbf{(A2)} was already proved in Theorem \ref{stability_class_nuclei}. 
The continuity property \textbf{(A4)} is easily verified as one has for any $\Omega',\Omega\in\cM$,
$$\Omega'\subseteq\Omega\Longrightarrow E(\Omega)\leq E(\Omega')$$
and a similar property for $F$ and $\uF$.

We turn to the proof of \textbf{(A5)} and \textbf{(A6)}. First we notice that 
\begin{align}
&F(\Omega,\beta,\mu)  \nonumber\\
&= \inf\Bigg\{\sum_{K\geq0}\frac1{K!}\int_{\Omega^K}dR_1\cdots dR_K\,\tr_{\cF(L^2(\Omega))}\bigg[\left(H_{\Omega,\{(R_k,z)\}}-\mu_1\cN-\mu_2K\right)\times  \nonumber\\
& \qquad\qquad\qquad\times\rho_K(R_1,...,R_K)+\frac1\beta\rho_K(R_1,...,R_K)\log\rho_K(R_1,...,R_K)\bigg],\nonumber\\
& \qquad  \rho_K\in L^1_s\left(\Omega^K,S_+(\cF(L^2(\Omega)))\right),\  \sum_{K\geq0}\frac1{K!}\int_{\Omega^K}\tr_{\cF(L^2(\Omega))}\rho_K=1\Bigg\}
\end{align}
where $S_+(\gH)$ denotes the cone of positive semi-definite self-adjoint operators acting on $\gH$ and the subscript $s$ on $L^1_s$ means that we restrict ourselves to symmetric functions.
A similar formula holds for $\uF$. An adequate formalism is provided in Appendix, Section \ref{sec:SSA_quantum_and_classical}, for such functionals.
For $F$, an optimal state $\rho=(\rho_K)$ is given by the ($R$-dependent) Gibbs state
$$\rho_K(R_1,...,R_K)=Z^{-1}\text{exp}\Big[-\beta\left(H_{\Omega,\{(R_k,z),\ k=1..K\}}-\mu_1\cN-\mu_2K\right)\Big],$$ 
$$Z=\sum_{K\geq0}\frac{1}{K!}\int_{\Omega^K}dR_1\cdots dR_K\tr_{\cF(L^2(\Omega))}\left[e^{-\beta\left(H_{\Omega,\{(R_i,z)\}}-\mu_1\cN-\mu_2K\right)}\right].$$
The average number of nuclei is given by the formula
$$\pscal{K\rho}:=\sum_{K\geq1}\frac{1}{(K-1)!}\int_{\Omega^K}\tr_{\cF(L^2(\Omega))}[\rho_K(R_1,...,R_K)]dR_1\cdots dR_K,$$
whereas the average number of electrons is given by
$$\pscal{\cN\rho}:=\sum_{K\geq1}\frac{1}{K!}\int_{\Omega^K}\tr_{\cF(L^2(\Omega))}[\cN\rho_K(R_1,...,R_K)]dR_1\cdots dR_K.$$
The same formulas hold for $\uF$.
Next we need a result similar to Lemma \ref{bound_nb_part_quantum}.
\begin{lemma}[Upper bound on the number of particles]\label{bound_nb_part_mixed} There exists a constant $C'$ such that for any $\Omega\in\cM$ the following holds: any ground state $\Psi\in \bigwedge_1^NL^2(\Omega)$ and any configuration $\{(R_k,z),\ k=1..K\}$ for the nuclei optimizing $E(\Omega)$ satisfy
\begin{equation}
N+\int_\Omega\rho_\Psi^{5/3}+K\leq C'|\Omega|_{\rm r}.
\label{estim_nb_particles_mixed1} 
\end{equation}
Similarly any ground state $\rho$ for $F(\Omega,\beta,\mu)$ or $\uF(\Omega,\beta,\mu)$ satisfies
\begin{equation}
\pscal{\cN^{5/3}\rho}+\pscal{K\rho}\leq C'|\Omega|.
\label{estim_nb_particles_mixed2} 
\end{equation}
\end{lemma}
\begin{proof}
The proof of \eqref{estim_nb_particles_mixed2} is exactly similar to the proof of Lemma \ref{bound_nb_part_quantum}, Equation \eqref{estim_free_energy_chemical_potential}. The estimate on the number of electrons and $\int\rho_\Psi^{5/3}$ is as usual derived by means of the Lieb-Thirring inequality. For \eqref{estim_nb_particles_mixed1}, we use \eqref{stability_of_matter_equation} and $E(\Omega)\leq0$ to get
$$\frac{z^2}{8}\sum_{k=1}^K\frac{1}{\delta_R(R_k)}\leq C|\Omega|$$
Let us denote by $K_1=\#\{k\ |\ \delta_R(R_k)\leq1\}$ and $K_2=K-K_1$. The above estimate shows that $K_1\leq 8C/z^2|\Omega|$. On the other hand we can place a ball of radius $1/2$ at the center $R_k$ of each nucleus which satisfies $\delta_R(R_k)>1$. These balls are disjoint and they all intersect $\Omega$, hence clearly $K_2\leq C|\Omega|_{\rm r}$. This gives \eqref{estim_nb_particles_mixed1}.
\end{proof}

\begin{remark}\rm
Like for the crystal case and contrary to the previous section, the number of nuclei for a ground state of $E(\Omega)$ can only be estimated by a constant times $|\Omega|_{\rm r}$ when $\Omega$ is not a regular set. This is because we do not have any kinetic energy for the nuclei which would have allowed us to use Proposition \ref{lemma:estim_nuclei}.
\end{remark}

Then, using Lemma \ref{bound_nb_part_mixed} and following the proof of the previous section, one can prove that the energy $E$ satisfies \textbf{(A5)} and \textbf{(A6)}. Localization induces a change of the charges of the nuclei which are at a distance $\sqrt{\ell}$ to the boundary of each $g\ell\triangle$ (like for the crystal case), but this is not a problem thanks to equality \eqref{Lieb_Daubechies_mov}, $E=\uE$. For the free energies $F$ or $\uF$, we need a localization method for both the nuclei and the electrons and the associated strong subadditivity of the entropy, as this is explained in details in Appendix, Section \ref{sec:SSA_quantum_and_classical}.

\appendix
\section{Localization of states in Fock spaces and strong subadditivity of entropy}
The purpose of this appendix is to provide an adequate setting for localization in Fock spaces, and use this to state the strong subadditivity of the entropy. 

\subsection{Localization in Fock space}\label{sec:localization}
In this section, we recall the concept of localization in Fock space, as was introduced for bosons by Derezi\'nski and Gérard in \cite{DG} and generalized to fermions by Ammari in \cite{Ammari}. We consider one Hilbert space $\cH$ and denote by $\cF(\cH)$ the associated (fermionic or bosonic) Fock space. 

\subsubsection{Restriction and extension of states}
We will use the important isomorphism between Fock spaces
\begin{equation}
\cF(\cH_1\oplus\cH_2)\simeq\cF(\cH_1)\otimes\cF(\cH_2), 
\label{isometry_Fock_spaces}
\end{equation}
where $\cH_1$ and $\cH_2$ are Hilbert spaces. 
In the fermionic case, the map is simply\footnote{Note we have ordered the tensor product to have all $\phi_i\in\cH_1$ on the left and all $\psi_j\in\cH_2$ on the right.}:
$$
\cF(\cH_1)\otimes\cF(\cH_2)\ni\bigwedge_{i=1}^N\phi_i\otimes\bigwedge_{j=1}^M\psi_j
\mapsto\bigwedge_{i=1}^N\phi_i\wedge\bigwedge_{j=1}^M\psi_j\in\cF(\cH_1\oplus\cH_2).
$$
It is similarly defined in the bosonic case.

Given a state $\omega$ on $\cF(\cH_1\oplus\cH_2)$ we can define the 
{\it restriction} $\omega_1$ to $\cF(\cH_1)$ by the partial trace
i.e., 
$$
\omega_1(A)=\omega(A\otimes1_{\cF(\cH_2)}),
$$
for bounded operators $A$ on $\cF(\cH_1)$.
If $\omega$ is given by a density matrix $\Gamma$, i.e. $\omega(B)=\tr_{\cF(\cH_1\oplus\cH_2)}(\Gamma B)$, then the density matrix $\Gamma_1$ of $\omega_1$ is obtained by taking the partial trace of $\Gamma$, $\Gamma_1=\tr_{\cF(\cH_2)}(\Gamma)$.

Conversely, given a state $\omega_1$ on $\cF(\cH_1)$
we can {\it extend} it to a state $\omega$ on $\cF(\cH_1\oplus\cH_2)$ by
$$\omega(A) =\omega_1(\cI_1^*A\cI_1),$$ 
for all bounded operators $A$ on $\cF(\cH_1\oplus\cH_2)$. Here
$\cI_1:\cF(\cH_1)\to\cF(\cH_1\oplus\cH_2)=\cF(\cH_1)\otimes\cF(\cH_2)$ 
is the inclusion map $\cI_1(\phi)=\phi\otimes|0\rangle_{\cF(\cH_2)}$,
where $|0\rangle_{\cF(\cH_2)}$ is the vacuum state on $\cF(\cH_2)$.
Alternatively, $\cI_1$ is the second quantization of the inclusion map
$i_1:\cH_1\to\cH_1\oplus\cH_2$, i.e., $\cI_1=\Upsilon(i_1)$.
We say that a state on $\cF(\cH_1\oplus\cH_2)$ lives over 
$\cH_1$ if it is equal to the extension of its restriction to $\cF(\cH_1)$.

\subsubsection{Localization of states}
Consider now an operator $0\leq q\leq1$ on the Hilbert space $\cH$. We may
define a partial isometry 
$$
Q:\cH\ni\phi\mapsto q\phi\oplus(1-q^2)^{1/2}\phi\in\cH\oplus\cH.
$$ 
Using that $\cF(\cH\oplus\cH)\simeq\cF(\cH)\otimes\cF(\cH)$ we may lift this isometry to a partial isometry of Fock spaces
$$
\Upsilon(Q):\cF(\cH)\to\cF(\cH)\otimes\cF(\cH)
$$
where $\Upsilon(Q)$ is a notation for the second quantization of $Q$. Note it satisfies $\Upsilon(Q)^*\Upsilon(Q)=1_{\cF(\cH)}$.

Let us denote by $a^\dagger(f)$ the usual creation operator acting on $\cF(\cH)$ which creates a particle in the state $f\in\cH$. Similarly, we may define
$$c^\dagger(f)=a^\dagger(f)\otimes 1_{\cF(\cH)}\ \text{ and }\ d^\dagger(f)=(-1_{\cF(\cH)})^{\epsilon\cN}\otimes a^\dagger(f)$$ 
which are creation operators on $\cF(\cH)\otimes\cF(\cH)$, with $\epsilon=0$ for bosons \cite{DG} and $\epsilon=1$ for fermions \cite{Ammari}. 

It can be shown that the operator $\Upsilon(Q)$ satisfies the following \emph{intertwinning properties} (see Lemma 2.14 and 2.15 in \cite{DG})
\begin{equation}
\Upsilon(Q)a^\dagger(f)=\Big(c^\dagger(qf)+d^\dagger\left(\sqrt{1-q^2}f\right)\Big)\Upsilon(Q),
\label{intertwinning1}
\end{equation}
\begin{equation}
\Upsilon(Q)a(f)=\Big(c(qf)+d\left(\sqrt{1-q^2}f\right)\Big)\Upsilon(Q),
\label{intertwinning2}
\end{equation}
\begin{equation}
\Upsilon(Q)a(qf)=c(f)\Upsilon(Q),\quad \Upsilon(Q)a\left(\sqrt{1-q^2}f\right)=d(f)\Upsilon(Q),
\label{intertwinning3}
\end{equation}
\begin{equation}
a^\dagger(qf)\Upsilon(Q)^*=\Upsilon(Q)^*c^\dagger(f),\quad a^\dagger\left(\sqrt{1-q^2}f\right)\Upsilon(Q)^*=\Upsilon(Q)^*d^\dagger(f).
\label{intertwinning4}
\end{equation}
Notice if $|\Psi\rangle$ is a pure state in $\cF(\cH)$ which is a linear combination of products of creation operators acting on the vacuum, then \eqref{intertwinning1} shows that $\Upsilon(Q)|\Psi\rangle$ is simply obtained by replacing $a^\dagger(f)$ by $c^\dagger(qf)+d^\dagger\left((1-q^2)^{1/2}f\right)$ everywhere, and of course the vacuum of $\cF(\cH)$ by the corresponding vacuum in $\cF(\cH)\otimes\cF(\cH)$. For bosons $c^\dagger(f)$ and $d^\dagger(g)$ commute for any $f$ and $g$. For fermions, one may use that they anticommute to lift all the $d^\dagger$ to the right and use that $(-1_{\cF(\cH)})^{\epsilon\cN}|0\rangle=|0\rangle$.

If $\omega$ is a general state on $\cF(\cH)$ we define the \emph{$q$-extension} $\tilde\omega_q$ of $\omega$ to $\cF(\cH)\otimes\cF(\cH)$ by
$$
\tilde\omega_q(B)=\omega(\Upsilon(Q)^* B\Upsilon(Q)).
$$
When $\omega$ arises from a density matrix $\Gamma$ which is expressed as a linear combination of terms of the form
$$a^\dagger(f_{i_1})\cdots a^\dagger(f_{i_M})|0\rangle\langle0|a(f_{j_1})\cdots a(f_{j_{M'}})$$
(where for shortness we have denoted by $|0\rangle$ the vacuum in ${\cF(\cH)}$), then the density matrix $\tilde\Gamma_q$ of $\tilde{\omega}_q$ is obtained by simply replacing $a^\dagger(f)$ by $c^\dagger(qf)+d^\dagger\left((1-q^2)^{1/2}f\right)$ and $a(f)$ by $c(qf)+d\left((1-q^2)^{1/2}f\right)$ everywhere as before.

Next we define the \emph{$q$-localized state} $\omega_q$ on $\cF(\cH)$ as the restriction of $\tilde{\omega}_q$ to the first Hilbert space:
$$
\omega_q(A)=\tilde{\omega}_q(A\otimes1_{\cF(\cH)})=\omega(\Upsilon(Q)^* A\otimes1_{\cF(\cH)}\Upsilon(Q)),
$$
for all bounded operators $A$ on $\cF(\cH)$. Notice $\omega_q$ is really a state, i.e. is it positive-semidefinite and normalized:
$$\omega_q(1)=\omega(\Upsilon(Q)^*\Upsilon(Q))=\omega(1)=1.$$
In fact, $\omega_q$ lives over ${\rm Ran}(q)$. 
If $\omega$ is given by a density matrix $\Gamma$, then $\omega_q$ is also given by a density matrix which is the partial trace of $\tilde{\Gamma}_q$:
$\Gamma_q:=\tr_2(\tilde{\Gamma}_q)$ (the subscript 2 means that we take the partial trace with respect to the second Fock space).
As $\Upsilon(Q)^*\Upsilon(Q)=1$ we deduce using \eqref{intertwinning3} and \eqref{intertwinning4} that
\begin{align*}
\omega_q\left(a^\dagger(f)a(g)\right)&=\omega(\Upsilon(Q)^* a^\dagger(f)a(g)\otimes1_{\cF(\cH)}\Upsilon(Q))\\
&=\omega(\Upsilon(Q)^* c^\dagger(f)c(g)\Upsilon(Q)) =\omega(a^\dagger(qf)a(qg))
\end{align*}
which proves that the one-body density matrix $\gamma^{(1)}_{\omega_q}$ of ${\omega_q}$ is given by 
\begin{equation}
\gamma^{(1)}_{\omega_q}=q\gamma^{(1)}_\omega q. 
\label{1b_density_matrix_localization}
\end{equation}
The same argument shows that all the $k$-body density matrices of $\omega_q$ are given by
$$
\gamma^{(k)}_{\omega_q}=q^{\otimes k} \gamma^{(k)}_\omega q^{\otimes k}.
$$
In particular, this easily implies that $\omega_q$ is a quasi-free state when $\omega$ is a quasi-free state.
If we take $q=0$, then we find that $\omega_0$ is just the vacuum state in $\cF(\cH)$.
When $\cH=\cH_1\oplus\cH_2$ and $q$ is the orthogonal projector onto $\cH_1$, then $\omega_{q}$ is just the restriction $\omega_1$ of $\omega$ as defined above.

\subsection{Strong subadditivity of entropy}\label{sec:SSA}
\subsubsection{Quantum particles}\label{sec:SSA_quantum}
\paragraph*{One species of quantum particles.}
We consider one Hilbert space $\cH$ and the associated (fermionic or bosonic) Fock space $\cF(\cH)$.
Let $(q_i)_{i\in I}$ be a countable family of commuting positive-semidefinite operators on $\cH$ such that 
$$
\sum_{i\in I} (q_i)^2=1_{\cH}.
$$ 
Consider now a fixed state $\omega$ on the Fock space $\cF(\cH)$, which is assumed to be given from a density matrix $0\leq\Gamma\leq1$ acting on $\cF(\cH)$: $\omega(A)=\tr_{\cF(\cH)}(\Gamma A)$. Its entropy is assumed to be finite:
$$S(\omega):= -\tr_{\cF(\cH)}(\Gamma\log\Gamma)<\ii.$$
For any finite set $\cP\subset I$, we introduce the operator
\begin{equation}
 q_\cP:=\left(\sum_{i\in\cP} (q_i)^2\right)^{1/2}
\label{def_q_P}
\end{equation}
with the convention that $q_\emptyset=0$, 
and denote by $\omega_\cP:=\omega_{q_\cP}$ the $q_\cP$-localized state as defined in the previous section. It arises from some density matrix $0\leq\Gamma_\cP\leq 1$, hence we may denote by $S(\omega_\cP)=-\tr_{\cF(\cH)}(\Gamma_\cP\log\Gamma_\cP)$ its (possibly infinite) entropy.
The main result of this section is
\begin{prop}[Strong subadditivity of entropy]\label{prop_strong_subadd} Let $\cP_1$, $\cP_2$ and $\cP_3$ be disjoint finite subsets of $I$. Then
\begin{equation}
S(\omega_{\cP_1\cup\cP_2\cup\cP_3})+ S(\omega_{\cP_2})\leq S(\omega_{\cP_1\cup\cP_2})+S(\omega_{\cP_2\cup\cP_3}).
\label{strong_subadd} 
\end{equation}
\end{prop}

When the entropy of subsystems is defined via partial traces, strong subadditivity of the quantum entropy was proved for the first time by Lieb and Ruskai \cite{LR1,LR2}. Proposition \ref{prop_strong_subadd} is a consequence of this result, as we shall see.

\begin{remark}\rm If $\cP_2=\emptyset$, then $q_{\cP_2}=0$ and $\omega_{\cP_2}$ is just the vacuum state in $\cF(\cH)$, whose entropy vanishes. Hence, inserting this in \eqref{strong_subadd}, one obtains the subadditivity of the entropy
\begin{equation}
S(\omega_{\cP_1\cup\cP_3})\leq S(\omega_{\cP_1})+S(\omega_{\cP_3}).
\label{subadd} 
\end{equation}
\end{remark}

\begin{proof}[Proof of Proposition \ref{prop_strong_subadd}]
We introduce
$$Q_\cP:\cH\ni\phi\mapsto q_\cP\phi\oplus(1-q_\cP^2)^{1/2}\phi\in\cH\oplus\cH$$
and $\Upsilon(Q_\cP):\cF(\cH)\to \cF(\cH)\otimes\cF(\cH)$.
Now, consider $\cP_1$, $\cP_2$ and $\cP_3$ as in the statement of the proposition and introduce the operator
$$\begin{array}{rcl}
Q:\ \cH & \longrightarrow & \bigoplus_1^4\cH\\
\phi & \longmapsto & q_{\cP_1}\phi\oplus q_{\cP_2}\phi\oplus q_{\cP_3}\phi\oplus q_{\cP_4}\phi
\end{array}
$$
with $\cP_4=I\setminus(\cP_1\cup\cP_2\cup\cP_3)$, and its second-quantization $\Upsilon(Q):\cF(\cH)\to\bigotimes_1^4\cF(\cH)$. 
Let us define as before the extension $\hat\omega$ of $\omega$ by
$$\hat{\omega}(B)=\omega\left(\Upsilon(Q)^*B\Upsilon(Q)\right).$$
Now we define states $\hat\omega_{123}$, $\hat\omega_{2}$, $\hat\omega_{12}$ and $\hat\omega_{23}$, where $\hat\omega_P$ acts on $\bigotimes_1^{\#P}\cF(\cH)$ as the partial traces of $\hat\omega$:
$$\hat\omega_{123}(A_3)=\hat{\omega}(A_3\otimes 1_{\cF(\cH)})=\omega\left(\Upsilon(Q)^*A_3\otimes 1_{\cF(\cH)}\Upsilon(Q)\right),$$
$$\hat\omega_{2}(A_1)=\hat\omega\left(1_{\cF(\cH)}\otimes A_1\otimes 1_{\cF(\cH)}\otimes 1_{\cF(\cH)}\right),$$
$$\hat\omega_{12}(A_2)=\hat\omega\left(A_2\otimes 1_{\cF(\cH)}\otimes 1_{\cF(\cH)}\right),$$
$$\hat\omega_{23}(A_2)=\hat\omega\left(1_{\cF(\cH)}\otimes A_2\otimes 1_{\cF(\cH)}\right).$$
We have used the convention that $A_j$ acts on $\cF(\cH)^{\otimes j}$. By \cite[Thm 2]{LR2}, we know that
$$S(\hat\omega_{123})+S(\hat\omega_{2})\leq S(\hat\omega_{12})+S(\hat\omega_{23}).$$
It rests to prove that\footnote{Recall that the $q_\cP$-localized state $\omega_\cP:=\omega_{q_\cP}$ was introduced in the previous section. It is defined by considering an appropriate extension of $\omega$ over $\cF(\cH)\otimes\cF(\cH)$ and then taking its restriction to the first component. Here we have defined an extension $\hat{\omega}$ of $\omega$ over $\otimes_1^4\cF(\cH).$} $S(\hat\omega_{123})=S(\omega_{\cP_1\cup\cP_2\cup\cP_3})$, $S(\hat\omega_{2})=S(\omega_{\cP_2})$, 
$S(\hat\omega_{12})=S(\omega_{\cP_1\cup\cP_2})$ and $S(\hat\omega_{23})=S(\omega_{\cP_2\cup\cP_3})$. This will clearly show \eqref{strong_subadd}. 
Indeed, we only give the proof that 
\begin{equation}
S(\hat\omega_{123})=S(\omega_{\cP_1\cup\cP_2\cup\cP_3}),
\label{equality_entropy} 
\end{equation}
the other three equalities being proved in the same way.
Let us introduce the following operator:
$$\begin{array}{rcl}
\hat Q:\ \cH & \longrightarrow & \bigoplus_1^3\cH\\
\phi & \longmapsto & \frac{q_{\cP_1}}{q_\cP}\phi\oplus \frac{q_{\cP_2}}{q_\cP}\phi\oplus \frac{q_{\cP_3}}{q_\cP}\phi
\end{array}
$$
with $\cP=\cP_1\cup\cP_2\cup\cP_3$, i.e. $q_\cP=(q_{\cP_1}^2+q_{\cP_2}^2+q_{\cP_3}^2)^{1/2}$. Observe that $\hat Q$ is a (not necessarily injective) partial isometry. Note that
$$Q=(\hat Q\oplus 1_\cH)Q_\cP.$$
Thus
$$\Upsilon(Q)=\left(\Upsilon(\hat Q)\otimes 1_{\cF(\cH)}\right)\Upsilon(Q_\cP)$$
and we have
$$\hat\omega_{123}(A)=\omega_{\cP_1\cup\cP_2\cup\cP_3}\left(\Upsilon(\hat Q)^*A\Upsilon(\hat Q)\right).$$
Equality \eqref{equality_entropy} is then a consequence of the following simple lemma:
\begin{lemma}
Let $U:\cH_1\to\cH_2$ be a partial isometry between Hilbert spaces and $\Gamma\geq 0$ be a self-adjoint operator on $\cH_1$ with
  $\tr_{\cH_1}\Gamma=\tr_{\cH_2}U\Gamma U^*$. If $f$ is a continuous real valued function
  with $f(0)=0$ then
  $\tr_{\cH_1}f(\Gamma)=\tr_{\cH_2}(f(U\Gamma U^*))$ if $f(\Gamma)$ is
  traceclass.
\end{lemma}
\begin{proof}
Recall that $U$ is a unitary map from $U^*(\cH_2)$ to
$U(\cH_1)$.
Since $\Gamma\geq0$ and $\tr_{\cH_1}\Gamma=\tr_{\cH_2}U\Gamma U^*$ we
conclude that $\Gamma$ vanishes on the orthogonal complement
of $U^*(\cH_2)$. 
Moreover, $U\Gamma U^*$ vanishes on the orthogonal complement
of $U(\cH_1)$. Thus
$$\tr_{\cH_1}f(\Gamma)=\tr_{U^*(\cH_2)}f(\Gamma)=\tr_{U(\cH_1)}(f(U\Gamma
U^*))=\tr_{\cH_2}(f(U\Gamma U^*)).
$$
\end{proof}
This ends the proof of Proposition \ref{prop_strong_subadd}.
\end{proof}

\paragraph*{Several species of quantum particles.} The previous analysis is easily extended to the case of several species of particles. This is indeed an obvious application of the one-species case when all the particles share the same symmetry, thanks to \eqref{isometry_Fock_spaces}.

If we have two different kinds of particles, the Fock space has the form $\cF(\cH_1)\otimes\cF(\cH_2)$.
We can then define for any state $\omega$ and any operators $q^1\in\cB(\cH_1)$, $q^2\in\cB(\cH_2)$ with $0\leq q^1,q^2\leq 1$ the $\omega_{q^1,q^2}$ localized state similarly as above. The strong subadditivity is then expressed in the same way, taking this time two families $(q^1_i)_{i\in I}$ and $(q^2_i)_{i\in I}$. We let the details to the reader.

\subsubsection{Classical and quantum particles}\label{sec:SSA_quantum_and_classical}
We consider a mixed model containing both classical and quantum particles. We have in mind the case of quantum electrons and classical nuclei, hence we treat for simplicity a model with only one species of each kind.

Let $\cH$ be a Hilbert space and $\cF(\cH)$ the associated (fermionic or bosonic) Fock space. Let $\Omega$ be a bounded subset of $\R^N$, which is the configuration space for the classical particles.
A \emph{state} is a sequence $\rho=(\rho_K)_{K\geq0}$ where each $\rho_K\in L^1_s(\Omega^K,S^+(\cF(\cH)))$. Here $S^+(\gH)$ denotes the cone of self-adjoint positive semi-definite operators acting on a Hilbert space $\gH$ and the subscript $s$ on $L^1_s$ means that we restrict ourselves to symmetric functions. Additionally, we normalize our states by \cite{Ruelle-Robinson,Wehrl} 
$$\int\rho:= \tr_{\cF(\cH)}(\rho_0)+\sum_{K\geq1}\frac1{K!}\int_{\Omega^K} \tr_{\cF(\cH)}\left(\rho_K(x_1,...,x_K)\right) dx_1\cdots dx_K=1.$$
The entropy of such a state is defined by
\begin{multline*}
S(\rho):=-\tr_{\cF(\cH)}(\rho_0\log\rho_0)\\
-\sum_{K\geq1}\frac1{K!}\int_{\Omega^K} \tr_{\cF(\cH)}\left(\rho_K(x_1,...,x_K)\log\rho_K(x_1,...,x_K)\right) dx_1\cdots dx_K.
\end{multline*}
Let $q$ be a self-adjoint operator acting on $\cH$ such that $0\leq q\leq 1$ and $\theta$ be a real function defined on $\Omega$ such that $0\leq\theta\leq1$. We define the $(q,\theta)$-localized state $\rho_{q,\theta}$ as follows \cite{Ruelle-Robinson}:
\begin{multline*}
\left(\rho_{q,\theta}\right)_K(x_1,...,x_K):=\theta(x_1)^2\cdots\theta(x_K)^2\Bigg(\rho_{K}(x_1,...,x_K)_q+\\
+\sum_{M\geq1}\frac{1}{M!}\int_{\Omega^M}\eta(y_1)^2\cdots\eta(y_M)^2\rho_{K+M}(x_1,...,x_K,y_1,...,y_M)_q\,dy_1\cdots dy_M\Bigg)
\end{multline*}
where $\eta=\sqrt{1-\theta^2}$. We recall that the notation $\gamma_q$ is used for the $q$-localized state of $\gamma$ defined in the Fock space $\cF(\cH)$ as explained in Section \ref{sec:localization}.

Next we state the strong subadditivity of the entropy. Let $(q_i)_{i\in I}$ be a countable family of commuting positive-semidefinite operators on $\cH$ such that $\sum_{i\in I} (q_i)^2=1_{\cH}$, and $(\theta_i)_{i\in I}$ a partition of unity on $\Omega$, i.e.  such that $\sum_{i\in I}\theta_i^2=1$. For any finite set $\cP\subset I$, we introduce as before
\begin{equation}
 q_\cP:=\left(\sum_{i\in\cP} (q_i)^2\right)^{1/2},\qquad \theta_\cP:=\left(\sum_{i\in\cP} (\theta_i)^2\right)^{1/2}
\label{def_q_theta_P}
\end{equation}
with the convention that $q_\emptyset=0=\theta_\emptyset$. We also define the associated localized state $\rho_\cP:=\rho_{q_\cP,\theta_\cP}$.

\begin{prop}[Strong subadditivity of entropy for classical and quantum particles]\label{prop_strong_subadd_class_quant} Let $\cP_1$, $\cP_2$ and $\cP_3$ be disjoint finite subsets of $I$. Then
\begin{equation}
S(\rho_{\cP_1\cup\cP_2\cup\cP_3})+ S(\rho_{\cP_2})\leq S(\rho_{\cP_1\cup\cP_2})+S(\rho_{\cP_2\cup\cP_3}).
\label{strong_subadd_2} 
\end{equation}
\end{prop}
Notice this result contains both the purely classical case (as was studied by Robinson and Ruelle in \cite{Ruelle-Robinson}, assuming that all the $\theta_i$ are characteristic functions) and the purely quantum case treated in Section \ref{sec:SSA_quantum}. Indeed the proof relies on the purely quantum case, as we shall explain.

\begin{proof}
As the entropy $S$ is continuous for the norm  $$\norm{\rho-\rho'}=\sum_{K\geq0}\frac1{K!}\norm{\rho_K-\rho'_K}_{L^1(\Omega^K,S_+(\cF(\cH)))}$$
(we use here that $\Omega$ is a bounded set), we may assume by density that $\rho_K\equiv0$ for $K\geq K_0$, that each $\rho_K$ is smooth and satisfies $\text{d}(\text{supp}(\rho_K),\partial\Omega^K)\geq\delta$ for some $\delta>0$ and all $K=1,...,K_0$. Also we may assume that
$$\forall K\geq 2,\ \forall (x,x_k)\in\Omega^{K-1}, \quad \rho_K(x,x,x_3,...,x_K)=0.$$
Similarly we may assume by density that the localizing classical functions $\theta_{\cP}$ for $\cP=\cP_1,\cP_2,\cP_3$ are smooth.

Now we introduce a quantum model approximating the classical particles in an appropriate sense. We consider the grid $\cX_\ell:=\ell^{-1}\Z^3\cap \Omega$ which contains $N_\ell$ points. For simplicity, we denote them by $\{X_j\}_{j=1}^{N_\ell}=\cX_\ell$. The associated Hilbert space $V_\ell$ is just the space of functions taking complex values at these points, i.e. $V_\ell\simeq\C^{N_\ell}$. As each $\rho_K$ is symmetric and vanishes when two particles are in the same place, we may consider the associated Fock space $\cF(V_\ell)$ with either the bosonic or the fermionic symmetry. We take the bosonic one for simplicity. 
We associate to our state $\rho$ a state $\Gamma^\ell$ in the Fock space $\cF(\cH)\otimes\cF(V_\ell)$ defined by
\begin{multline}
\Gamma^\ell:= \frac1{t_\ell}\Bigg(\rho_0\otimes|0\rangle\langle0|+\sum_{K=1}^{K_0}\sum_{i_1<\cdots< i_K} \frac{1}{\ell^{N_\ell K}}\rho_K(X_{i_1},...,X_{i_K})\otimes\\
\otimes \left(a^\dagger_{i_1}\cdots a^\dagger_{i_K}|0\rangle\langle0|a_{i_K}\cdots a_{i_1}\right)\Bigg).
\end{multline}
Here $|0\rangle\langle0|$ is the projector on the vacuum state of $\cF(V_\ell)$ and $a^\dagger_j$ is the creation operator of a particle at point $X_j$. The number $t_\ell$ is a normalization factor:
\begin{align*}
t_\ell&=\sum_{K=0}^{K_0}\sum_{i_1<\cdots< i_K} \frac{1}{\ell^{N_\ell K}}\tr\left[\rho_K(X_{i_1},...,X_{i_K})\otimes \left(a^\dagger_{i_1}\cdots a^\dagger_{i_K}|0\rangle\langle0|a_{i_K}\cdots a_{i_1}\right)\right]\\
 &=\sum_{K=0}^{K_0}\sum_{i_1<\cdots< i_K} \frac{1}{\ell^{N_\ell K}}\tr_{\cF(\cH)}\left[\rho_K(X_{i_1},...,X_{i_K})\right],
\end{align*}
with an obvious convention for $K=0$.
We have denoted by `$\tr$' the trace in the full Fock space ${\cF(\cH)\otimes\cF(V_\ell)}$. The above formula is a Riemann sum which
converges to
$$\lim_{\ell\to\ii}t_\ell=\tr_{\cF(\cH)}\rho_0+\sum_{K=1}^{K_0}\frac1{K!}\int_{\Omega^K}\tr_{\cF(\cH)}\left[\rho_K(x_1,...,x_K)\right]dx_1\cdots dx_K=1.$$
Similarly, we may use that the projectors $a^\dagger_{i_1}\cdots a^\dagger_{i_K}|0\rangle\langle0|a_{i_K}\cdots a_{i_1}$ are orthogonal for different sets of indices $\{i_k\}$ and different $K$'s, and that for any projector $P$ on a Hilbert space, $\tr[(A\otimes P) \log (A\otimes P)]=\tr[A\log(A)]\times \tr[P]$. Denoting by $S_\ell$ the quantum entropy on $\cF(\cH)\otimes\cF(V_\ell)$, we obtain
\begin{align}
S_\ell(\Gamma^\ell)&=-\sum_{K=0}^{K_0}\sum_{i_1<\cdots< i_K} \tr_{\cF(\cH)}\left[\frac{\rho_K(X_{i_1},...,X_{i_K})}{t_\ell\ell^{N_\ell K}}\log\left(\frac{\rho_K(X_{i_1},...,X_{i_K})}{t_\ell\ell^{N_\ell K}}\right)\right]\nonumber\\
& =-\sum_{K=0}^{K_0}\sum_{i_1<\cdots< i_K} \frac{1}{t_\ell\ell^{N_\ell K}}\tr_{\cF(\cH)}\bigg[\rho_K(X_{i_1},...,X_{i_K})\log\rho_K(X_{i_1},...,X_{i_K})\bigg]\nonumber\\
 & \qquad\qquad \qquad\qquad\qquad\qquad+ \log t_\ell+N_\ell\log\ell + \pscal{K\Gamma^\ell}\log\ell\label{entropy_approx_state}
\end{align}
where we have introduced the notation
$$\pscal{K\Gamma^\ell}:=\sum_{K=0}^{K_0}K\sum_{i_1<\cdots< i_K} \frac{\tr_{\cF(\cH)}\left[\rho_K(X_{i_1},...,X_{i_K})\right]}{t_\ell\ell^{N_\ell K}}$$
for the total average number of particles in the Fock space $\cF(V_\ell)$. Formula \eqref{entropy_approx_state} is again a Riemann sum hence
$$S_\ell(\Gamma^\ell)-N_\ell\log\ell - \pscal{K\Gamma^\ell}\log\ell\to_{\ell\to\ii} S(\rho).$$

Next we consider localized states. Let $q$ be an operator acting on $\cH$ with $0\leq q\leq 1$ and $\theta$ a function on $\Omega$ such that $0\leq \theta\leq 1$. We may extend $\theta$ to a multiplication operator acting on $V_\ell$. Next we look at the localized state $\rho_{q,\theta}^\ell$. As explained before, it is obtained by replacing each $a^\dagger_j$ by $\theta(X_j)c^\dagger_j+\sqrt {1-\theta(X_j)^2}d^\dagger_j$ where $c^\dagger_j=a^\dagger_j\otimes 1_{\cF(V_\ell)}$ and $d^\dagger_j= 1_{\cF(V_\ell)}\otimes a^\dagger_j$ which are operators acting on $\cF(V_\ell)\otimes\cF(V_\ell)$, and then taking the partial trace for the $d^\dagger_j$'s operators. Introducing $\eta=\sqrt{1-\theta^2}$ for simplicity, we notice that
\begin{multline*}
\tr_2\left[\prod_{k=1}^K\left(\theta(X_{i_k})c^\dagger_{i_k}+ \eta(X_{i_k})d^\dagger_{i_k}\right)|0'\rangle\langle0'|\prod_{k=1}^K\left(\theta(X_{i_k})c_{i_k}+\eta(X_{i_k})d_{i_k}\right)\right]\\
=\sum_{k=1}^K\sum_{\substack{J\subset I\\ \#J=k}}\prod_{j\in J}\theta(X_{j})^2\prod_{r\in I\setminus J}\eta(X_r)^2 a^\dagger_J|0\rangle\langle0|a_J
\end{multline*}
where $I=\{i_1,...,i_K\}$ and $a_J^\dagger:=a^\dagger_{j_1}\cdots a^\dagger_{j_k}$ when $J=\{j_1<\cdots< j_k\}$. We have used the notation $|0'\rangle=|0\rangle\otimes|0\rangle$ to denote the vacuum state in $\cF(V_\ell)\otimes\cF(V_\ell)$.
Hence we obtain
\begin{equation}
\Gamma^\ell_{q,\theta}:= \frac1{t_\ell}\sum_{K=0}^{K_0}\sum_{i_1<\cdots< i_K} \frac{1}{\ell^{N_\ell K}}f^\ell_K(X_{i_1},...,X_{i_K})\otimes \left(a^\dagger_{i_1}\cdots a^\dagger_{i_K}|0\rangle\langle0|a_{i_K}\cdots a_{i_1}\right)
\end{equation}
where
\begin{multline*}
f^\ell_K(X_{i_1},...,X_{i_K})
=\prod_{k=1}^K\theta(X_{i_k})^2\sum_{M\geq0}\sum_{\substack{j_1<\cdots <j_M\\ j_r\notin \{i_k\}}}\frac{\eta(X_{j_1})^2\cdots\eta(X_{j_M})^2}{\ell^{N_\ell M}}\times\\
\times\rho_{K+M}(X_{i_1},...,X_{i_K},X_{j_1},...,X_{j_M})_q.
\end{multline*}
We notice that since $\rho_K(x,x,...)$ vanishes, we can remove the constraint $j_r\notin \{i_k\}$ in the above sum. As $\rho_K$ is smooth for all $K$, we see that
$$\lim_{\ell\to\ii}\sup_{K=1...K_0}\sup_{Y\in(\cX_\ell)^K}\left|f^\ell_K(Y)-\rho_{q,\theta}(Y)\right|=0.$$
This allows to prove similarly as before that
$$S_\ell(\Gamma_{q,\theta}^\ell)-N_\ell\log\ell - \pscal{K\Gamma_{q,\theta}^\ell}\log\ell\to_{\ell\to\ii} S(\rho_{q,\theta}).$$
Notice by \eqref{1b_density_matrix_localization}
$$\pscal{K\Gamma_{q,\theta}^\ell}=\tr_{V_\ell}\left[\gamma_{\Gamma_{q,\theta}^\ell}^{(1)}\right]=\tr_{V_\ell}\left[\theta^2\gamma_{\Gamma_{q,1}^\ell}^{(1)}\right]$$
where $\gamma_{\Gamma}^{(1)}$ is the one-body density matrix of $\Gamma$. By the strong subadditivity for the quantum case, we get
\begin{align*}
0&\leq \lim_{\ell\to\ii}\left(S_\ell(\rho_{\cP_1\cup\cP_2}^\ell)+S_\ell(\rho_{\cP_2\cup\cP_3}^\ell)-S_\ell(\rho_{\cP_2}^\ell)-S_\ell(\rho_{\cP_1\cup\cP_2\cup\cP_3}^\ell)\right)\\
&=S(\rho_{\cP_1\cup\cP_2})+S(\rho_{\cP_2\cup\cP_3})-S(\rho_{\cP_2})-S(\rho_{\cP_1\cup\cP_2\cup\cP_3})
\end{align*}
which ends the proof of Proposition \ref{prop_strong_subadd_class_quant}.
\end{proof}

\addcontentsline{toc}{section}{References}
\bibliographystyle{amsplain}

\end{document}